\documentclass[a4paper,11pt]{article}
\pdfoutput=1 % if your are submitting a pdflatex (i.e. if you have
\usepackage{jheppubmod} 

\usepackage[T1]{fontenc} 

\usepackage{graphicx} 
\usepackage{physics}
\usepackage{amssymb}
\usepackage{mathrsfs}
\usepackage{hyperref}
\usepackage{xcolor}
\usepackage{soul}
\usepackage{comment}
\usepackage{tikz}
\usepackage{tikz-cd}
\usetikzlibrary{positioning}
\usepackage[shortlabels]{enumitem}

\newcommand{\cB}{{\cal B}}

\newcommand{\cD}{{\cal D}}

\newcommand{\cF}{{\cal F}}

\newcommand{\cO}{{\cal O}}

\newcommand{\be}{\begin{equation}}
\newcommand{\ee}{\end{equation}}
\newcommand{\bea}{\begin{eqnarray}}
\newcommand{\eea}{\end{eqnarray}}
\newcommand{\nn}{\nonumber}
\newcommand{\scri}{\mathscr{I}}

\newcommand{\D}{\mathrm{d}}

\newtheorem{proposition}{Proposition}
\newtheorem{lemma}{Lemma}

\newcommand{\lap}{\boldsymbol{\Delta}}
\renewcommand{\dim}{\Delta}

\newcommand{\pr}{\partial}
\newcommand{\prd}{\partial \cdot}
\title{\boldmath{Flat from AdS: in any even dimension and for any spin}}

\author[a]{Xavier Bekaert,}
\author[b,1]{Andrea Campoleoni,\note{Research Associate of the Fund for Scientific Research – FNRS, Belgium.}}
\author[c,d]{Simon Pekar}
\author[a]{and S.I. Aadharsh Raj}

\affiliation[a]{Institut Denis Poisson, Université de Tours, Université d’Orléans,\\ CNRS, IDP, UMR 7013, Parc de Grandmont, 37200 Tours, France}
\affiliation[b]{Service de Physique de l'Univers, Champs et Gravitation, Universit\'e de Mons -- UMONS,\\
20 place du Parc, 7000 Mons, Belgium}
\affiliation[c]{International School for Advanced Studies (SISSA),\\ Via Bonomea 265, 34136 Trieste, Italy}
\affiliation[d]{Istituto Nazionale di Fisica Nucleare, Sezione di Trieste,\\ Via Valerio 2, 34127 Trieste, Italy}

\emailAdd{xavier.bekaert@univ-tours.fr}
\emailAdd{andrea.campoleoni@umons.ac.be}
\emailAdd{spekar@sissa.it}
\emailAdd{aadharsh.raj@etu.univ-tours.fr}

\abstract{The space of solutions to the free equations of motion for massless fields of arbitrary integer spin in Minkowski spacetime is recovered as a smooth limit of the anti-de Sitter solution space for any even spacetime dimension. The infinite set of boundary data near null infinity that characterise solutions in Minkowski spacetime is obtained from an expansion of the anti-de Sitter source and vev in powers of the cosmological constant. In particular, the source gives rise to the analogue of the gravitational shear tensor, while the vev yields the analogues of the mass and angular-momentum aspects, as well as the subleading infinite tower of boundary data. These identifications are further supported by the branching of the source and vev into representations of the Lorentz algebra identified with the conformal algebra of the celestial sphere.}

\begin{document} 

\maketitle

\flushbottom

%%%%%%%%%%%%%%%%%%%%%%%%%%%%%%%%%%%%%%%%%%%%%%%%%%%%%%%
\section{Introduction and summary}\label{sec:intro}

In the presence of a negative cosmological constant $\Lambda$, 
a metric satisfying vacuum Einstein's equations can be reconstructed asymptotically by fixing the conformal class of the boundary metric (aka the source) and the vacuum expectation value (vev) of the boundary energy-momentum tensor \cite{AST_1985__S131__95_0, Anderson:2004yi, Fefferman:2007rka}. This basic result on the structure of the space of solutions to the equations of motion has far-reaching implications in the AdS/CFT correspondence, see, e.g., \cite{Skenderis:2002wp}.
In the limit of vanishing cosmological constant, Ricci-flat metrics can still be reconstructed asymptotically from boundary data, but their structure differs significantly from that of asymptotically anti-de Sitter (AdS) spacetimes.
Near null infinity, one has to specify a shear tensor encoding gravitational radiation, as well as mass and angular-momentum aspects, together with an infinite tower of functions on the celestial sphere, which are typically subleading in a radial expansion \cite{Bondi:1962px, Sachs:1962wk, Barnich:2010eb}. If the spacetime dimension is greater than four, one can also include overleading pure-gauge contributions, to be interpreted as the soft modes associated with Bondi-Metzner-Sachs (BMS) asymptotic symmetries \cite{Kapec:2015vwa}.

In view of the importance of the source and the vev in the AdS/CFT framework, it is natural to ask whether these differences signal that flat-space holography should rely on a completely different setup, or whether the independent data encoded in the source and the vev simply reorganise themselves in a subtle way in the limit of vanishing cosmological constant. This question fits within the ongoing efforts to exploit the $\Lambda \to 0$ limit of AdS/CFT to define (or at least constrain) flat-space holography (see, e.g., \cite{Pasterski:2021raf, Donnay:2023mrd, Nguyen:2025zhg, Ruzziconi:2026bix} for reviews).
Such a limit has been investigated, for instance, for solution spaces and their asymptotic symmetries \cite{Ciambelli:2018wre, Poole:2018koa, Compere:2019bua, Compere:2020lrt, Fiorucci:2020xto, Campoleoni:2023fug, Ciambelli:2024kre, Arenas-Henriquez:2025rpt}, for boundary operators and correlators \cite{Bagchi:2023fbj, Marotta:2024sce, Alday:2024yyj, Kulp:2024scx, Kulkarni:2025qcx, Adamo:2025bfr}, and for matter fields \cite{Bekaert:2024itn, Berenstein:2025tts}. Most notably, in four dimensions the Bondi shear has been related to the AdS boundary metric \cite{Compere:2019bua, Campoleoni:2023fug, Ciambelli:2024kre, Arenas-Henriquez:2025rpt}, while mass and angular-momentum aspects, as well as the infinite tower of independent subleading data, have been shown to originate from an expansion of the boundary energy-momentum tensor in powers of the cosmological constant \cite{Campoleoni:2023fug}. Here, we consolidate these ideas by studying the $\Lambda \to 0$ limit in any even spacetime dimension and for free massless fields of arbitrary integer spin, whose spaces of solutions exhibit a structure analogous to that recalled above for gravity. The main outcome of our analysis is that, \emph{at least in even spacetime dimensions, all boundary data at null infinity originate from an expansion of the AdS source and vev in powers of the cosmological constant}.\footnote{\label{integerpowers}We restrict our analysis to the self-consistent asymptotic expansion in integer powers of the radial coordinate in Bondi gauge. As a result, we shall not discuss the AdS origin, if any, of the logarithmic terms that one has to include in order to describe all
relevant Ricci-flat metrics near null infinity \cite{osti_5768010, Chrusciel:1993hx, Valiente-Kroon:2002xys, Friedrich:2017cjg, Geiller:2024ryw}.} In the rest of this introduction we shall summarise the details of this identification, while commenting on the subtleties it reveals. Its possible implications for flat-space holography will be discussed in section~\ref{sec:conclusions}. 

\subsection{Protagonists, objectives and strategy}

To establish the dictionary between the two sets of boundary data, let us first briefly recall the standard AdS$_{d+2}$/CFT$_{d+1}$ dictionary for massless fields of integer spin $s>0$.
In AdS$_{d+2}$ there are two possible choices (Dirichlet vs Neumann) of boundary conditions for on-shell fields, which in Poincar\'e coordinates read $\varphi\sim z^{\Delta_\pm-s}\varphi_{_{\Delta_\pm}}$ with $\Delta_+=d+s-1$ and $\Delta_-=2-s$ the respective scaling dimensions of the so-called normalisable vs non-normalisable mode. Here we omitted for simplicity transverse indices to focus on the conformal dimensions of the tensors encoding the boundary data, and we refer to \cite{Mikhailov:2002bp, Metsaev:2008ks, Campoleoni:2016uwr} and to the main body of the text for more details (see also \cite{Berenstein:2025qhb} for a complementary approach).
The boundary tensor $\varphi_{_{\Delta_+}}$ in the boundary condition $\varphi\sim z^{d-1}\varphi_{_{\Delta_+}}$ is interpreted as the \textit{vacuum expectation value} of an operator $\mathcal{O}_{\Delta_+}$, corresponding to a spin-$s$ conformal conserved current, with scaling dimension \mbox{$\Delta_+=d+s-1$} saturating the unitarity bound (which corresponds to the boundary energy-momentum tensor for $s=2$).
As a result, the space of solutions with this 
boundary condition spans a unitary irreducible representation of \mbox{$SO(d+1,2)$}, hence the terminology ``normalisable'' modes.
In the boundary condition $\varphi\sim z^{2-2s}\varphi_{_{\Delta_-}}$, the boundary tensor $\varphi_{{}_{\Delta_-}}$ is interpreted as the \textit{source} coupling to the vev and, accordingly, is a spin-$s$ conformal gauge field (which corresponds to the conformal class of boundary metrics for $s=2$). This space of solutions carries a non-unitary representation (hence the terminology ``non-normalisable'' modes) which, for $d$ even, is irreducible.

To describe the AdS asymptotic solution space, we referred to Poincar\'e coordinates. By introducing a retarded time $u$ on the Minkowski slices at constant $z$ and by rescaling appropriately the coordinates, the metric on the Poincar\'e patch can be cast in the form
\be \label{AdS_Bondi}
\D s^2 = \frac{1}{z^2} \left[ 2\, {\rm d}u {\rm d}z - \frac{1}{R^2}\, {\rm d}u^2 + \delta_{ij} {\rm d} x^i {\rm d} x^j \right] ,
\ee
where $R$ denotes the AdS radius, related to the cosmological constant by $\Lambda = - \frac{d(d+1)}{2R^2}$. This metric manifestly admits an $R \to \infty$ limit, in which one obtains the Minkowski metric in the so-called flat Bondi coordinates (see, e.g., \cite{Alday:2024yyj} for more details). In this limit, null infinity $\mathscr I^+$ is reached by sending \mbox{$z \to 0$} at $u$ constant, so that boundary data near null infinity can be classified according to their $z$ behaviour as in AdS. 
Even if the background metric admits a smooth flat limit, the boundary data organise themselves in a very different way in Minkowski spacetime \cite{Campoleoni:2017qot, Campoleoni:2020ejn}. \emph{Radiative (aka hard) modes} appear at order $z^{\frac{d}2-s}$ in the components of the field along the transverse plane $\mathbb{R}^d$ and span Wigner's spin-$s$ massless unitary irreducible representation of the Poincar\'e group $ISO(d+1,1)$. The complete set of boundary data contains however other contributions which may be subleading as well as overleading in $z$.  
\emph{Soft modes} related to asymptotic symmetries may appear at radiative or overleading orders (until order $z^{2-2s}$) \cite{Campoleoni:2017mbt, Campoleoni:2020ejn}. The temporal components of the field encode instead, at order $z^{d-1}$, Coulombic data generalising the charge or mass and angular-momentum \emph{aspects}, satisfying evolution equations with respect to retarded time. The subleading contributions to the transverse components are similarly governed by evolution equations. Consequently, at every order in the $z$ expansion, arbitrary integration ``constants'' appear (these are actually transverse tensors independent of retarded time). 
Following \cite{Mittal:2022ywl, Campoleoni:2023fug}, we denote these subradiative modes as \emph{Chthonians} (since they arise below the radiative order). 
The system of evolution equations they satisfy has a triangular structure, implying that they correspond to an indecomposable (and, thus, non-unitary) representation of the Poincar\'e group.\footnote{A representation is \textit{indecomposable} if it is reducible but not fully reducible. Any reducible unitary representation is fully reducible, therefore indecomposable representations are always non-unitary. The Poincar\'e group being non semi-simple, its typical representations are indecomposable.}
In \cite{Bekaert:2024tkv}, it was further argued that the Chthonians can be encoded into a single function defined on a bulk null cone and it was shown that, in four dimensions where massless fields with spin $s>0$ exhibit conformal symmetry, the radiative and Chthonian modes are related by an inversion symmetry. 

The question we wish to address is \textit{how the degrees of freedom encoded in the AdS source and vev reorganise themselves in the flat limit}, and in particular, \textit{how to recover the radiative modes corresponding to the massless unitary irreducible representation of the Poincar\'e group?} This turns out to be subtle: 
on the one hand, we find that the aspects originate from the vev while the soft modes originate from the source, as could be expected from the order at which they appear in the $z$ expansion. On the other hand, our analysis (restricted, on this issue, to $d$ even) also shows that, somewhat counter-intuitively, the radiative modes originate from the flat limit of the source. Therefore, one gets the massless unitary representation of the Poincar\'e group out of a non-unitary representation of the AdS isometry group, while the massless unitary representation of the AdS isometry group becomes a non-unitary representation of the Poincar\'e group in the flat limit. 
This is summarised in table~\ref{tableprops}, where, for simplicity, the soft modes and a set of Chthonians originating from the source are referred to as ``others''.\footnote{The role of the ``other'' boundary data is important from a group-theoretical perspective because, strictly speaking, the flat limit of the source is also an indecomposable representation, that becomes irreducible only if one sets all the ``other'' contributions to zero. \label{note1}}  
A more detailed description of the identification is given in section~\ref{sec:summary}.
\begin{center}
\begin{table}[ht]
    \centering
    \renewcommand*{\arraystretch}{1.7}
    \begin{tabular}{c|c||c|c}
  Symmetry algebra & Boundary data & Irreducibility & Unitarity \\\hline\hline
    $\mathfrak{so}(d+1,2)$ & Source (gauge field) & Irreducible & Non-unitary\\[-5pt]
    $\downarrow$ & $\downarrow$ & $\downarrow$ & $\downarrow$\\[-5pt]
    $\mathfrak{iso}(d+1,1)$ & Radiation (+ others)  & Irreducible\textsuperscript{\ref{note1}} & Unitary\\\hline 
    $\mathfrak{so}(d+1,2)$ & Vev (conserved current) & Irreducible & Unitary\\[-5pt]
    $\downarrow$ & $\downarrow$ & $\downarrow$ & $\downarrow$\\[-5pt]
    $\mathfrak{iso}(d+1,1)$ & Aspects + Subleading & Indecomposable & Non-unitary\\\hline
    \end{tabular}
\caption{The flat limit of the boundary data of the two branches of solutions on AdS$_{d+2}$ (with $d$ even), 
and their properties as representations of the isometry group.\label{tableprops}}
\end{table}
\end{center}

How does one compare these so different-looking spaces of solutions?
We shall follow two complementary strategies: 
\begin{enumerate}
    \item \textbf{Flat limit of solutions:}  
    we start from the  
    standard asymptotic expansion of the solutions to the AdS equations of motion in the Poincar\'e patch \eqref{AdS_Bondi}. Following \cite{Campoleoni:2023fug}, we then assume an expansion of the boundary data in a Taylor series in the inverse curvature radius and impose that the poles in $1/R$ induced in the subleading terms in $z$ vanish in the limit $R \to \infty$. The constraints arising from this requirement correspond to the equations of motion in Minkowski spacetime, and guarantee that the information encoded in the source and the vev is redistributed among the resulting constrained Taylor coefficients. In this way, we trace the AdS origin of all boundary data in Minkowski spacetime.
    \item \textbf{Group theory:}  we restrict the AdS and Minkowski isometry groups to their common Lorentz subgroup $SO(d+1,1)$ in order to compare the representations that appear before and after the $R \to \infty$ limit. In CFT language, this means that the primary fields of the AdS isometry group $SO(d+1,2)$ and the conformal Carroll group \mbox{$ISO(d+1,1)$} are decomposed in celestial primary fields of the Lorentz group $SO(d+1,1)$, seen as the conformal group on the $d$-dimensional transverse manifold at conformal infinity (the celestial sphere $S^d$, or in the case of the metric \eqref{AdS_Bondi}, the plane $\mathbb R^d$). In practice, we decompose the representations of $SO(d+1,2)$ in irreducible representations of $SO(d+1,1)$ using the known branching rules, and we check the agreement with the celestial primary fields appearing in the Minkowski solution space. This confirms that our previous way to take the flat limit, involving an expansion of the source and the vev complemented by constraints on the Taylor coefficients, does not lead to any loss or doubling of information. It merely reorganises the independent AdS components in a way that is adapted to the $R \to \infty$ limit.
\end{enumerate}

\subsection{AdS origin of the Minkowski solution space} \label{sec:summary}

We have organised the boundary data in Minkowski spacetime in four groups: radiation, soft modes, aspects and Chthonians. The last three groups comprise however celestial fields ---i.e., boundary tensors independent of retarded time--- with rather different features, which are instrumental to guarantee the matching with the AdS degrees of freedom. 

To summarise the structure of each group of boundary data, let us first better frame the setup that we shall use to study the solution spaces. We describe the dynamics of a field of integer spin $s$ using Fronsdal's equations in AdS \cite{Fronsdal:1978vb}. These admit a gauge symmetry and, to study their solution space, we fix the gauge so as to work with traceless fields. The equations of motion thus reduce to the Maxwell-like form \cite{Skvortsov:2007kz, Campoleoni:2012th}\footnote{The equations of motion of \cite{Campoleoni:2012th} also contain a double divergence of the field in order to project \eqref{Maxwell-like} on its traceless component. On the other hand, one can show that the double divergence vanishes on shell up to polynomial terms that do not affect the propagated degrees of freedom. } 
\be \label{Maxwell-like}
\cF_{\mu(s)} := \Box \varphi_{\mu(s)} - s\, \nabla^\alpha \nabla_{\!\mu} \varphi_{\mu(s-1)\alpha} - \frac{2(s-1)(d+s-1)}{R^2}\,\varphi_{\mu(s)} = 0 \,,
\ee
where $R$ denotes the AdS curvature radius and $\Box := g^{\mu\nu} \nabla_{\!\mu} \nabla_{\!\nu}$. 
An index followed by a label between round brackets denotes a set of $n$ symmetrised indices. Moreover, repeated covariant or contravariant indices denote a symmetrisation, where dividing by the number of terms in the symmetrisation is understood, e.g., $\nabla_{\!\mu} \varphi_{\mu(2)} = \frac{1}{3} \left( \nabla_{\!\mu_1} \varphi_{\mu_2\mu_3} + \nabla_{\!\mu_2} \varphi_{\mu_3\mu_1} + \nabla_{\!\mu_3} \varphi_{\mu_1\mu_2} \right)$. 

More specifically, we work in the Poincar\'e patch with the metric \eqref{AdS_Bondi} and, following \cite{Campoleoni:2017mbt, Campoleoni:2017qot, Campoleoni:2020ejn}, we actually impose even stronger on-shell gauge-fixing conditions greatly simplifying the study of the space of solutions to eq.~\eqref{Maxwell-like}:
\be \label{Bondi-like_intro}
\varphi_{z\mu(s-1)} = 0 \, , \qquad
\delta^{ij} \varphi_{ij\mu(s-2)} = 0 
\qquad \Rightarrow \qquad
g^{\alpha\beta} \varphi_{\alpha\beta\mu(s-2)} = 0 \, .
\ee
For $s=1$ this amounts to working in radial gauge, while for  
$s \geqslant 2$ the number of conditions we are imposing is bigger than the number of independent components of Fronsdal's gauge parameter. For instance, for $s = 2$ eqs.~\eqref{Bondi-like_intro} correspond to the linearisation of the conditions defining the Bondi gauge, plus the additional constraint $h_{uz} = 0$.
As we shall show in section~\ref{sec:bondi-like_gauge}, in linearised gravity this additional constraint can be however imposed by performing an additional gauge-fixing on each solution of the equations of motion. 
For arbitrary values of $s$, we similarly assume that eqs.~\eqref{Bondi-like_intro} are reachable with an on-shell gauge fixing, in analogy with the customary ``TT gauge'' employed for AdS, e.g., in \cite{Mikhailov:2002bp, Campoleoni:2016uwr}.

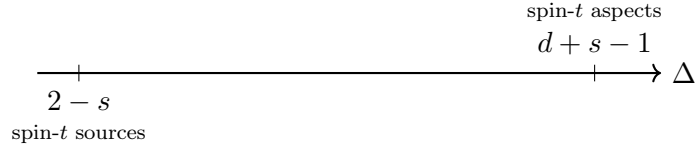
\begin{figure}[h!]
\centering
\begin{tikzpicture}[scale=1.1]
  % Axis
  \draw[->, thick] (-1.5,0) -- (6,0) node[right] {$\Delta$};

  % Major tick positions and labels
  \draw (-1.0,0.1) -- (-1.0,-0.1);
    \node[below] at (-1.0,-0.1) {$2-s$};
  \draw (5.2,0.1) -- (5.2,-0.1);
    \node[above] at (5.2,0.1) {$d+s-1$};
  
  \node[below=11pt] at (-1.0,-0.15)
    {\scriptsize spin-$t$ sources};
  %\node[below=20pt] at (-1.0,-0.15)
  %  {\scriptsize (frozen to zero)};
  \node[above=11pt] at (5.2,+0.15)
    {\scriptsize spin-$t$ aspects};
\end{tikzpicture}
\caption{The celestial field appearing in the asymptotic expansion of the temporal component $\varphi_{u(s-t) i(t)}$
is represented on the right-hand side of the figure. It can be interpreted as the  spin-$t$ celestial field belonging to the spin-$s$ aspect multiplet. On the left-hand side stands the shadow field on the celestial sphere that would couple to it, but that we gauge-fix to zero. 
\label{boring_graph}}
\end{figure}

\begin{figure}[h!] 
\centering
\tikzset{
  myabove/.style={above,yshift=2pt},
  mybelow/.style={below,yshift=-2pt}
  }
\begin{tikzpicture}[scale=1.1]
  % Axis
  \draw[->, thick] (-1.5,0) -- (11.7,0) node[right] {$\Delta$};

  % Major tick positions and labels
  \foreach \x/\label/\pos in {
    -1.0/{$2-s$}/mybelow,
    1.0/{$1$}/mybelow,
    1.4/{$2$}/myabove,
    3.2/{$\tfrac{d}{2} - 1$}/myabove,
    4.0/{$\tfrac{d}{2}$}/mybelow,
    4.8/{$\tfrac{d}{2} + 1$}/myabove,
    6.6/{$d-2$}/myabove,
    7.2/{$d-1$}/mybelow,
    9.2/{$d+s-2$}/mybelow,
    10.2/{$d+s-1$}/myabove,
  }{
    \draw (\x,0.1) -- (\x,-0.1);
    \node[\pos] at (\x,0) {\label};
  }

  % Ellipses between groups
  \node at (0.1,-0.3) {$\cdots$};
  \node at (2.1,+0.3) {$\cdots$};
  \node at (5.8,+0.3) {$\cdots$};
  \node at (8.1,-0.3) {$\cdots$};
  \node at (11.3,+0.25) {$\cdots$};

  % Underbraces to indicate grouped ranges
  \node[below=10pt] at (-0.2,-0.1)
    {$\underbrace{\hspace{2.7cm}}_{\text{celestial gauge fields}}$};
  \node[above=12pt] at (2.5,+0.15)
    {$\overbrace{\hspace{2.7cm}}^{\text{Erdmenger--Osborn fields}}$};
  \node[below=10pt] at (4.0,-0.35)
    {\scriptsize radiation};
  \node[above=12pt] at (5.6,+0.15)
    {$\overbrace{\hspace{3.0cm}}^{\text{source Chthonians}}$};
  \node[below=10pt] at (8.4,-0.1)
    {$\underbrace{\hspace{3.4cm}}_{\text{conserved celestial currents}}$};
  \node[above=12pt] at (10.6,+0.15)
    {$\overbrace{\hspace{2.4cm}}^{\text{vev Chthonians}}$};
\end{tikzpicture}
\caption{The spin-$s$ celestial fields of  scaling dimension $\Delta$ appearing at order $z^{\Delta-s}$
in the asymptotic expansion of the transverse components of a massless field of integer spin $s>0$
in spacetime dimension $d+2$ with $d$ even. \label{nice_graph}}
\end{figure}
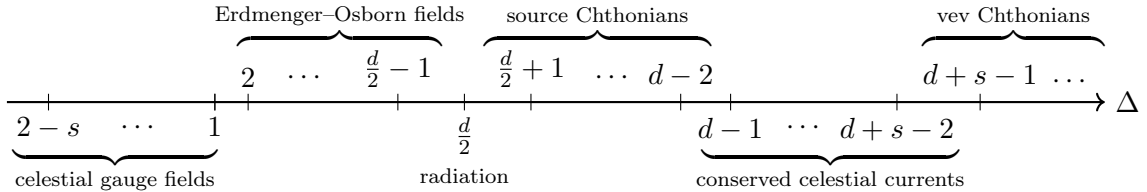

Before imposing the conditions \eqref{Bondi-like_intro}, in AdS the source is encoded in the traceless transverse tensors $\varphi_{u(s-t) i(t)}$ with $0 \leqslant t \leqslant s$ at order $z^{\Delta_- - s}$, while the vev is encoded in the traceless transverse tensors $\varphi_{u(s-t) i(t)}$ with $0 \leqslant t \leqslant s-1$ at order $z^{\Delta_+ - s}$.
The Minkowski solution space can instead be summarised in the following diagrams, representing the new data entering at the order $z^{\Delta-s}$ in the components $\varphi_{u(s-t) i(t)}$
(for $0 \leqslant t \leqslant s-1$) in figure~\ref{boring_graph}, and at various orders in $\varphi_{i(s)}$ in figure~\ref{nice_graph}, respectively.
In these diagrams, the boundary data are organised in \textit{celestial fields} except for the shear tensor encoding radiation, which is the only boundary datum that depends on retarded time. The other boundary data are traceless tensors which live in $d$ dimensions, and on which the Lorentz group $SO(d+1,1)$ acts as conformal transformations.

We now explain the terminology used in figures~\ref{boring_graph} and \ref{nice_graph}, presenting the various celestial fields appearing in the asymptotic expansion of an on-shell massless spin-$s$ gauge field in Minkowski spacetime, and distinguishing them according to their AdS origin.

\subsubsection*{Celestial fields originating from the AdS vev}

\begin{enumerate}[(a)]
        \item The temporal components $\varphi_{u(s-t) i(t)}$
        for $0 \leqslant t \leqslant s-1$ contain spin-$t$ celestial primary fields appearing at the Coulombic order $z^{d-1}$. For a fixed value of the bulk spin $s$, they form a finite collection of celestial primary fields with equal scaling dimension \mbox{$\Delta=s+d-1$} and spin $t$ ranging from $0$ to $s-1$; together they span an indecomposable Poincar\'e representation that we %one may 
        call \textit{spin-$s$ aspect multiplet} since, for $s=2$, they correspond to the mass and angular momentum aspects at a given cut of $\scri^+$.
        \item The transverse components $\varphi_{i(s)}$
        contain spin-$s$ celestial primary fields appearing at orders $z^n$ with $n\geqslant d-1$ (Coulombic and below). For a fixed value of the bulk spin $s$, this is an infinite tower of spin-$s$ celestial fields with growing scaling dimension $\Delta_n=s+n$ (with $n\geqslant d-1$). They span an indecomposable Poincar\'e representation that we refer to as 
        \textit{vev Chthonians}.\footnote{In \cite{Mittal:2022ywl, Campoleoni:2023fug}, the term ``Chthonians'' refers to certain $u$-dependent coefficients in the asymptotic radial expansion subject to evolution equations in retarded time, in analogy with the standard nomenclature for the aspects. Here, we shall often use the same term to denote instead the celestial fields corresponding to the values of these Chthonians on a given cut of $\scri^+$.} They are represented in the portion of figure~\ref{nice_graph} to the right of $\Delta = d-1$.
\end{enumerate}

\subsubsection*{Celestial fields originating from the AdS source}

The temporal components $\varphi_{u(s-t)i(t)}$ between the orders $z^{2-2s}$ and $z^{1-s-t}$ can be gauge-fixed to zero, profiting from the gauge symmetry of the conformal spin-$s$ field playing the role of the source. Without this gauge fixing that we shall exploit in the following, they would have given the spin-$t$ sources in figure~\ref{boring_graph} (with $0 \leqslant t \leqslant s-1$) which are the shadow fields of the aspects.  
The transverse components $\varphi_{i(s)}$ contain instead spin-$s$ celestial primary fields represented in figure~\ref{nice_graph} and appearing at 
        \begin{enumerate}[(a),resume]
            \item \emph{Orders from $z^{2-2s}$ till $z^{1-s}$ (Soft modes):}  They form a finite tower of \textit{celestial gauge fields} which satisfy differential equations that are invariant under gauge transformations generalising to arbitrary spin super-translations and super-rotations \cite{Campoleoni:2017mbt, Campoleoni:2020ejn}, and that read here $\delta h_{i_1\cdots i_s}=
            \partial_{\langle i_1}\cdots\partial_{i_{t}}\varepsilon_{i_{t+1}\cdots i_s\rangle}$ with $0 \leqslant t \leqslant s-1$. Under some global regularity conditions on the corresponding fields defined on the celestial sphere $S^d$, one can show that they must be pure-gauge. 
            \item \emph{Orders from $z^{2-s}$ till $z^{\frac{d}2-s-1}$ (Over-radiative modes):} They form a finite tower of celestial primary fields of spin $s$ and scaling dimension $\Delta$ from $2$ to $\frac{d}2-1$ obeying conformally-invariant equations of order from $d-2$ to $2$, respectively. We call them \textit{Erdmenger--Osborn fields} because these equations are the generalisation and Riemannian analogue of the fields first discussed in \cite{Erdmenger:1997wy}. Under some global regularity conditions for the corresponding fields defined on the celestial sphere, one can show that they must vanish.
            \item \emph{Order $z^{\frac{d}2-s}$ (Radiation):} There is an infinite tower of celestial primary fields that can be reorganised as the Taylor coefficients for the power series in retarded time $u$ of a single Carrollian primary field living at null infinity, which receives the natural interpretation of \textit{radiative data}. 
            This Carrollian primary field at null infinity carries a massless unitary irreducible representation of the Poincar\'e group.
            \item \emph{Orders from $z^{\frac{d}2-s+1}$ till $z^{d-2-s}$ (Sub-radiative modes):} They form a finite tower of celestial primary fields with growing scaling dimension from $\frac{d}2-1$ till $d-2$. We call them \textit{source Chthonians} since they originate from the source and are below radiation.
            \item \emph{Orders from $z^{d-1-s}$ till $z^{d-2}$ (Over-Coulombic modes):} They form a tower of \textit{celestial conserved currents}, with partial conservation law \mbox{$\partial_{i_1}\cdots\partial_{i_t}j^{i_1\cdots i_s}=0$} with $1 \leqslant t \leqslant s$. 
    \end{enumerate}

\subsubsection*{Four spacetime dimensions} 

\begin{figure}[h!]
\centering
\tikzset{
  myabove/.style={above,yshift=2pt},
  mybelow/.style={below,yshift=-2pt}
  }
\begin{tikzpicture}[scale=1.1]
  % Axis
  \draw[->, thick] (-1.5,0) -- (6,0) node[right] {$\Delta$};

  % Major tick positions and labels
  \foreach \x/\label/\pos in {
    -1.0/{$2-s$}/mybelow,
    1.0/{$1$}/mybelow,
    3.0/{$s$}/mybelow,
    4.0/{$s+1$}/myabove,
  }{
    \draw (\x,0.1) -- (\x,-0.1);
    \node[\pos] at (\x,0) {\label};
  }

  % Ellipses between groups
  \node at (0.1,-0.3) {$\cdots$};
  \node at (2.1,-0.3) {$\cdots$};
  \node at (5.3,+0.25) {$\cdots$};

  % Underbraces to indicate grouped ranges
  \node[below=10pt] at (-0.2,-0.1)
    {$\underbrace{\hspace{2.7cm}}_{\text{celestial gauge}}$};
  \node[below=10pt] at (-0.2,-0.65)
    {\scriptsize modes (memory)};
  \node[above=8pt] at (1.0,-0.05)
    {\scriptsize radiation};
  \node[below=10pt] at (2.1,-0.1)
    {$\underbrace{\hspace{2.4cm}}_{\text{conserved currents}}$};
  \node[below=10pt] at (2.1,-0.65)
    {\scriptsize (celestial)};
  \node[above=10pt] at (4.6,+0.15)
    {$\overbrace{\hspace{2.4cm}}^{\text{Chthonians}}$};
\end{tikzpicture}
\caption{The spin-$s$ celestial fields of  scaling dimension $\Delta$ appearing at order $z^{\Delta-s}$ in the asymptotic expansion of the transverse components of a massless field of integer spin $s>0$ in spacetime dimension four. \label{4d_graph}}
\end{figure}
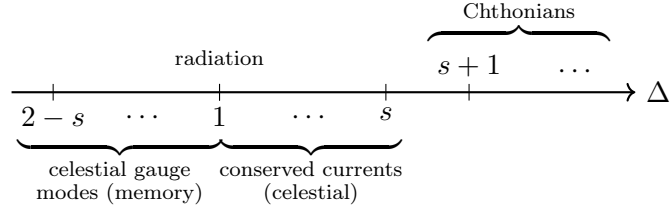

When $s > 0$, this rich picture collapses into the familiar case of four-dimensional spacetime (i.e., $d=2$). The diagram for the transverse components in figure~\ref{nice_graph} becomes the much simpler diagram in figure~\ref{4d_graph}. 

In four dimensions, only the following features of the general picture remain. In the transverse components, the source gives rise to the celestial gauge modes at orders from $z^{2-2s}$ till $z^{1-s}$, to the radiative modes at order $z^{1-s}$ and to the celestial conserved currents at orders from $z^{1-s}$ till $z^{0}$, while the vev gives rise to the infinite tower of Chthonians at Coulombic order $z$ and below. In the temporal components, one only finds the charge aspects (electric charge, mass and angular momentum aspects, etc.) originating from the vev; they  remain as they were in figure~\ref{boring_graph}.  
Note that, for any value of $s$, there is a collision at the scaling dimension $\Delta = 1$ (equivalently, at the radiative order $z^{1-s}$) between the orders at which super-translations (or their spin-$s$ generalisations) act, hard modes like the graviton appear and the super-translation celestial current is defined.
This collision makes the so-called infrared triangle particularly transparent in four dimensions \cite{Strominger:2017zoo}, although its higher-dimensional generalisation was discussed, e.g., in \cite{Kapec:2014zla, Kapec:2015vwa, Pate:2017fgt, He:2019jjk, Campoleoni:2019ptc, He:2023bvv}. The analogous relation between supertranslations and soft theorems was also studied for higher-spin fields in \cite{Campoleoni:2017mbt, Campoleoni:2020ejn}.

\subsection{Plan of the paper}

The paper is organised as follows. In section~\ref{sec:spin0}, we discuss the warm-up example of a scalar field with the specific mass determined by setting $s=0$ in eq.~\eqref{Maxwell-like}, whose solution space shares many qualitative features with that of gauge fields.  
We detail the structure of the solutions of this Klein-Gordon equation in AdS, contrast it with that in Minkowski spacetime, and show how to recover the latter as a smooth limit of the former. In this way, we introduce in a simplified context most of the tools that will also be employed for fields of arbitrary spin. In section~\ref{sec:spin1}, we extend the analysis to electromagnetism to highlight the additional features specific to gauge fields. In section~\ref{sec:HS}, we move to gauge fields of any spin, emphasising the implications of our results for linearised gravity.
Finally, in section~\ref{sec:group-theory} we confirm the picture that emerges from the analysis of the solution spaces via a group-theoretical analysis of the corresponding representations. More precisely, we  decompose the representations of $SO(d+1,2)$ corresponding to the vev and source (for any $s\in\mathbb N$) as a sum of irreducible representations of the Lorentz subgroup $SO(d+1,1)$ and check the agreement with the representations corresponding to the celestial fields presented in section~\ref{sec:summary}. Furthermore, we study the contraction $SO(d+1,2)\to ISO(d+1,1)$ of the vev (for $s=0,1,2$) in CFT language, in terms of (relativistic, Carrollian and celestial) primary fields. Section~\ref{sec:conclusions} discusses some future research directions, while details of technical computations and proofs are deferred to a number of appendices.

%%%%%%%%%%%%%%%%%%%%%%%%%%%%%%%%%%%%%%%%%%%%%%%%%%%%%%%
\section{Scalar field}\label{sec:spin0}

In this section, we illustrate the main ideas in the example of a free scalar field on AdS$_{d+2}$ with squared mass $m^2 = - 2(d-1)/R^2$. We detail the structure of the space of solutions of the Klein-Gordon equation in AdS$_{d+2}$ and we contrast it with that of a massless scalar in Minkowski spacetime $\mathbb{R}^{d+1,1}$. Eventually, we show how to recover the latter space of solutions as a smooth limit of the former.

%%%%%%%%%%%%%%%%%%%%%%%%%%%%%%%%%%%%%%%%%%%%%%%%%%%%%%%
\subsection{Asymptotic solutions of the Klein-Gordon equation} \label{sec:comparison_scalar}

For $s=0$, the Fronsdal equation \eqref{Maxwell-like} in AdS$_{d+2}$ reduces to the Klein-Gordon equation for a scalar field of squared mass $m^2  = -2(d-1)/R^2$. For any value of $d$, this is the mass of the bulk scalar entering Vasiliev's equations \cite{Vasiliev:2003ev} that, in higher-spin holography, couples to a boundary operator of scaling dimension $\dim = d-1$, built as a single trace of two $O(N)$ scalars (see, e.g., the review \cite{Giombi:2016ejx}). Besides its applications in holography, a scalar of this mass is interesting because its space of solutions has a structure very close to that we shall discuss later for gauge fields. 

In the flat Bondi coordinates \eqref{AdS_Bondi}, the equation of motion for a scalar of the above mass reads
\be \label{scalareom}
    z \left(2\, z\partial_z - d \right) \partial_u \varphi + z^2 \lap \varphi + \frac{1}{R^2} \left[ z^2 \partial_z^2 - d\, z \partial_z + 2(d-1) \right] \varphi = 0 \,,
\ee 
where $\lap$ denotes the Laplacian on the transverse plane (not to be confused with the scaling dimension $\dim$). Expanding the field in powers of the radial coordinate $z$,
\be  \label{scalar_expansion}
    \varphi(z,u,\mathbf{x}) = \sum_{n \,=\, \bar{n}}^\infty z^n \varphi^{(n)}(u,\mathbf{x}) \,,
\ee 
it implies
\be  \label{scalar_eom}
    \lap \varphi^{(n-2)} + (2n-d-2)\, \partial_u \varphi^{(n-1)} + \frac{(n-2)(n-d+1)}{R^2}\, \varphi^{(n)} = 0 \,.
\ee 
Its structure is therefore very different in AdS and Minkowski backgrounds: in AdS it fixes algebraically the components $\varphi^{(n)}$ in terms of the previous ones, while in the limit $R \to \infty$ it fixes only $\partial_u \varphi^{(n)}$ in terms of the previous components. This difference between algebraic and differential equations persists also for gauge fields and is at the origin of the different structure of the solution spaces with and without cosmological constant.

\paragraph{Solution space in AdS.}

Asking for a finite leading power in the expansion \eqref{scalar_expansion} fixes $\bar{n}$ to be the smallest value for which the coefficient of the last term in \eqref{scalar_eom} vanishes. Therefore, for $d > 2$ one has $\bar{n} = 2$, while for $d = 1, 2$ one gets $\bar{n} = 0,1$ respectively. For $d > 2$, $\varphi^{(2)}$ is a free function, corresponding to the source (or non-normalisable mode) in the AdS/CFT jargon,\footnote{\label{AdSCFTjargon}Remember that the unitarity bound on the scaling dimension of a scalar primary field in a CFT$_{d+1}$ is $\Delta\geqslant \frac{d-1}2$. For a bulk scalar of squared mass $m^2  = -\Delta_+\Delta_-/R^2$ with $\Delta_++\Delta_-=d+1$ and $\Delta_+\geqslant\Delta_-$, there are two possible choices of boundary condition $\varphi\sim z^{\Delta_\pm}\phi_{\Delta_\pm}$ corresponding respectively to the so-called normalisable/non-normalisable modes. This terminology is unfortunate for the scalar because, sometimes, both $\Delta_+$ and $\Delta_-$ are above the unitarity bound. In this case, both boundary conditions are admissible and one speaks of \textit{holographic degeneracy}. In the case at hand $\Delta_\pm\in\{2,d-1\}$, so that this phenomenon happens for $1\leqslant d\leqslant 5$ (see, e.g., the review \cite{Bekaert:2012ux} for a detailed discussion).} so we will follow this terminology.
The subleading components $\varphi^{(n>2)}$ are fixed recursively as
\be  \label{recursion_scalar}
    \varphi^{(n)} = - \frac{R^2}{{(n-2)(n-d+1)}} \left[(2n-d-2)\, \partial_u\varphi^{(n-1)} + \lap \varphi^{(n-2)} \right] ,
\ee 
with the exception of $\varphi^{(d-1)}$ which corresponds to the vev in the AdS/CFT jargon.
For $n = d-1$, eq.~\eqref{scalar_eom} reduces to
\be  \label{n=d-1}
    \lap \varphi^{(d-3)} + (d-4)\, \partial_u \varphi^{(d-2)} = 0 \,,
\ee 
and in section~\ref{sec:solution0} we shall prove that, when $d$ is even, this equation is identically satisfied when one takes eq.~\eqref{recursion_scalar} for $2<n<d-1$ into account. As a result, the above equation does not impose any constraint on the source.\footnote{For $d = 4$ this equation is also automatically satisfied since $\varphi^{(d-3)} = \varphi^{(1)} = 0$. The fact that eq.~\eqref{n=d-1} is identically satisfied is a peculiarity of the scaling dimension $\Delta_- = 2$: for generic integer scaling dimensions the source is constrained by equations similar to \eqref{n=d-1} and it describes higher-order singletons, see e.g. \cite{Bekaert:2013zya}.} In AdS$_{d+2}$, the asymptotic solutions of the Klein-Gordon equation with squared mass $m^2 = - 2(d-1)/R^2$ are thus specified by the two free functions $\varphi^{(2)}(u,\mathbf{x})$ and $\varphi^{(d-1)}(u,\mathbf{x})$, playing the role of source and vev.

\paragraph{Solution space in Minkowski.}

In the limit $R \to \infty$, one usually fixes $\bar n$ in the expansion \eqref{scalar_expansion} to be the value for which the coefficient in front of $\partial_u \varphi^{(n-1)}$ in \eqref{scalar_eom} vanishes (see, e.g., \cite{Satishchandran:2019pyc}).
This implies $\bar{n} = d/2$, which is an integer since we consider only spacetimes of even dimension.\footnote{If one were to start the expansion at a lower $\bar n$, one would encounter fields that are ``on-shell'' from the point of view of the transverse space $\mathbb R^d$, i.e., that satisfy certain differential equations.
The solutions of these equations, however, cannot vanish at $|\mathbf{x}| \to \infty$, and are therefore usually discarded. We shall comment about these fields around eq.~\eqref{eq_erd-osborn_scalar}.}
The corresponding coefficient $\varphi^{(d/2)}(u,\mathbf{x})$ is then a free function encoding radiation reaching null infinity, while the subleading components (with $n>d/2$) are partially fixed by \eqref{scalar_eom} as
\be \label{flat_rec_scalar}
\varphi^{(n)}(u,\mathbf{x}) = \phi^{(n)}(\mathbf{x}) - \frac{1}{2n-d} \int_0^u {\rm d} u' \lap \varphi^{(n-1)}(u',\mathbf{x}) \, .
\ee
Notice that we have chosen to define the ``integration functions'' $\phi^{(n)}(\mathbf{x}):=\varphi^{(n)}(0,\mathbf{x})$ on a cut of $\scri^+$ at $u = 0$, rather than at $u \to - \infty$. This choice will prove convenient in the following.

In $(d+2)$-dimensional Minkowski spacetime, the solutions of the massless Klein-Gordon equation are thus specified by a single free function $\varphi^{(d/2)}(u,\mathbf{x})$, that we shall sometimes denote as $C(u,\mathbf{x})$ in the following, and by an infinite tower of $u$-independent functions $\phi^{(n)}(\mathbf{x})$ with $n > d/2$, that we shall collectively refer to as ``Chthonians''. To avoid confusion, let us stress that this term was originally introduced in \cite{Mittal:2022ywl, Campoleoni:2023fug} to denote the gravitational analogue of the full $\varphi^{(n)}(u,\mathbf{x})$ coefficients subject to evolution equations in retarded time.

%%%%%%%%%%%%%%%%%%%%%%%%%%%%%%%%%%%%%%%%%%%%%%%%%%%%%%%
\subsection{AdS solution space} \label{sec:solution0}

We now discuss in more detail the AdS solution space. As we have seen, assuming the asymptotic expansion \eqref{scalar_expansion}, for $d$ even and greater than two the general solution of the equation of motion \eqref{scalar_eom} reads
\be 
    \varphi = z^2 \varphi^{(2)} - z^3 R^2 \partial_u \varphi^{(2)} + \dots + z^{d-1} \varphi^{(d-1)} + \cdots \, , 
\ee 
where we emphasised that the solution depends on $\varphi^{(2)}(u,\mathbf{x})$ and $\varphi^{(d-1)}(u,\mathbf{x})$. The other orders are expressed algebraically in terms of these two functions via eq.~\eqref{recursion_scalar}. The restriction to $d>2$ excludes a four-dimensional spacetime. Nevertheless, we do not detail the modifications required to treat the $d=2$ case, because for gauge fields the source does not become subleading with respect to the vev in four dimensions, and the main goal of this section is to introduce the tools needed to study fields with $s>0$.

We shall now use the recursion relation \eqref{recursion_scalar} to express the orders between the source and vev ---i.e., the $\varphi^{(n)}$ with $3 \leqslant n \leqslant d-2$ in eq.~\eqref{scalar_expansion}--- in terms of the source. We focus on this portion of the asymptotic expansion because, for even values of $d$, it has a peculiar structure induced by the precise form of the coefficients entering eq.~\eqref{recursion_scalar}. This explains why eq.~\eqref{n=d-1} is identically satisfied and has an impact on the $R \to \infty$ limit. 
To this end, it is convenient to introduce the differential operators
\be \label{scalar_operators}
f^{(n)} = - \frac{(2n-d-2)}{(n - 2)(n - d + 1)}\, \partial_u \,, \qquad
g^{(n)} = -\frac{1}{(n - 2)(n - d + 1)}\, \lap \, .
\ee
For $n \neq 2$ and $n \neq d-1$, one can thus rewrite eq.~\eqref{recursion_scalar} as 
\be \label{recursion_path}
\varphi^{(n)} = R^2 \left( f^{(n)} \varphi^{(n-1)} + g^{(n)} \varphi^{(n-2)} \right) .
\ee
We now introduce a diagrammatic representation for this relation. We assign a weight $w_{n+1} := R^2 f^{(n+1)}$ to the single path\\
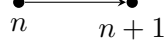
\begin{center}
\begin{tikzpicture}[->,>=stealth]
  % Dot nodes with labels below
  \node[circle, fill=black, inner sep=1.5pt, label=below:$n$] (n) at (0,0) {};
  \node[circle, fill=black, inner sep=1.5pt, label=below:$n+1$] (n1) at (1.5,0) {};

  % Arrow from n to n+1
  \draw (n) -- (n1);
\end{tikzpicture}
\end{center}
and we assign the weight $w_{n+2} := R^2 g^{(n+2)}$ to the double path\\
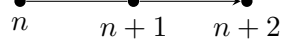
\begin{center}
\begin{tikzpicture}[->,>=stealth]
  % Dot nodes with labels below
  \node[circle, fill=black, inner sep=1.5pt, label=below:$n$] (n) at (0,0) {};
  \node[circle, fill=black, inner sep=1.5pt, label=below:$n+1$] (n1) at (1.5,0) {};
  \node[circle, fill=black, inner sep=1.5pt, label=below:$n+2$] (n2) at (3,0) {};

  % Arrow from n to n+1
  \draw (n) -- (n1) -- (n2);
\end{tikzpicture}
\end{center}
Expressing $\varphi^{(n)}$ in terms of $\varphi^{(2)}$ and $\varphi^{(d-1)}$ is equivalent to summing over all paths connecting the nodes $2$ and $d-1$ with the $n$-th node. 
For instance, eq.~\eqref{recursion_path} implies that $\varphi^{(4)}$ depends on $\varphi^{(2)}$ as
\be
\varphi^{(4)} = \left(R^4 f^{(3)} f^{(4)} + R^2 g^{(4)}  \right) \varphi^{(2)} \, ,
\ee
where the two addenda correspond to the two inequivalent diagrams with two intermediate nodes:
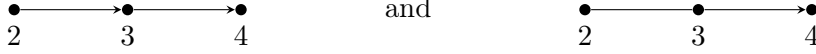
\begin{center}
\begin{minipage}{0.45\textwidth}
\vspace{0.51cm}
\centering
\begin{tikzpicture}[->,>=stealth]
  % Dot nodes with labels below
  \node[circle, fill=black, inner sep=1.5pt, label=below:$2$] (n) at (0,0) {};
  \node[circle, fill=black, inner sep=1.5pt, label=below:$3$] (n1) at (1.5,0) {};
  \node[circle, fill=black, inner sep=1.5pt, label=below:$4$] (n2) at (3,0) {};

  % Arrows
  \draw (n) -- (n1);
  \draw (n1) -- (n2);
\end{tikzpicture}
\end{minipage}% 
and
\begin{minipage}{0.45\textwidth}
\vspace{0.51cm}
\centering
\begin{tikzpicture}[->,>=stealth]
  % Dot nodes with labels below
  \node[circle, fill=black, inner sep=1.5pt, label=below:$2$] (n) at (0,0) {};
  \node[circle, fill=black, inner sep=1.5pt, label=below:$3$] (n1) at (1.5,0) {};
  \node[circle, fill=black, inner sep=1.5pt, label=below:$4$] (n2) at (3,0) {};

  % Arrow in one line
  \draw (n) -- (n1) -- (n2);
\end{tikzpicture}
\end{minipage}
\end{center}
In general, we can write $\varphi^{(n)}$ as
\be \label{sum_diagrams}
\varphi^{(n)} = \sum_{\substack{\textrm{distinct diagrams}\\\textrm{connecting $2$ to $n$}}} \left( \prod_k w_k \right)\varphi^{(2)} \, + \!\!\sum_{\substack{\textrm{distinct diagrams}\\\textrm{connecting $d-1$ to $n$}}}\!\! \left( \prod_k w_k \right)\varphi^{(d-1)}\,,
\ee
where the $w_k$ are the weights of the single or double paths contributing to each diagram, with the convention $w_{d-1} = 0$, accounting for the fact that eq.~\eqref{recursion_path} does not hold when $n = d-1$.

Notice that $f^{(n)}$ never vanishes for $n \leqslant \frac{d}{2}$, so that the diagram\\
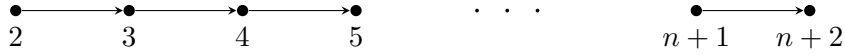
\begin{center}
\begin{tikzpicture}[->,>=stealth]
  % Dot nodes with labels below
  \node[circle, fill=black, inner sep=1.5pt, label=below:$2$] (n) at (0,0) {};
  \node[circle, fill=black, inner sep=1.5pt, label=below:$3$] (n1) at (1.5,0) {};
  \node[circle, fill=black, inner sep=1.5pt, label=below:$4$] (n2) at (3,0) {};
  \node[circle, fill=black, inner sep=1.5pt, label=below:$5$] (n3) at (4.5,0) {};
  \node[circle, fill=black, inner sep=0.5pt] (n-) at (6.1,0) {};
  \node[circle, fill=black, inner sep=0.5pt] (n--) at (6.5,0) {};
  \node[circle, fill=black, inner sep=0.5pt] (n---) at (6.9,0) {};
  \node[circle, fill=black, inner sep=1.5pt, label=below:$n +1$] (n4) at (9,0) {};
  \node[circle, fill=black, inner sep=1.5pt, label=below:$n+2$] (n5) at (10.5,0) {};

  % Arrow in one line
  \draw (n) -- (n1);
  \draw (n1) -- (n2);
  \draw (n2) -- (n3);
  \draw (n4) -- (n5);
\end{tikzpicture}
\end{center}
always contributes to the sum \eqref{sum_diagrams} in this range. As a result, for $n \leqslant \frac{d}{2}$ the maximal number of retarded-time derivatives $\partial_u$ of the source increases with $n$. More precisely, for $n \leqslant \frac{d}{2}$,
\be \label{single_paths_scalar}
\varphi^{(n)} = R^{2(n-2)} \prod_{k = 3}^{n} f^{(k)} \varphi^{(2)} + \cdots = \frac{(-1)^n (d-n-2)! R^{2(n-2)}}{(n-2)!(d-2n)!!(d-5)!!}\, \partial_u^{n-2} \varphi^{(2)} + \cdots \, ,
\ee
where we omitted terms with at least one $g^{(k)}$ weight, that contain a lower number of $\partial_u$ derivatives.

On the other hand, since $f^{\left(\frac{d}{2} + 1\right)} = 0$, all diagrams that contain the following single path do not contribute to the sum:\\
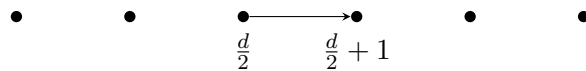
\begin{center}
\begin{tikzpicture}[->,>=stealth]
  % Dot nodes with labels below
  \node[circle, fill=black, inner sep=1.5pt] (n) at (0,0) {};
  \node[circle, fill=black, inner sep=1.5pt] (n1) at (1.5,0) {};
  \node[circle, fill=black, inner sep=1.5pt, label=below:$\frac{d}{2}$] (n2) at (3,0) {};
  \node[circle, fill=black, inner sep=1.5pt, label=below:$\frac{d}{2}+1$] (n3) at (4.5,0) {};
  \node[circle, fill=black, inner sep=1.5pt] (n4) at (6,0) {};
  \node[circle, fill=black, inner sep=1.5pt] (n5) at (7.5,0) {};

  % Arrow in one line
  \draw (n2) -- (n3);
\end{tikzpicture}\\
\end{center}
As a result, we find
\be
\varphi^{\left(\frac{d}{2}+1\right)} = R^2 g^{\left(\frac{d}{2}+1\right)} \varphi^{\left(\frac{d}{2}-1\right)} \,, \qquad 
\varphi^{\left(\frac{d}{2} + 2\right)} = R^4 g^{\left(\frac{d}{2}+2\right)} g^{\left(\frac{d}{2}\right)}\varphi^{\left(\frac{d}{2} - 2\right)} \,,
\ee
and, as we prove in appendix~\ref{sec:mirror_scalar}, this structure generalises also to the next orders thanks to cancellations induced by the precise form of the coefficients entering \eqref{scalar_operators}:
\be \label{mirror_scalar}
\varphi^{\left(\frac{d}{2} + k\right)} = R^{2k} \prod_{l=0}^{k-1} g^{\left(\frac{d}{2}+k-2l\right)} \varphi^{\left(\frac{d}{2} - k\right)} \qquad \textrm{for}\ 1 \leqslant k \leqslant \frac{d}{2}-2 \, .
\ee
In the following, we refer to this identity as the \emph{mirror relation}. Diagrammatically, it means that all $\varphi^{(\frac{d}{2}+k)}$ with $1 \leqslant k \leqslant \frac{d}{2} - 2$ are connected to $\varphi^{(\frac{d}{2}-k)}$ by double paths only. As a result, the number of $\partial_u$ derivatives of the source decreases with $n$ in the coefficients $\varphi^{(n)}$ contained in the range $\frac{d}{2}+1 \leqslant n \leqslant d-2$. No particular cancellations take place below the order at which the vev appears, and both types of paths contribute to each order.

The mirror relation \eqref{mirror_scalar} plays a crucial role in determining the structure of the space of solutions above the vev and, consequently, of its $R\to\infty$ limit. It also allows one to prove that eq.~\eqref{n=d-1} is identically satisfied when $d$ is even and therefore does not impose any constraint on the source.
Indeed, using \eqref{mirror_scalar}, one finds
\be
\lap \varphi^{(d-3)} + (d-4)\, \partial_u \varphi^{(d-2)} 
 = (4-d)R^{d-4} g^{(d-2)} \left[ \prod_{l=0}^{\frac{d}{2}-4}  g^{(d-2l-3)} - \prod_{l=0}^{\frac{d}{2}-4} g^{(d-2l-4)} \right] \partial_u \varphi^{(2)} \, ,
\ee
where we imposed $\lap \varphi^{(3)} = (4-d) R^2 g^{(d-2)} \partial_u \varphi^{(2)}$ as implied by \eqref{recursion_path} at $n=3$. To conclude, by renaming the indices one can reorganise the above products as
\be
\prod_{l=0}^{\frac{d}{2}-4} g^{(d-2l-3)} - \prod_{l=0}^{\frac{d}{2}-4} g^{(d-2l-4)} = \prod_{l=0}^{\frac{d}{2}-4} g^{(d-2l-3)} - \prod_{l=0}^{\frac{d}{2}-4} g^{(4+2l)}
\ee
and notice that the difference on the right-hand side vanishes thanks to the symmetry $g^{(\frac{d}{2}+k)} = g^{(\frac{d}{2}-k+1)}$ of the operators defined in \eqref{scalar_operators}.

%%%%%%%%%%%%%%%%%%%%%%%%%%%%%%%%%%%%%%%%%%%%%%%%%%%%%%%
\subsection{Flat limit of the AdS solution space} \label{sec:flat-limit_scalar}

Following \cite{Campoleoni:2023fug}, to define a smooth flat limit of the previous space of solutions we expand the source and the vev in powers of the AdS radius $R$ as follows:
\be \label{R_expansion}
    \varphi^{(2)} = \sum_{k = 0}^{\frac{d}{2}-2} \frac{1}{R^{2k}}\, \varphi^{(2,\,k)} \,,\qquad \varphi^{(d-1)} = \sum_{k = 0}^\infty \frac{1}{R^{2k}}\, \varphi^{(d-1,\,k)} \,,
\ee 
where the upper bound on the expansion of the source will become clear in the following.
Substituting these expansions in the general asymptotic solution \eqref{sum_diagrams} of the equation of motion, gives both divergent and regular contributions in the limit $R \to \infty$:
\be \label{def_notation}
\varphi^{(n)} = \sum_{k} \frac{1}{R^{2k}}\,\varphi^{(n,\,k)} \,,\quad \text{for}\ 3\leqslant n\leqslant d-2\ \text{or}\ n\geqslant d\,,
\ee
where $k<0$ terms in this sum correspond to divergent contributions, while the regular part corresponds to $k\geqslant 0$ terms.
To obtain a smooth flat limit, we then impose the vanishing of the divergent part of the on-shell field, and observe that this corresponds to imposing the Minkowski equations of motion on the regular components $\varphi^{(n,\,0)}$. Indeed, eq.~\eqref{recursion_path} implies
\be \label{recursion_divergences}
\varphi^{(n,\,k)} = f^{(n)}\, \varphi^{(n-1,\,k+1)} + g^{(n)}\,\varphi^{(n-2,\,k+1)} \quad \text{for}\ n \neq d-1
\ee
and this ensures that all divergent contributions, corresponding to the $\varphi^{(n,\,k)}$ with $k < 0$, vanish once the $k=-1$ term is set to zero, i.e., once
\be \label{flat_constraints}
(n - 2)(n - d + 1)\, \varphi^{(n,\,-1)} = - (2n-d-2)\, \partial_u \varphi^{(n-1,\,0)} - \lap \varphi^{(n-2,\,0)} \overset{!}{=} 0 \, .
\ee
We recovered in this way the flat spacetime equations of motion (cf.~\eqref{scalar_eom}), while identifying the coefficients of the asymptotic $z$-expansion in Minkowski spacetime with the $\varphi^{(n,\,0)}$ terms in the expansion \eqref{def_notation}.

The previous discussion shows that imposing the Minkowski equation of motion on the components $\varphi^{(n,\,0)}$ in the expansion \eqref{def_notation} ensures the regularity of the $R \to \infty$ limit, but it does not explain yet how these components are related to the AdS source and vev. In the following, we make explicit the relation between the components $\varphi^{(n,\,0)}$ and the terms in the expansions \eqref{R_expansion} of the source and the vev, so as to track the AdS origin of all free boundary data in the solutions of the massless Klein-Gordon equation in Minkowski spacetime.  As manifested by the group-theoretical analysis of section~\ref{sec:group-theory}, the constraints induced by eqs.~\eqref{flat_constraints} on the coefficients of the expansion of the source and the vev are indeed such that the information they contain is distributed among the whole set of $\varphi^{(2,\,k)}$ and $\varphi^{(d-1,\,k)}$ in eq.~\eqref{R_expansion}.

Before moving to the detailed analysis of the limit, let us notice that the mirror relation \eqref{mirror_scalar} implies
\be \label{phi_d-2}
\varphi^{(d-2)} = \frac{R^{d-4}}{(d-4)!}\, \lap^{\frac{d}{2}-2} \varphi^{(2)} = \frac{1}{(d-4)!} \sum_{n=0}^{\frac{d}{2}-2} R^{d-2n-4} \lap^{\frac{d}{2}-2} \varphi^{(2,\,n)} \,,
\ee
so that regularity of the $R \to \infty$ limit requires
\be \label{higher-order-Lap}
\lap^{\frac{d}{2}-2} \varphi^{(2,\,n)} = 0 \quad \textrm{for} \ 0 \leqslant n \leqslant \frac{d}{2}-3 \, ,
\ee
while the last term in the expansion \eqref{R_expansion}, i.e., $\varphi^{(2,\,\frac{d}{2}-2)}$, remains free.
One could impose boundary conditions on the transverse plane $\mathbb{R}^d$ (i.e.\ fall-off conditions when $|\mathbf x|\to \infty$) for the $\varphi^{(2,\,n)}$ with $n < \frac{d}{2}-2$ in such a way eqs.~\eqref{higher-order-Lap} do not admit any solution. Choosing this option, in the flat limit one obtains the space of solutions considered in \cite{Satishchandran:2019pyc, Campoleoni:2020ejn} and briefly reviewed in section~\ref{sec:comparison_scalar}. For this reason, we shall discuss the role of the $\varphi^{(2,\,n)}$ with $n < \frac{d}{2}-2$ only at the end of this section. To begin with, we shall set them to zero and discuss the other terms in the expansions \eqref{R_expansion} starting from the leading contribution in $z$, i.e., in the order in which they enter figure~\ref{nice_graph}.

\paragraph{Radiation.}

Following the above discussion, 
we begin by imposing $\varphi^{(2,\,n)} = 0$ for $0 \leqslant n \leqslant \frac{d}{2}-3$. In the limit $R \to \infty$, one then obtains
\be \label{leading-order_scalar}
\varphi(z,u,\mathbf{x}) = z^{\frac{d}{2}} \frac{(-1)^{\frac{d}{2}}}{(d-5)!!}\, \partial_u^{\frac{d}{2}-2}  \varphi^{(2,\,\frac{d}{2}-2)}(u,\mathbf{x}) + \cO(z^{\frac{d}{2}+1}) \, .
\ee
The leading order thus becomes $z^\frac{d}{2}$ as expected from the direct analysis in flat spacetime that we recalled above \eqref{flat_rec_scalar}. At the subleading orders in the expansion in powers of $z$, a priori one could encounter terms in $\varphi^{(2,\,\frac{d}{2}-2)}$ that diverge in the limit $R \to \infty$ because they contain positive powers of $R^2$. On the other hand, above the vev all these contributions vanish identically thanks to the mirror relation \eqref{mirror_scalar}. For instance, before taking the limit $R \to \infty$, the mirror relation implies
\be
\varphi^{(\frac{d}{2}+1)} = R^2 g^{(\frac{d}{2}+1)} \varphi^{(\frac{d}{2}-1)} 
\quad \textrm{with} \quad
\varphi^{(\frac{d}{2}-1)} = R^{-2}\, \prod_{k=3}^{\frac{d}{2}-1} f^{(k)} \varphi^{(2,\,\frac{d}{2}-2)} + \cO(R^{-4})
\ee
and similar considerations apply to the subsequent orders up to that at which the vev appears. Eventually, one obtains
\be \label{above-vev}
\begin{split}
\varphi &= \sum_{k=0}^{\frac{d}{2}-2} z^{\frac{d}{2}+k}\,\frac{(-1)^{\frac{d}{2}+k}}{(2k)!!(d-5)!!}\, \lap^k \partial_u^{\frac{d}{2}-k-2}  \varphi^{(2,\,\frac{d}{2}-2)} + z^{d-1} \varphi^{(d-1,\,0)} \\
&\qquad - z^d \left[ R^2 \left( \pr_u \varphi^{(d-1,\,0)} + \frac{\lap^{\frac{d}{2}-1} \varphi^{(2,\,\frac{d}{2}-2)}}{(d-2)!!(d-5)!!} \right) + \pr_u \varphi^{(d-1,\,1)}  \right] + \cdots
\end{split}
\ee
where $(\cdots)$ encompasses terms that are subleading either in $R^{-2}$ or in $z$. Cancelling the divergence thus requires the constraint
\be \label{first_constraint}
\pr_u \varphi^{(d-1,\,0)} + \frac{1}{(d-2)!!(d-5)!!}\, \lap^{\frac{d}{2}-1} \varphi^{(2,\,\frac{d}{2}-2)} = 0 \, ,
\ee
which fixes $\varphi^{(d-1,\,0)}$ in terms of $\varphi^{(2,\,\frac{d}{2}-2)}$ up to an arbitrary function of the transverse coordinates. As a result, this relation does not impose any constraint on $\varphi^{(2,\,\frac{d}{2}-2)}$. The $R$-divergent terms involving the source appearing at the subleading orders in the $z$-expansion 
fulfil the recursive relation \eqref{recursion_divergences}, and thus vanish once the constraint \eqref{first_constraint} is imposed.

Therefore, the function $\varphi^{(2,\,\frac{d}{2}-2)}(u,\mathbf{x})$ is completely free.
In Minkowski spacetime, it contributes to the leading order in the $z$ expansion via 
\be \label{radiation_scalar}
    C(u,\mathbf{x}) := \frac{(-1)^{\frac{d}{2}}}{(d-5)!!}\, \partial_u^{\frac{d}{2}-2} \varphi^{(2,\,\frac{d}{2}-2)}(u,\mathbf{x}) \,,
\ee
which encodes radiative modes and is the scalar field analogue of the gravitational shear. Note that for $d=4$, this definition has to be understood as $C = \varphi^{(2,\,0)}$ directly, while we recall that in this section we impose $d > 2$. For $d=2$, the shear should instead be identified with $\varphi^{(d-1,\,0)}$. 

\paragraph{Source Chthonians.} 

The terms in the first line of eq.~\eqref{above-vev} can be rewritten as
\be \label{above-vev-full}
\varphi^{(\frac{d}{2}+k,\,0)}(u, \mathbf x) = \frac{(-1)^k}{(2k)!!} \lap^k \!\left( \partial_u^{-k} C(u,\mathbf{x}) + \frac{(-1)^\frac{d}{2}}{(d-5)!!} \sum_{l=0}^{k-1} \frac{u^l}{l!} \partial_u^{\frac{d}{2}-2-k+l} \varphi^{(2,\,\frac{d}{2}-2)} (u=0,\mathbf x) \right) ,
\ee
where $0 \leqslant k \leqslant \frac{d}{2}-2$ and we defined the inverse of the retarded-time derivative as
\be\label{uintegral}
\partial_u^{-1} f(u,\mathbf{x}) := \int_{0}^u {\rm d}u'\, f(u',\mathbf{x}) \, .
\ee
Notice that, as in eq.~\eqref{flat_rec_scalar}, we fixed the integration domain so as to conveniently evaluate the integration ``constants'' at $u = 0$.
Comparing with the explicit solution \eqref{flat_rec_scalar} of the equations of motion in Minkowski spacetime, we can identify a first set of Chthonians as
\be \label{1st-Cht-scalar}
\phi^{(\frac{d}{2}+k)}(\mathbf x) = \frac{(-1)^{\frac{d}{2}+k}}{(2k)!!(d-5)!!}\, \lap^k \partial_u^{\frac{d}{2}-2-k} \varphi^{(2,\,\frac{d}{2}-2)} (u=0,\mathbf x)
\quad \textrm{with} \quad 1 \leqslant k \leqslant \frac{d}{2}-2 \, .
\ee
Eqs.~\eqref{radiation_scalar} and \eqref{1st-Cht-scalar} display the full contribution of the AdS source to the independent boundary data in Minkowski spacetime (under the hypothesis $\varphi^{(2,\,n)} = 0$ for $n < \frac{d}{2}-2$). 
Indeed, the mirror relation \eqref{mirror_scalar} implies
\be \label{scalar_before_vev}
    \varphi^{(d-2)} = R^{d-4} \prod_{k = 0}^{\frac{d}{2}-3} g^{(d-2-2k)} \varphi^{(2)} = \prod_{k = 0}^{\frac{d}{2}-3} g^{(d-2-2k)} \varphi^{(2,\, \frac{d}{2}-2)} \,,
\ee
because we truncated the $R^2$-expansion of the source in eq.~\eqref{R_expansion} and, for the time being, we are imposing $\varphi^{(2,\,n)} = 0$ for $n < \frac{d}{2}-2$. With these assumptions, $\varphi^{(d-2)}$ therefore contains only a single term of order $\cO(R^0)$. If we had not imposed the latter constraints, we would have obtained additional divergent terms, but no subleading terms of $\cO(R^{-2})$. All contributions of the source to the terms $\varphi^{(n)}$ below the vev, that is with $n > d-1$, result from a sum over all single and double paths connecting $\varphi^{(d-2)}$ to $\varphi^{(n)}$. Each path brings an additional power of $R^2$ and $\varphi^{(d-2)} \sim \cO(R^0)$. As a result, the source can only appear directly in divergent terms that vanish thanks to the constraint \eqref{first_constraint}, and it affects the explicit expression of the $\varphi^{(n,\,0)}(u,\mathbf{x})$ with $n > d-1$ only via the solution of the constraint \eqref{first_constraint}: 
\be 
\varphi^{(d-1,\,0)}(u,\mathbf x) - \varphi^{(d-1,\,0)}(u=0,\mathbf x) = - \frac{1}{(d-2)!!(d-5)!!}\, \lap^{\frac{d}{2}-1} \pr_u ^{-1} \varphi^{(2,\,\frac{d}{2}-2)}(u,\mathbf x) \,.
\ee
As such, it does not contribute directly to the Chthonians below the vev. Had we continued the expansion of the source in eq.~\eqref{R_expansion} beyond that order, we would have recovered additional contributions from the source in the vev Chthonians, which would have been spurious. This is actually the reason why we chose to truncate the expansion to that order. We refer the interested reader to appendix~\ref{app:scalar_chthonians} for a proof of why this is the case.
Notice also that the decomposition of $\varphi^{(\frac{d}{2}-2)}(u,\mathbf{x})$ into the radiation $C(u,\mathbf{x}) \propto \partial_u^{\frac{d}{2}-2} \varphi^{(2,\,\frac{d}{2}-2)}(u,\mathbf{x})$ (defined in \eqref{radiation_scalar}) and the $d/2-2$ Chthonians $\phi^{(n)}(\mathbf{x})$ (defined in \eqref{1st-Cht-scalar}) reflects the branching of the source discussed in section~\ref{sec:group-theory}.

\paragraph{Vev Chthonians.}

As explained before, it is enough to cancel the $\varphi^{(d-1+n,\,-1)}$ divergence in order to make the solution finite in the limit $R \to \infty$. This is encoded in eq.~\eqref{flat_constraints}, signalling that the regular fields $\varphi^{(d-1+n,\,0)}$ are subject to the flat-space evolution equations.
This leaves exactly one integration function on the transverse plane for each integer $n\geqslant 0$, providing another set of Chthonians originating from the vev, to be distinguished from the previous set of Chthonians originating from the expansion of the source. The evolution equations are informed of both the radiation $C(u,\mathbf{x})$ and the source Chthonians through eq.~\eqref{first_constraint}, but a simple counting of the powers of $R$ shows that the source does not appear in the coefficients $\varphi^{(d-1+n,\,0)}$ in the limit $R \to \infty$.

In order to express the contribution to $\varphi^{(d-1+n,\,0)}$ of the components $\varphi^{(d-1,\,k)}$, we can use the recursion \eqref{recursion_divergences}, summing over an increasing number of diagrams connecting $d-1$ to $d-1+n$. The explicit expression becomes increasingly cumbersome as $n$ increases; however, there is a simple way to encode the Chthonian degrees of freedom. Notice that the single-path contribution is always present in the diagrams, giving the term with the maximal number of $\partial_u$ derivatives:
\be \label{vev_Chth}
    \varphi^{(d-1+n,\,0)} = \prod_{k=1}^n f^{(d-1+k)} \varphi^{(d-1,\,n)} + \cdots = \frac{(-1)^n(d-3)!!(d+2n-4)!!}{n!(d+n-3)!} \, \partial_u^n \varphi^{(d-1,\,n)} + \cdots \,,
\ee
where $(\cdots)$ represent contributions of the form
\be \label{Chthonian_other_contributions}
    \partial_u^{n-2k} \lap^k \varphi^{(d-1,\,n-k)}\,, \quad \text{for } 1 \leqslant k \leqslant \left\lfloor \frac{n}{2} \right\rfloor .
\ee
In order to identify where the vev Chthonians sit in $\varphi^{(d-1)}$, we will adopt the following strategy. At each order $n$, the expression of $\varphi^{(d-1+n,\,0)}$ involves the new coefficient $\varphi^{(d-1,\,n)}$, entering with $\partial_u^n$ derivatives, as well as the $\varphi^{(d-1,\,k < n)}$, entering with less derivatives (cf.~\eqref{vev_Chth}). The evolution equation of $\varphi^{(d-1+n,\,0)}$ then implies that the $\partial_u^{n+1}$ derivative of $\varphi^{(d-1,\,n)}$ is constrained, so that it effectively provides only a new function on the transverse plane. We can identify it with the Chthonian $\phi^{(d-1+n)}$ entering eq.~\eqref{flat_rec_scalar} by setting $\partial_u^k \varphi^{(d-1,\,n)}(u=0,\mathbf x) = 0$ for all $0 \leqslant k \leqslant n-1$. As discussed in appendix~\ref{app:scalar_chthonians}, this condition is consistent because these terms always enter the expression for $\varphi^{(d-1+n,\,0)}$ together with the new term $\partial_u^n \varphi^{(d-1,\,n)}$, so that they would just introduce a redundancy in the parameterisation of the solution space.
Eventually, this condition allows one to obtain the following simple expression for the flat space Chthonians below the order $\cO(z^{d-1})$:
\be \label{eq_vev-Chthonians_realisation}
    \phi^{(d-1+n)}(\mathbf x) = \varphi^{(d-1+n,\,0)}(u=0,\mathbf x) = \frac{(-1)^n(d-3)!!(d+2n-4)!!}{n!(d+n-3)!}\, \partial_u^n \varphi^{(d-1,\,n)}(u=0,\mathbf x) \,.
\ee
The second equality comes about because the other contributions of the form \eqref{Chthonian_other_contributions} vanish on account of the strategy presented above and further detailed in appendix~\ref{app:scalar_chthonians}.

\paragraph{Erdmenger--Osborn fields.}

We now turn back to the issue of the components $\varphi^{(2,\,n)}$ with $0 \leqslant n \leqslant \frac{d}{2}-3$ that we had originally set to zero. Whenever these are non-zero, the flat limit features components $\varphi^{(2+n,\,0)}$ leading to an asymptotic fall-off which is overleading with respect to radiation, see the discussion in section~\ref{sec:comparison_scalar}. Going on-shell, not only do the overleading contributions verify an evolution equation in retarded time, but they are also ``on-shell'' from the point of view of the transverse space.\footnote{For instance, turning on $\varphi^{(0,\,\frac{d}{2}-3)}$ leads to introducing just one field $\varphi^{(\frac{d}{2}-1,\,0)}$ at order $\mathcal O(z^{\frac{d}{2}-1})$ in the flat limit. This field must satisfy $\partial_u \varphi^{(\frac{d}{2}-1,\,0)} = 0 = \lap \varphi^{(\frac{d}{2}-1,\,0)}$. In the case where the transverse space is the celestial sphere $S^d$, the latter equation would be replaced by $\left(\nabla^2_{S^d} - \frac{d}{2} (\frac{d}{2}-1)\right) \varphi^{(\frac{d}{2}-1,\,0)} = 0$, where the differential operator on the left-hand-side is invertible in the functional space of square-integrable solutions, leading to $\varphi^{(\frac{d}{2}-1,\,0)}$ itself being zero. This argument extends to all components $\varphi^{(n,\,0)}$ for $2 \leqslant n \leqslant \frac{d}{2}-1$.}

As in the case of vev Chthonians, regular contributions $\varphi^{(2+n,\,0)}$ with $0 \leqslant n \leqslant \frac{d}{2}-3$ are nourished by $\varphi^{(2,\,k)}$ with $0 \leqslant k \leqslant \frac{d}{2}-3$ and satisfy the equations of motion of flat spacetime by virtue of eq.~\eqref{recursion_divergences}. In order to identify where these \emph{on-shell scalars} sit in the expansion of the source in powers of $R^2$, we can proceed in a way similar to what we did for the vev Chthonians. First, notice that  
\be
    \varphi^{(2+n,\,0)} = \prod_{k=1}^n f^{(2+k)} \varphi^{(2,\,n)} + \cdots = \frac{(-1)^n (d-n-4)!}{n!(d-2n-4)!!(d-5)!!}\,\partial_u^n \varphi^{(2,\,n)} + \cdots \,,
\ee
where $(\cdots)$ represent again contributions of the form \eqref{Chthonian_other_contributions}. Looking at the most divergent contributions in the fields below radiative order
\be
    \varphi^{(\frac{d}{2}+k,\,-k)} = \prod_{l=0}^{k-1} g^{(\frac{d}{2}+k-2l)} \varphi^{(\frac{d}{2}-k,\,0)} \,,\quad \text{for } 1 \leqslant k \leqslant \frac{d}{2}-2 \,,
\ee
we obtain the constraints
\be \label{eq_erd-osborn_scalar}
    \lap^{\frac{d}{2}-2-n} \varphi^{(2+n,\,0)} = 0 \,,\quad \text{for } 0 \leqslant n \leqslant \frac{d}{2}-3 \,.
\ee
These constraints follow from the Minkowski equations of motion, but they are the only consequences of the latter that do not involve any derivative in retarded time.
By adopting the same requirement as for the Chthonians, i.e.\ that $\partial_u^k\varphi^{(2,\,n)}(u=0,\mathbf x) = 0$ for $0 \leqslant k < n$ and $0 \leqslant n \leqslant \frac{d}{2}-3$, we conclude that the $z$ expansion in Minkowski spacetime also contains at the orders $z^{2+n}$ with $0 \leqslant n \leqslant \frac{d}{2}-3$ the coefficients
\be \label{eq_erd-osborn_realisation}
    \phi^{(2+n)}(\mathbf x) = \varphi^{(2+n,\,0)}(u=0,\mathbf x) = \frac{(-1)^n(d-n-4)!}{n!(d-2n-4)!!(d-5)!!}\, \partial_u^n \varphi^{(2,\,n)}(u=0,\mathbf x) \,,
\ee
satisfying eqs.~\eqref{eq_erd-osborn_scalar}. We dub these ``on-shell'' coefficients Erdmenger--Osborn fields because these equations are the generalisation and Riemannian analogue of the fields first discussed in \cite{Erdmenger:1997wy}. Note also that introducing these fields does not alter our previous discussion of the vev Chthonians. Indeed, they enter $\varphi^{(d-2)}$ only via divergent terms that vanish on account of \eqref{eq_erd-osborn_scalar}, and the fields $\varphi^{(d-1+n,\,-1)}$ could depend on the source only via that term because \eqref{recursion_divergences} is not defined for $n = d-1$. The higher-order equations \eqref{eq_erd-osborn_scalar} coincide with the constraints on these celestial fields on account of the group-theoretical analysis of section~\ref{sec:group-theory}.

%%%%%%%%%%%%%%%%%%%%%%%%%%%%%%%%%%%%%%%%%%%%%%%%%%%%%%%
\section{Maxwell field}\label{sec:spin1}

In this section, we show how the space of solutions of Maxwell's equations in Minkowski spacetime can be recovered as a smooth limit of the AdS one. We follow a similar strategy as that discussed for the scalar field, highlighting the subtleties brought by the presence of more field components and of a gauge symmetry. 

%%%%%%%%%%%%%%%%%%%%%%%%%%%%%%%%%%%%%%%%%%%%%%%%%%%%%%%
\subsection{AdS solution space} \label{sec:solution1}

We parametrise the background as in eq.~\eqref{AdS_Bondi} and we work in the radial gauge, i.e., we impose
\be \label{radial_gauge}
A_z = 0 \, .
\ee
As for scalars, we expand the remaining field components in integer powers of $z$,
\be \label{z-expansion_spin-1}
A_\mu = \sum_n z^n A^{(n)}_\mu \, ,
\ee
where the lower extremum of each sum may depend on the field component under consideration.

\paragraph{Equations of motion.}

Plugging this expansion in the Maxwell equations gives
\begin{align} 
\mathcal{F}_z^{(n)} & := -n(n-d+1)\, A_u^{(n)}- (n-1)\, \partial \cdot A^{(n-1)} =0 \,, \label{max_1} \\[10pt] 
\mathcal{F}_u^{(n)} & := \frac{n(n-d+1)}{R^2}\, A_u^{(n)} + (n-1)\, \partial_u A_u^{(n-1)} + \lap A_u^{(n-2)} -  \partial_u \partial\cdot A^{(n-2)} = 0 \,, \label{max_2} \\[10pt]
\mathcal{F}_i^{(n)} & := \frac{n(n-d+1)}{R^2}\, A_i^{(n)} + (2n - d)\, \partial_u A_i^{(n-1)} - (n - d+1)\, \partial_iA_u^{(n-1)} \nonumber \\
& \phantom{=}\ + \lap A_i^{(n-2)} - \partial_i \partial \cdot A^{(n-2)} = 0 \label{max_3} 
\end{align}
(see eqs.~\eqref{Fz1} and \eqref{Fz2} for the explicit expression of Maxwell's equations in the coordinates \eqref{AdS_Bondi} before the substitution of the asymptotic expansion \eqref{z-expansion_spin-1}). Eq.~\eqref{max_1} has the same form in AdS and Minkowski backgrounds, while eqs.~\eqref{max_2} and \eqref{max_3} are algebraic in AdS and differential in Minkowski, in full analogy with the scalar case.
The equations of motion satisfy the Bianchi identity
\be \label{Bianchi_spin1}
\left(n-d\right) \left(\mathcal{F}_u^{(n)} + \frac{1}{R^2}\,  \mathcal{F}_z^{(n)}\right) = -\partial_u \mathcal{F}_z^{(n-1)} - \partial \cdot \mathcal{F}^{(n-1)} \, ,
\ee
implying that if one imposes $\mathcal{F}_z^{(n)} = 0$ and $\mathcal{F}_i^{(n)} = 0$ for all $n$, then the equation $\mathcal{F}_u^{(n)} = 0$ is identically satisfied for all $n \neq d$. Eq.~\eqref{max_1} thus allows one to rewrite all components of $A_u$, with the exception of $A_u^{(0)}$ and $A_u^{(d-1)}$, in terms of $A_i$, while eq.~\eqref{max_2} for $n=d$ can be reduced to the evolution equation
\be \label{Vev_maxwell}
 \partial_u A_u^{(d-1)} - \frac{1}{R^2}\, \partial \cdot A^{(d-1)} = - \frac{d-3}{(d-1)(d-2)}\, \lap \partial\cdot A^{(d-3)} \, .
\ee

\paragraph{Source and vev.}

Asking for a finite leading power in the expansion of $A_u$ and $A_i$ fixes both of them to be of order $z^0$. The leading-order terms $A_u^{(0)}$ and $A_i^{(0)}$ are completely free and encode the source in the AdS/CFT jargon. On the other hand, our gauge choice \eqref{radial_gauge} leaves the residual gauge symmetry $\delta A_\mu = \partial_\mu \lambda$ with $\lambda = \lambda(u,\mathbf{x})$. We can use this residual gauge symmetry to also impose
\be \label{extra_bnd-cond}
A_u^{(0)} = 0 \, .
\ee
This is a standard choice in the asymptotic analysis of Maxwell's equations in Minkowski spacetime (see, e.g., \cite{He:2014cra, Kapec:2014zla}). Given the useful simplifications it brings, we adopt it for AdS as well, and we refer, e.g., to \cite{Barnich:2013sxa, Campoleoni:2023eqp} for a more complete discussion of the solution space in radial gauge.  
With this choice, the independent boundary data are therefore the free function $A_i^{(0)}(u,\mathbf{x})$, which corresponds to the partially gauge-fixed source, the free function $A^{(d-1)}_i(u,\mathbf{x})$, and the charge aspect\footnote{This name comes from the fact that boundary charges, including the electric charge, can be written as $Q = \int_{\mathbb{R}^d} \lambda F_{uz} \propto \int_{\mathbb{R}^d} \lambda A_u^{(d-1)}$.} $A^{(d-1)}_u(u,\mathbf{x})$, which satisfies the evolution equation \eqref{Vev_maxwell}.
The components appearing at order $z^{d-1}$ encode the AdS/CFT vev. 

Source and vev are usually presented in Poincaré coordinates, where they are respectively given by a transverse vector subject to the gauge symmetry $\delta A^{\textrm{source}}_a(y^b) = \partial_a\lambda(y^b)$, where $y^a=(u,\mathbf{x})$ with $a \in \{0,1,\dots,d\}$, and a conserved current satisfying $\partial_a J^a(y^b) = 0$. Our ``source'' has one component less because we partially fixed its gauge symmetry imposing \eqref{extra_bnd-cond}, while a conserved current $J_a =(J_u,J_i)$ can be defined starting from $A^{(d-1)}_u$ and $A^{(d-1)}_i$, e.g., as follows: 
\be
J_u= A_u^{(d-1)}\,,\quad J_i= A^{(d-1)}_i - \frac{R^2(d-3)}{(d-1)(d-2)}\, \lap A_i^{(d-3)} \,,
\ee
where $A_i^{(d-3)}$ can be expressed completely in terms of the source using the equations of motion. In the following, with a slight abuse of terminology, we shall refer to $A_i^{(0)}$ as the source and $A^{(d-1)}_u$, $A^{(d-1)}_i$ as the vev.

\paragraph{Solution space.}

The equations of motion have a structure very close to that we encountered in eq.~\eqref{recursion_path} when studying a scalar with mass $m^2 R^2 = -2(d-1)$. Indeed, using eq.~\eqref{max_1} to eliminate the $A_u^{(n)}$ with $n \neq d-1$ from eq.~\eqref{max_3} gives, for $n > 0$ and $n \neq d-1$,
\be \label{eom_maxwell}
A_{i}^{(n)} = R^2 \left( f^{(n)}A_i^{(n-1)} + g^{(n)\,j}_{\phantom{(n)}\!\!  i} A_j^{(n-2)} \right) + \frac{R^2}{d}\, \delta_{n,d}\, \partial_i A_u^{(d-1)} \, , 
\ee
where we introduced the differential operators
\begin{align} 
f^{(n)} = - (2n - d) b_{n}\,  \partial_u \, , \qquad 
g_{ij}^{(n)} = - b_{n} \big(\delta_{ij} \lap + (d-2) b_{n-1}\, \partial_i\partial_j \big) , \label{def2} 
\end{align}
with
\be
b_n = \frac{1}{n(n-d+1)} \, .
\ee
Moreover, eq.~\eqref{eom_maxwell} takes into account that \eqref{max_1} implies $\partial\cdot A^{(d-2)} = 0$, so that one can substitute the ill-defined $g^{(d)}_{ij}$ in \eqref{def2} with $g^{(d)}_{ij} = - \frac{1}{d}\, \delta_{ij} \lap$.

As in section~\ref{sec:solution0}, we can introduce a diagrammatic representation for this relation and express $A_i^{(n)}$ with $n \notin \{0, d-1, d\}$ as
\be \label{sum_diagrams_spin1}
\begin{split}
A_i^{(n)} & = \sum_{\substack{\textrm{distinct diagrams}\\\textrm{connecting $0$ to $n$}}} \Big( \prod_k w_k \Big)_{\! i}^{\ j}  A_j^{(0)} \, + \!\!\sum_{\substack{\textrm{distinct diagrams}\\\textrm{connecting $d-1$ to $n$}}}\!\! \Big( \prod_k w_k \Big)_{\! i}^{\ j} A_j^{(d-1)} \\
& + \sum_{\substack{\textrm{distinct diagrams}\\\textrm{connecting $d$ to $n$}}} \frac{R^2}{d} \Big( \prod_k w_k \Big)_{\! i}^{\ j}  \partial_j A_u^{(d-1)} \,,
\end{split}
\ee
where the $w_k$ are the weights of the single or double paths contributing to each diagram, i.e., $(w_k)_{ij} = R^2 f^{(k)} \delta_{ij}$ for a single path and $(w_k)_{ij} = R^2 g^{(k)}_{ij}$ for a double path, with the convention $(w_{d-1})_{ij} = 0$.

For $n = d-1$, eqs.~\eqref{max_1} and \eqref{max_3} imply
\begin{align}
& \prd A^{(d-2)} = 0 \, , \label{conserved_maxwell_1} \\[5pt]
& (d -2)\, \partial_u A_i^{(d-2)} + \lap A_i^{(d-3)} - \partial_i \prd A_j^{(d-3)} = 0 \, . \label{conserved_maxwell_2}
\end{align}
These are the analogues of eq.~\eqref{n=d-1} for the scalar: a priori these equations could imply a constraint on the source, but they are actually identically satisfied when \eqref{eom_maxwell} holds.
This is the case because the components above the vev satisfy an analogue of the mirror relation \eqref{mirror_scalar}:
\be \label{mirror_maxwell}
A_i^{\left(\frac{d}{2} - 1 + k\right)} = R^{2k}\Big( \prod_{l=0}^{k-1} g^{\left(\frac{d}{2}-1+k-2l\right)} \Big)_{\! i}^{\ j}  A_j^{\left(\frac{d}{2} - 1 -k\right)} , \qquad \text{for } 1 \leqslant k \leqslant \frac{d}{2} - 1 \, ,
\ee
where the product of the $g^{(n)}$ operators has to be understood as matrix multiplication. A proof of this relation is given in appendix~\ref{sec:mirror_anyspin}. In appendix~\ref{sec:identities}, we also prove that the spin-one mirror relation can be rewritten as
\be \label{Mirror_Spin_1_new}
A_i^{\left(\frac{d}{2} - 1 +k\right)} = \frac{R^{2k}}{\left( \frac{d}{2} - k \right)_{2k}}\, \lap^{k-1} \left[ \delta_{i}{}^{j}\lap  - \frac{2k}{\frac{d}{2} + k - 1}\, \partial_i \partial^j \right] A_j^{\left(\frac{d}{2} - 1 -k\right)} , \quad \text{for } 1 \leqslant k \leqslant \frac{d}{2} - 1 \, ,
\ee
where $(x)_n=(x+n-1)\cdots(x+1)x=\frac{\Gamma(x+n)}{\Gamma(x)}$ denotes the Pochhammer symbol (aka. rising factorial).

The second form of the mirror relation simplifies the proof that eq.~\eqref{conserved_maxwell_1} does not impose any constraints on the source. Indeed, for $k = \frac{d}{2} - 1$ eq.~\eqref{Mirror_Spin_1_new} becomes 
\be \label{A(d-2)}
A_i^{(d-2)} = \frac{R^{d-2}}{(d-2)!}\, \lap^{\frac{d}{2}-2} \left[ \lap A_i^{(0)} - \partial_i \partial\cdot A^{(0)} \right] .
\ee
Within parentheses the Maxwell operator appears, and its divergence vanishes. 
It is worth noting that this is the only order for which the celestial divergence identically vanishes, and in the limit $R \to \infty$ this condition gives rise to the single celestial conserved current entering figure~\ref{nice_graph} for $s=1$. The proof that  eq.~\eqref{conserved_maxwell_2} is identically satisfied is slightly more technical and we refer directly to the $s=1$ instance of the general proof presented in appendix~\ref{sec:source_constraints}.

%%%%%%%%%%%%%%%%%%%%%%%%%%%%%%%%%%%%%%%%%%%%%%%%
\subsection{Flat limit of the AdS solution space} \label{sec:flat-limit_maxwell}

To take the flat limit of the solution space, we follow the same steps as in section~\ref{sec:flat-limit_scalar}. We thus expand the gauge-fixed source and the transverse components of the vev in powers of the AdS radius $R$ as in \cite{Campoleoni:2023fug}:
\be \label{R_expansion_maxwell}
A_i^{(0)} = \sum_{k = 0}^{\frac{d}{2} - 1}\frac{1}{R^{2k}}\, A_i^{(0,\,k)} \,, \qquad 
A_{i}^{(d-1)} = \sum_{k = 0}^{\infty} \frac{1}{R^{2k}}\, A_i^{(d-1,\,k)} \, .
\ee
As for scalars, the upper bound on the expansion of the source avoids a redundant parametrisation of the boundary data in the $R \to \infty$ limit and it will be discussed below. 

The time component of the vev $A_u^{(d-1)}$ satisfies the evolution equation \eqref{Vev_maxwell}. As a result, the independent information it contains is captured by the $u$-independent ``integration constant''. A priori, the latter could also admit an expansion in powers of $R^{-2}$, but it is sufficient to keep only a term of order $R^0$ to reproduce the Minkowski solution space in the limit $R \to \infty$. We shall therefore make this assumption in the following and substitute $A_u^{(d-1)}$ with
\be \label{solution_Au}
\begin{split}
A_u^{(d-1)}(u ,\mathbf{x}) 
& = R^{0} \left( Q(\mathbf{x}) - \frac{d-3}{(d-1)(d-2)} \int_0^u du'\, \lap \partial\cdot A^{(d-3)}(u',\mathbf{x}) \right) \\
&\quad + \sum_{k=0}^\infty R^{-2k-2}\int_0^u du'\, \partial \cdot A^{(d-1,\,k)}(u',\mathbf{x}) \, ,
\end{split}
\ee
where
\be
    Q(\mathbf{x}) := A_u^{(d-1,\,0)}(u=0,\mathbf{x}) 
\ee
and where we stress that $A_i^{(d-3)}$ contributes only to the term of order $R^0$ thanks to the truncation in the expansion of the source (see also eq.~\eqref{d-3} below). As in section~\ref{sec:spin0}, when solving the evolution equation \eqref{Vev_maxwell} we defined for convenience the ``integration constant'' $Q(\mathbf{x})$ on
a cut of $\scri^+$ at $u = 0$, rather than at $u \to - \infty$.

Substituting these expansions in the general asymptotic solution \eqref{sum_diagrams_spin1}
of the equations of motion gives both divergent and regular contributions in the limit $R \to \infty$,
\be \label{def_notation_spin1}
A_i^{(n)} = \sum_{k} \frac{1}{R^{2k}}\,A_i^{(n,\,k)} \,,\quad \text{for}\;\; 1\leqslant n\leqslant d-2\;\; \text{or}\;\; n\geqslant d \,,
\ee
while eq.~\eqref{solution_Au} implies that $A_u^{(d-1)}$ only admits regular contributions,
\be
A_u^{(d-1)} = \sum_{k=0}^\infty \frac{1}{R^{2k}}\,A_u^{(d-1,\,k)} \,.
\ee
Imposing the vanishing of the divergent part of the on-shell field as in section~\ref{sec:flat-limit_scalar} still corresponds to imposing the Minkowski equations of motion on the regular components $A_i^{(n,\,0)}$. Indeed, eq.~\eqref{eom_maxwell} implies 
\be \label{recursion_divergences_spin1}
A_i^{(n,\,k)} = f^{(n)} A_i^{(n-1,\,k+1)} + g^{(n)\,j}_{\phantom{(n)}\!\!  i} A_j^{(n-2,\,k+1)} + \frac{1}{d}\,  \delta_{n,d}\, \partial_i A_u^{(d-1,\,k+1)} \quad \text{for}\ n \neq d-1
\ee
and this ensures that all divergent contributions, corresponding to the $A_i^{(n,\,k)}$ with $k < 0$, vanish once the $k=-1$ term is set to zero, i.e., once
\be \label{flat_constraints_spin1}
(2n-d+2)\, \partial_u A_i^{(n,\,0)} + \lap A_i^{(n-1,\,0)} + \frac{d-2}{n(n-d+1)}\, \partial_i \partial \cdot A^{(n-1,\,0)} - \delta_{n,d-1}\, \partial_i A_u^{(d-1,\,0)}\overset{!}{=} 0 \, .
\ee
This corresponds to the flat limit of eq.~\eqref{max_3}, taking into account the $R$-independent eq.~\eqref{max_1}. As for scalars, we thus identified the coefficients of the asymptotic $z$-expansion in Minkowski spacetime with the $A_i^{(n,\,0)}$ terms in the expansion \eqref{def_notation_spin1}. All in all, the $R \to \infty$ limit of the relation \eqref{eom_maxwell}, fixing recursively the $A_i^{(n)}$, works as the limit of the corresponding relation \eqref{recursion_path} for a scalar of mass $m^2 R^2 = -2(d-1)$. The main difference is that 
one also has to handle the component $A_u^{(d-1)}$ of the vev, which brings the additional $u$-independent function $Q(\mathbf{x})$, related to the charge aspect in the limit $R \to \infty$.

We now discuss how the components $A_i^{(n,\,0)}$ are related to the terms in the expansions of the source and the vev in \eqref{R_expansion_maxwell} and what constraints the Minkowski equations of motion \eqref{flat_constraints_spin1} impose on the latter. As for scalars, we follow the order in which the various terms enter figure~\ref{nice_graph}.

\paragraph{Overleading pure-gauge mode.}

We begin by noticing that, again in full analogy with the scalar case (cf.~eq.~\eqref{phi_d-2}), the mirror relation \eqref{Mirror_Spin_1_new} implies at $k = \frac{d}{2} - 1$ and $k = \frac{d}{2} - 2$, respectively,
\begin{align}
A_i^{(d-2)} & = \frac{1}{(d-2)!} \sum_{k=0}^{\frac{d}{2} -1}R^{d-2 - 2k}\left(\delta_i{}^j \lap - \partial_i\partial^j \right) \lap^{\frac{d}{2} -2} A_j^{(0,\,k)} \, , \label{d-2} \\
A_i^{(d-3)} & = -\,\frac{1}{(d-3)!}  \sum_{k = 0}^{\frac{d}{2} - 1}R^{d - 2 - 2k } \left(\delta_i{}^j \lap - \frac{d-4}{d-3}\, \partial_i\partial^j \right)  \lap^{\frac{d}{2}-3} \partial_u A_j^{(0,\,k)}\, , \label{d-3}
\end{align}
where the first relation can also be recovered by substituting the expansion \eqref{R_expansion_maxwell} in eq.~\eqref{A(d-2)}.
Asking for a smooth flat limit therefore forces the components $A_i^{(0,\,k)}$ in the expansion of the source with $0 \leqslant k \leqslant \frac{d}{2} -2$  to satisfy the constraints
\begin{align} 
& \left(\delta_i{}^j\lap - \partial_i\partial^j \right) \lap^{\frac{d}{2} -2} A_j^{(0,\,k)} = 0 \,, \label{d-2_const} \\[5pt]
& \lap^{\frac{d}{2}-2} \,\partial_u \partial \, \cdot \, A^{(0,\,k)} = 0 \,, \label{d-3_const}
\end{align}
where the second one follows by computing the divergence of the differential operator entering \eqref{d-3}.

As for scalars, one could then impose boundary conditions on the transverse plane $\mathbb{R}^d$ corresponding to regularity on the celestial sphere, so that 
\be \label{bdy_cond}
\lap h(\mathbf{x}) = 0 \implies h(\mathbf{x}) = 0 \, .
\ee
Imposing this condition allows one to recover in the flat limit the Minkowski solution space of \cite{Kapec:2014zla, Campoleoni:2020ejn}. We therefore make this assumption in the following, and defer to the end of this section a discussion on how the Minkowski solution space is modified if one relaxes it.
When \eqref{bdy_cond} holds, the constraints \eqref{d-2_const} and \eqref{d-3_const} imply that the admissible $A_i^{(0,\,k)}$ must satisfy
\be \label{consts_both}
\left(\delta_i{}^j \lap - \partial_i\partial^j \right)A_j^{(0,\,k)} = 0 \,,\qquad \partial_u A_i^{(0,\,k)} = 0\,, \qquad \text{for} \ 0 \leqslant k \leqslant \frac{d}{2} -2 \, ,
\ee
where the second constraint is obtained by combining the derivative in retarded time of \eqref{d-2_const} with \eqref{d-3_const}. 
We stress that the previous analysis does not impose any constraint on $A_i^{(0,\,\frac{d}{2}-1)}$ because, thanks to the mirror relation \eqref{mirror_maxwell}, it does not contain divergent terms above the order $z^{d-1}$.

In the notation of eq.~\eqref{eom_maxwell}, these constraints can be translated into
\be \label{consts_both_2}
g^{(2)\,j}_{\phantom{(2)}\!\!  i} A_j^{(0,\,k)} = 0 \,,\qquad f^{(n)} A_i^{(0,\,k)} = 0\,, \qquad \text{for} \ 0 \leqslant k \leqslant \frac{d}{2} -2 \, ,
\ee
so that, under the assumption \eqref{bdy_cond}, only the component $A_i^{(0,\,\frac{d}{2}-1)}$ in the expansion of the source can contribute to the $A_i^{(n,\,0)}$ with $n > 0$: their expression in terms of the source contains indeed either a $f^{(n)}$ or a chain of double paths only. The latter would however always contain $g^{(2)}_{ij}$. Therefore, even if the conditions \eqref{consts_both} admit non-trivial solutions for any value of $k$, the corresponding non-vanishing $A_i^{(0,\,k)}$ cannot appear at any order below that of the source. Note also that in the range $1 \leqslant n \leqslant \frac{d}{2}-1$, the numerical coefficient in the operator $f^{(n)}$ never vanishes, so that the leading term in $R$ in the expansion of $A_i^{(n)}$ contains only single paths:
\be
A_i^{(n)} = R^{2n} \prod_{k = 1}^{n} f^{(k)} A_i^{(0)} + \cO(R^{2n-2}) \propto R^{2n-d+2}\, \partial_u^{n} A_i^{(0,\,\frac{d}{2}-1)} + \cO(R^{2n-d}) \, ,
\ee
where we took into account that the other terms in the expansion of the source cannot appear in the $A_i^{(n)}$ with $n > 0$ when \eqref{bdy_cond} holds.
This expression manifests that in the range $1 \leqslant n \leqslant \frac{d}{2}-2$, i.e.\ above the order in the $z$-expansion where radiation is located in Minkowski spacetime, $A_i^{(n)}$ only contains negative powers of $R^2$ and thus vanishes in the limit $R \to \infty$. This argument does not apply at order $z^0$, where the regular term in the expansion \eqref{R_expansion_maxwell} remains there. 

In conclusion, imposing boundary conditions on the transverse plane $\mathbb{R}^d$ such that \eqref{bdy_cond} holds, we obtain in the flat limit a pure-gauge mode at order $z^0$ satisfying
\be \label{pure gauge_spin1}
\partial_u A_i^{(0,\,0)} = 0 \,, \qquad 
\left(\delta_i{}^j\lap - \partial_i\partial^j \right)A_j^{(0,\,0)} = 0 \, ,
\ee
while the subleading orders till $z^{\frac{d}{2}-1}$ vanish:
\be \label{no-EO-spin1}
A_i^{(n,\,0)}(u,\mathbf{x}) = 0  \qquad \text{for}\ 1 \leqslant n \leqslant \frac{d}{2} - 2 \, .
\ee
We called $A_i^{(0,\,0)}$ pure-gauge mode because, again under the assumption \eqref{bdy_cond}, the only solution of the constraints \eqref{pure gauge_spin1} is $A_i^{(0,\,0)}(u,\mathbf{x}) = \partial_i \lambda(\mathbf{x})$. Up to this stage, its presence in the limit $R \to \infty$ is the main difference with respect to the scalar example. As we anticipated, this Minkowski solution space, including only a pure-gauge contribution above radiation, matches that of references \cite{Kapec:2014zla, Campoleoni:2020ejn}. The pure-gauge mode corresponds to the residual gauge symmetry of the radial gauge after imposing the boundary condition \eqref{extra_bnd-cond}, and in the same references this asymptotic symmetry has been related to soft theorems.

\paragraph{Radiation.} 

Taking into account the previous discussion, in the limit $R \to \infty$, one obtains 
\be
A_i(z,u,\mathbf{x}) = \partial_i\lambda(\mathbf{x}) + z^{\frac{d}{2}-1} \, \frac{(-1)^{\frac{d}{2} - 1}}{(d-3)!!}\, \partial_u^{\frac{d}{2} - 1}A_i^{\left(0,\,\frac{d}{2} - 1\right)}(u,\mathbf{x}) + \cO(z^{\frac{d}{2}}) \,.
\ee
At order $z^{\frac{d}{2}-1}$ we find a function of retarded time, in agreement with the structure of the Minkowski equation of motion \eqref{flat_constraints_spin1}. In analogy with the analysis of section~\ref{sec:flat-limit_scalar}, the function $A_i^{(0,\,\frac{d}{2} - 1)}(u,\mathbf{x})$ is completely free thanks to the mirror relation \eqref{mirror_maxwell}. For instance, at the next order, where this function could potentially enter a divergent term, one has
\be
A_i^{\left(\frac{d}{2}\right)} = R^2\, g^{\left(\frac{d}{2}\right)\,j}_{\phantom{(n)}\!\!  i}A_j^{\left(\frac{d}{2} -2\right)} \quad \textrm{with} \quad A_j^{\left(\frac{d}{2} -2 \right)} = R^{-2} \prod_{k=1}^{\frac{d}{2}-2} f^{(k)} A_j^{\left(0,\,\frac{d}{2}-1\right)} + \cO(R^{-4}) \, .
\ee
In general, for $1 \leqslant k \leqslant \frac{d}{2}-1$, i.e.\ above the order of the vev, the mirror relation implies 
\be
A_i^{\left(\frac{d}{2} -1 +k\right)} =  \Big( \prod_{m=1}^{\frac{d}{2}-1 -k}f^{\left(m\right)} \Big) \Big( \prod_{l=0}^{k-1} g^{\left(\frac{d}{2} -1 +k-2l \right)} \Big)_{\! i}^{\ j} A_j^{\left(0,\,\frac{d}{2} -1 \right)} + \cO(R^{-2}) \,,
\ee
so that, using \eqref{Mirror_Spin_1_new},
\begin{align}
&A_i(z,u,\mathbf{x}) = \partial_i\lambda(\mathbf{x}) + z^{\frac{d}{2}-1} \frac{(-1)^{\frac{d}{2} - 1}}{(d-3)!!}\, \partial_u^{\frac{d}{2} - 1}A_i^{(0,\,\frac{d}{2} - 1)} \nonumber \\
&+\sum_{k = 1}^{\frac{d}{2}-1} \frac{z^{\frac{d}{2}-1+k} (-1)^{\frac{d}{2}-1+k}}{ (2k)!! (d-3)!!} \, \partial_u^{\frac{d}{2}-1-k} \left( \delta_i{}^j \lap - \frac{2k}{\frac{d}{2} +k-1}\, \partial_i\partial^j \right) \lap^{k-1}A_j^{(0,\,\frac{d}{2} - 1)} + z^{d-1}A_i^{(d-1,\,0)} \nonumber \\[8pt]
&- z^d \Bigg[ R^2 \left(\partial_u A_i^{(d-1,\,0)} + \frac{1}{d(d-2)!} \left(\delta_i{}^j \lap - \partial_i\partial^j \right) \lap^{\frac{d}{2} -1} A_j^{(0,\,\frac{d}{2} -1)}- \frac{1}{d}\,\partial_i A_u^{(d-1,\,0)}  \right) \nonumber\\
& \qquad\qquad + \partial_u A_i^{(d-1,\,1)} - \frac{1}{d}\, \int_0^u du' \partial_i\partial \cdot A^{(d-1,\,0)}(u',\mathbf{x}) + \cO(R^{-2})\Bigg] \, + \cO(z^{d+1}) \,. 
 \label{above_vev_spin1}
\end{align}
As in eq.~\eqref{above-vev}, the component $A_i^{(0,\,\frac{d}{2} - 1)}$ of the source enters the divergent term in the third line, but imposing its cancellation only fixes $A_i^{(d-1,\,0)}$ in terms of the source, up to an arbitrary one-form of the transverse coordinates. Once this condition is satisfied, the recursion relation \eqref{recursion_divergences_spin1} guarantees that all divergent contributions involving the source at the subleading orders in $z$ vanish identically, thus confirming that $A_i^{(0,\,\frac{d}{2} - 1)}$ is not constrained. 
Since this function of retarded time is completely free, we can define the spin-one shear vector
\be \label{radiation_spin1}
C_i(u,\mathbf{x}) := \frac{(-1)^{\frac{d}{2} - 1}}{(d-3)!!}\, \partial_u^{\frac{d}{2} - 1}A_i^{\left(0,\,\frac{d}{2} - 1\right)}(u,\mathbf{x}) \,, 
\ee
which encodes radiative modes in Minkowski spacetime.

\paragraph{Source Chthonians.} 

In terms of the shear vector, one can rewrite the terms between radiative and Coulombic order in \eqref{above_vev_spin1} as 
\be \label{above-vev-full_spin1}
\begin{split}
A_i^{(\frac{d}{2} -1 + k ,\,0 )} &= \frac{(-1)^k}{(2k)!!} \left( \delta_i{}^j \lap -  \frac{2k}{\frac{d}{2} +k-1}\, \partial_i\partial^j \right) \lap^{k-1} a^{(k)}_j(u,\mathbf{x}) \, , \\
\textrm{with}\ \,  a_j^{(k)}(u,\mathbf{x}) & = \partial_u^{-k} C_j + \frac{(-1)^{\frac{d}{2}-1}}{(d-3)!!} \sum_{l=0}^{k-1} \frac{u^l}{l!} \partial_u^{\frac{d}{2}-1-k+l} A_j^{(0,\,\frac{d}{2}-1)} (u=0,\mathbf x) \,,
\end{split}
\ee
where we used the operator $\partial_u^{-1}$ defined in \eqref{uintegral}. The Minkowski equation of motion \eqref{flat_constraints_spin1} fixes recursively the derivative in retarded time of $A_i^{(n)}$ in terms of the previous terms in the $z$-expansion. As such, at each order a new function independent of retarded time appears. These are the Chthonians in figure~\ref{nice_graph}, and eq.~\eqref{above-vev-full_spin1} shows that those appearing from the order $z^{\frac{d}{2}}$ till $z^{d-2}$ can be expressed solely in terms of the source as
\be \label{1st-Cht-spin1}
A_i^{(\frac{d}{2}-1+k,\,0)}\eval_{u=0} = \frac{(-1)^{\frac{d}{2}-1+k}}{(2k)!!(d-3)!!} \left( \delta_{i}{}^{j} \lap - \frac{2k}{\frac{d}{2} +k-1}\, \partial_i\partial^j \right) \lap^{k-1}  \partial_u^{\frac{d}{2}-1-k} A_j^{(0,\, \frac{d}{2}-1)} (0,\mathbf x) \, ,
\ee
for $1 \leqslant k \leqslant \frac{d}{2}-1$. As a result, those Chthonians appearing from the order $z^\frac{d}{2}$ till $z^{d-3}$, that do not have any peculiar property, are denoted as ``source Chthonians'' in figure~\ref{nice_graph}.

\paragraph{Celestial conserved current.}

While eq.~\eqref{1st-Cht-spin1} also holds for $k = \frac{d}{2}-1$ or, equivalently, at order $z^{d-2}$, the structure of the differential operator it contains is such that the divergence of $A_i^{\left(\frac{d}{2}-1+k\right)}$ only vanishes for $k = \frac{d}{2}-1$. In this case, the source Chthonian satisfies the celestial conservation equation
\be
\partial \cdot A^{(d-2,\,0)} = 0 \,,
\ee
which can be evaluated at $u =0$ in order to have a condition for the celestial field itself. Hence this source Chthonian corresponds to the single celestial conserved current that enters figure~\ref{nice_graph} for $s=1$.

\paragraph{Charge aspect.} We already solved the evolution equation \eqref{Vev_maxwell} for the temporal component of the vev in eq.~\eqref{solution_Au}. In the limit $R \to \infty$ and expressing everything in terms of the shear vector \eqref{radiation_spin1} and the source Chthonians \eqref{1st-Cht-spin1}, this gives 
\be \label{Q_final_C_expression}
\begin{split}
A_u^{(d-1,\,0)}(u,\mathbf{x}) & = Q(\mathbf{x}) + \frac{1}{(d-1)!}\, \lap^{\frac{d}{2}-1} \left(  \partial \cdot A^{\left(0,\,\frac{d}{2} - 1\right)}(u,\mathbf{x}) - \partial \cdot A^{(0,\,\frac{d}{2} - 1)}(0,\mathbf{x}) \right) \\
&= Q(\mathbf{x}) + \frac{(-1)^{\frac{d}{2}-1}}{(d-1)(d-2)!!}\, \lap^{\frac{d}{2}-1}  \partial_u^{-\frac{d}{2}+1} \partial \cdot C(u,\mathbf{x})  
\\&\qquad+ \frac{1}{(d-1)!} \sum_{l=1}^{\frac{d}{2}-2} \frac{u^l}{l!}\, \lap^{\frac{d}{2}-1} \partial_u^l \partial \cdot  A^{(0,\,\frac{d}{2}-1)}(0,\mathbf{x}) \, .
\end{split}
\ee
This is also the explicit solution of the Minkowski evolution equation for the charge aspect, corresponding to the flat limit of eq.~\eqref{Vev_maxwell}. The independent boundary datum $Q(\mathbf{x})$ thus corresponds to the single celestial field entering the right-hand side of figure~\ref{boring_graph} for $s=1$.

\paragraph{Vev Chthonians.}

The terms below the order $z^{d-1}$ of the vev are organised as in the scalar case. The regular terms $A_i^{(d-1+n,\,0)}$ with $n > 0$ are subject to the flat-space evolution equations \eqref{flat_constraints_spin1}.
This leaves exactly one integration function on the transverse plane for each integer $n\geqslant 0$, providing another set of Chthonians originating from the vev, to be distinguished from the previous set of Chthonians originating from the expansion of the source. The evolution equations are informed recursively of both the radiation $C_i(u,\mathbf{x})$ and the source Chthonians through the evolution equation for $A_i^{(d-1,\,0)}$, but the source cannot contribute directly to the Chthonians (that is to the ``integration constants'') below the vev, from a simple counting of the involved powers of $R^2$. The source can indeed only enter through divergent terms $A_i^{(n,\,k<0)}$, that vanish on account of the recursion relation \eqref{recursion_divergences_spin1} once the cancellation of the third line in eq.~\eqref{above_vev_spin1} has been enforced. Had we continued the expansion of the source in eq.~\eqref{R_expansion_maxwell} beyond the order $R^{2-d}$, we would have recovered additional contributions from the source in the vev Chthonians. However, these would have been spurious for the same reason as that detailed for scalars in appendix~\ref{app:scalar_chthonians}: at each order below $z^{d-1}$ a new contribution coming from the expansion of the vev also appears, so that the independent boundary data in Minkowski spacetime would be given by a linear combination of the source and vev contributions, with the two never appearing separately in the solution space. As such, it is convenient to set to zero the unnecessary contributions from the source with the truncation in the expansion \eqref{R_expansion_maxwell}. This choice is also in agreement with the group-theoretical analysis of the limit in section~\ref{sec:group-theory}.

In order to express the contribution to $A_i^{(d-1+n,\,0)}$ of the components $A_i^{(d-1,\,k)}$ in the expansion of the vev, we can use the recursion \eqref{recursion_divergences_spin1}, summing over an increasing number of diagrams connecting $d-1$ to $d-1+n$. As for scalars, the explicit expression becomes increasingly cumbersome as $n$ increases, but the Chthonians can be identified by noticing that the single-path contribution is always present in the diagrams:
\be \label{vev_Chth_spin1}
    A_i^{(d-1+n,\,0)}\, =\, \prod_{k=1}^n f^{(d-1+k)} A_i^{(d-1,\,n)} + \cdots\; \propto\; \partial_u^n A_i^{(d-1,\,n)} + \cdots \,,
\ee
where dots represent contributions involving at least one $g^{(n)}_{ij}$ and a smaller number of derivatives in retarded time.
Also in this case, at each order $n$, the expression of $A_i^{(d-1+n,\,0)}$ involves the new coefficient $A_i^{(d-1,\,n)}$, entering with $\partial_u^n$ derivatives, as well as the $A_i^{(d-1,\,k < n)}$, entering with less retarded-time derivatives (cf.~\eqref{vev_Chth_spin1}). The evolution equation of $A_i^{(d-1+n,\,0)}$ then implies that the $\partial_u^{n+1}$ derivative of $A_i^{(d-1,\,n)}$ is constrained, so that it effectively provides only a new function on the transverse plane. We can identify it with the Chthonian $A_i^{(d-1+n,\,0)}\eval_{u=0}$ by setting $\partial_u^k A_i^{(d-1,\,n)}(u=0,\mathbf x) = 0$ for all $0 \leqslant k \leqslant n-1$. In analogy with the analysis performed for scalars in appendix~\ref{app:scalar_chthonians}, this condition is consistent because these terms always enter the expression for $A_i^{(d-1+n,\,0)}$ together with the new term $\partial_u^n A_i^{(d-1,\,n)}$, so that they would just introduce a redundancy in the parameterisation of the solution space.
Eventually, this condition allows one to obtain the following simple expression for the flat space Chthonians below the order $z^{d-1}$: 
\be \label{eq_vev-Chthonians_realisation_spin1}
    A_i^{(d-1+n,\,0)}\eval_{u=0} =(-1)^n \frac{(d-1)!! (d+2n-2)!!}{n! (d+n-1)!} \, \partial_u^n A_i^{(d-1,\,n)}(u=0,\mathbf x) \,.
\ee

\paragraph{Erdmenger--Osborn fields.} If we relax the assumption \eqref{bdy_cond}, it is no longer true that in the limit $R \to \infty$ only the pure-gauge mode at order $z^0$ survives, and non-vanishing regular terms $A_i^{(n,\,0)}$ that satisfy the Minkowski equations of motion \eqref{flat_constraints_spin1} appear. In analogy with the case of the vev Chthonians discussed above, a new integration function depending only on the transverse coordinates thus arises for each $0 \leqslant n \leqslant \frac{d}{2} - 2$. On the other hand, the mirror relation \eqref{Mirror_Spin_1_new} implies that the regular term must satisfy the condition
\be \label{eq_erd-osborn_spin1}
    \left( \delta_{i}{}^j \lap - \frac{d-2-2n}{d-2-n}\, \partial_i\partial^j \right) \lap^{\frac{d}{2}-2-n} A_j^{(n,\,0)} = 0 \qquad \text{for } 0 \leqslant n \leqslant \frac{d}{2} - 2 \,,
\ee
that guarantees the cancellation of divergences of order $R^{2k}$ in the terms below radiation. 
Moreover, in analogy with the discussion of the vev Chthonians, at each subleading order a new term in the expansion of the source enters with the maximal number of retarded-time derivatives,
\be
    A_i^{(n,\,0)} \,=\, \prod_{k=1}^n f^{(k)} A_i^{(0,\,n)} + \cdots\;\propto\; \partial_u^{n} A_i^{(0,\,n)} + \cdots
\ee
By adopting the same strategy as for the Chthonians, i.e.\ imposing $\partial_u^k A_i^{(0,\,n)}\eval_{u=0} = 0$  
for $k < n$, we can conclude that the $z$ expansion in Minkowski spacetime also contains at the orders $z^{n}$ with $0 \leqslant n \leqslant \frac{d}{2}-2$ the coefficients
\be \label{eq_erd-osborn_realisation_spin1}
    A_i^{(n,\,0)}\eval_{u=0} = (-1)^n \frac{(d-n-2)!}{n! (d-3)!! (d-2n-2)!!} \, \partial_u^n A_i^{(0,\,n)}\eval_{u=0} \,,
\ee
satisfying however also eq.~\eqref{eq_erd-osborn_spin1}. These are the Erdmenger--Osborn celestial fields in figure~\ref{nice_graph}. Note that the celestial field appearing at order $z^0$  still has a gauge symmetry, but one should now consider all solutions of eq.~\eqref{eq_erd-osborn_spin1}, and not only the pure-gauge mode that we discussed above.

%%%%%%%%%%%%%%%%%%%%%%%%%%%%%%%%%%%%%%%%%%%%%%%%%%%%%%%
\section{Spin two and higher}\label{sec:HS}

In this section, we extend the techniques developed for spin-one fields to fields of arbitrary spin, emphasising the features that already emerge for linearised gravity. Before discussing the flat limit of the AdS solution space, we first justify the on-shell gauge fixing we employ in the following with a detailed analysis of the spin-two case.

%%%%%%%%%%%%%%%%%%%%%%%%%%%%%%%%%%%%%%%%%%%%%%%%%%%%%%%
\subsection{Bondi-like gauge: the spin-2 example} \label{sec:bondi-like_gauge}

In most of the literature describing gravitational fluctuations near (future) null infinity, a Bondi--Sachs gauge \cite{Bondi:1962px, Sachs:1962wk}, or generalisations thereof \cite{Barnich:2010eb, Geiller:2022vto}, is employed. This gauge also finds a counterpart when the cosmological constant is negative \cite{Poole:2018koa,Compere:2019bua,Campoleoni:2023fug}. For linearised perturbations around the metric \eqref{AdS_Bondi}, the analogue of the gauge-fixing conditions defining the Bondi--Sachs gauge read
\be \label{Bondi-Sachs}
    h_{zz} = h_{zi} = 0 \quad \text{and} \quad \delta^{ij} h_{ij} = 0 \,.
\ee
As anticipated in section~\ref{sec:summary}, this imposes $(d+2)$ conditions, that require to involve all independent components of the gauge parameter to be reached. The component $h_{uz}$, which is set to zero in eq.~\eqref{Bondi-like_intro}, is instead left free in the Bondi--Sachs gauge. 
On the other hand, we now show that, on the space of solutions to the linearised vacuum Einstein equations, one can impose the additional constraint $h_{uz} = 0$ using the residual gauge symmetry left by \eqref{Bondi-Sachs}, thus completing the set of conditions \eqref{Bondi-like_intro}, that have been dubbed ``Bondi-like gauge'' in \cite{Campoleoni:2017mbt, Campoleoni:2017qot, Campoleoni:2020ejn}.

To reach this conclusion, it is sufficient to impose the Bondi--Sachs gauge \eqref{Bondi-Sachs} and to consider the following components of the equations of motion:
\begin{align}
    R_{zz} & = 2d \left( z\partial_z + 2 \right) h_{uz}=0 \,, \\[10pt]
    R_{uz} - \frac{2(d+1)}{R^2}\, h_{uz} & = - \left( z^2 \partial_z^2 - (d-2)\, z\partial_z \right) h_{uu} - z^2\, \partial_z \partial^i h_{ui} \nonumber \\
    &\quad + \frac{2}{R^2} \left( z\partial_z - 2d \right) h_{uz} + 2z \left( z\partial_z + 2 \right) \partial_u h_{uz} + z^2 \lap h_{uz}=0 \,.
\end{align}
The first equation already sets to zero almost all terms in the expansion of $h_{uz}$ in powers of $z$, with the exception of the one at order $z^{-2}$:
\be
    h_{uz}(z,u,\mathbf{x}) = z^{-2} h_{uz}^{(-2)}(u,\mathbf{x}) \, .
\ee
Substituting this information in the second equation, and assuming an expansion in powers of $z$ starting at order $z^{-2}$ for all components, we obtain
\be \label{huu-huz}
    h_{uu}^{(-2)}(u,\mathbf{x})  = - \frac{2}{R^2}\, h_{uz}^{(-2)}(u,\mathbf{x})  \, .
\ee
This coincides with the linearisation of the solution space obtained in \cite{Compere:2019bua, Geiller:2022vto, Geiller:2024amx}.

The residual linearised diffeomorphisms of the Bondi--Sachs gauge are generated by
\begin{subequations}
\begin{align}
    \epsilon_z & = z^{-2} f(u,\mathbf{x}) \, , \\[5pt]
    \epsilon_i & = z^{-2} v_i(u,\mathbf{x}) - z^{-1} \partial_i f(u,\mathbf{x}) \, , \\
    \epsilon_u & = - \frac{z^{-2}}{R^2} f + \frac{z^{-1}}{d}\, \partial\cdot v - \frac{1}{d}\, \lap f \, ,
\end{align}
\end{subequations}
and they induce the variations
\begin{subequations} \label{eq_variation_spin2}
\begin{align} 
    \delta h_{uz} & = z^{-2} \left( \partial_u f - \frac{1}{d}\, \partial\cdot v \right) , \\
    \delta h_{uu} & = - \frac{2z^{-2}}{R^2} \left( \partial_u f - \frac{1}{d}\, \partial\cdot v \right) + \frac{2z^{-1}}{d} \left( \partial_u \partial\cdot v - \frac{1}{R^2}\, \lap f \right) - \frac{2}{d}\, \partial_u \lap f \, , \\
    \delta h_{ui} & = z^{-2} \left( \partial_u v_i - \frac{1}{R^2}\, \partial_i f \right) - z^{-1} \partial_i \left( \partial_u f - \frac{1}{d}\, \partial\cdot v \right) - \frac{1}{d}\, \lap \partial_i f \, . \label{delta_hui} 
\end{align}
\end{subequations}
Therefore, the surviving contribution at order $z^{-2}$ in $h_{uz}$ can be set to zero with a further gauge fixing that, consistently with eq.~\eqref{huu-huz}, also sets to zero the contribution of order $z^{-2}$ in $h_{uu}$. This completes the proof that the ``Bondi-like gauge'' \eqref{Bondi-like_intro} can be reached on-shell. 

Let us stress that eqs.~\eqref{eq_variation_spin2} show that one can also impose the vanishing of $h^{(-2)}_{ui}$ with an on-shell gauge fixing. Setting to zero the order $z^{-2}$ in all field components with the exception of $h_{ij}$, we perform a partial gauge fixing of the AdS source in the same spirit as what we did in section~\ref{sec:spin1} for a Maxwell field (cf.~eq.~\eqref{extra_bnd-cond}). This also corresponds to the linearisation of the $\Lambda$-BMS$_{d+2}$ boundary conditions imposed in \cite{Fiorucci:2020xto}. In this case, the residual gauge symmetries are generated by two functions of the angles but independent of $u$, obtained as the solutions of the conditions
\be \label{constr-gauge_spin2}
\partial_u f - \frac{1}{d}\, \partial\cdot v  = 0 \, , \qquad
\partial_u v_i - \frac{1}{R^2}\, \partial_i f = 0 \, ,
\ee
whose non-linear version already appeared in \cite{Compere:2019bua} for $d=2$. Note that setting to zero the orders $z^{-2}$ in $h_{uu}$, $h_{uz}$ and $h_{ui}$ can be interpreted as specifying the boundary conditions on these components in the Bondi--Sachs gauge \eqref{Bondi-Sachs}. From this perspective, the vanishing of the whole $h_{uz}$ would follow from the analysis of the equations of motion. 

For fields of spin $s > 2$ we refrain from performing such a detailed analysis, and we assume that the Bondi-like gauge \eqref{Bondi-like_intro}, which greatly simplifies the study of the space of solutions, can be reached on-shell. Furthermore, we shall impose additional boundary conditions, corresponding to gauge-fixing to zero all components of the source sitting in the field components with at least one $u$ index.

\subsection{AdS solution space in Bondi-like gauge}

We parametrise the background as in eq.~\eqref{AdS_Bondi} and, following the discussion in the previous subsection, we work in the Bondi-like gauge \eqref{Bondi-like_intro}, that we recall here for the reader's convenience:
\be \label{Bondi-like}
    \varphi_{z\mu(s-1)} = 0 \, , \qquad \delta^{ij} \varphi_{ij\mu(s-2)} = 0 \, .
\ee
We expand the remaining field components in integer powers of $z$,
\be \label{z-expansion_HS}
    \varphi_{u(s-k)i(k)} = \sum_n z^n \varphi^{(n)}_{u(s-k)i(k)} \, ,
\ee
where the lower extremum of each sum may depend on the transverse and traceless tensor under consideration.

\paragraph{Equations of motion.} The conditions \eqref{Bondi-like} imply that the only components of the Fronsdal tensor \eqref{Maxwell-like} that do not vanish identically are those with at most one index $z$. Plugging the expansion \eqref{z-expansion_HS} in the non-vanishing components gives
\be \label{eom_constraint_HS}
    \mathcal{F}^{(n)}_{z u(s-k-1) i(k)} = (n+2k)(d-1-n)\, \varphi^{(n)}_{u(s-k)i(k)} - (n+2k-1)\, \partial \cdot \varphi^{(n-1)}_{u(s-k-1)i(k)} = 0
\ee
and
\begin{align}
    \mathcal{F}^{(n)}_{u(s-k) i(k)} & = \frac{1}{R^2}\Big[\big(n-d+1\big)\big( n-(s-k)(s-k-1)+2(s-1) \big) \Big] \varphi^{(n)}_{u(s-k)i(k)} \nn \\
    &+ k(k-1)(n-d+1)\, \delta_{ii} \varphi^{(n)}_{u(s-k+2)i(k-2)} \nn \\
    & - \Big[ (s-k-2)(n-d+1)-(2s+d-4) \Big] \partial_u \varphi^{(n-1)}_{u(s-k)i(k)} \nn\\
    & - \frac{(s-k)(s-k-1)}{R^2}\, \partial \cdot \varphi^{(n-1)}_{u(s-k-1)i(k)} - k(n-d+1)\, \partial_i \varphi^{(n-1)}_{u(s-k+1)i(k-1)} \quad \nn \\
    & + k(k-1)\, \delta_{ii} \partial \cdot \varphi^{(n-1)}_{u(s-k+1)i(k-2)}+ \Delta \varphi^{(n-2)}_{u(s-k)i(k)} - k\, \partial_i \partial \cdot \varphi^{(n-2)}_{u(s-k)i(k-1)} \nn \\
    &- (s-k)\, \partial_u \partial \cdot \varphi^{(n-2)}_{u(s-k-1)i(k)} = 0
\label{eom_HS}
\end{align}
(see appendix~\ref{sec:eom_Bondi} for more details).
In analogy with the spin-one case, the $s$ equations in \eqref{eom_constraint_HS} have the same form in both AdS and Minkowski backgrounds, while the $s+1$ equations in \eqref{eom_HS} are algebraic in AdS and differential in Minkowski. 

The Fronsdal tensor \eqref{Maxwell-like} satisfies the Bianchi identities 
\be \label{Bianchi}
\cB_{\mu(s-1)} := \nabla\cdot \cF_{\mu(s-1)} - \frac{s-1}{2}\, \nabla_{\!\mu} \cF_{\mu(s-2)\alpha}{}^\alpha = 0 \, ,
\ee
where the right-hand side vanishes because we work with traceless fields (see, e.g., \cite{Francia:2002pt} for the full expression). The expansion of these identities in flat Bondi coordinates is given in appendix~\ref{sec:eom_Bondi}. Their components with a single $z$ index imply that, when eqs.~\eqref{eom_constraint_HS} are satisfied at all orders, i.e.\ assuming $\mathcal{F}^{(n)}_{z u(s-k-1) i(k)} = 0$ for all $n$ and for $0 \leqslant k \leqslant s-1$, almost all remaining components of the Fronsdal tensor are bound to be traceless:
\be 
(n + 2k)\, \mathcal{F}^{(n)\,\prime}_{u(s-k-2) i(k)} = 0 \, ,
\ee
where a prime denotes a transverse trace, i.e., a contraction with $\delta^{ij}$. The surviving traces impose the conditions
\be \label{extra_constraint_HS}
    \mathcal{F}^{(-2k)\,\prime}_{u(s-k-2) i(k)} = 2(-2k-d+1)(d+2k)\, \varphi^{(-2k)}_{u(s-k)i(k)} - 2\, \partial \cdot \partial \cdot \varphi^{(-2k-2)}_{u(s-k-2)i(k)} = 0 \,,
\ee
where we discarded all terms involving a transverse trace of the field, on account of the Bondi-like gauge, and the terms in $\partial \cdot \varphi^{(-2k-1)}_{u(s-k-1)i(k)}$ using eq.~\eqref{eom_constraint_HS} for $n = -2k$. Therefore, eq.~\eqref{extra_constraint_HS} expresses $\varphi^{(-2k)}_{u(s-k)i(k)}$ in terms of the double divergence of $\varphi^{(-2k-2)}_{u(s-k-2)i(k+2)}$, for $0 \leqslant k \leqslant s-2$.
Taking into account both eqs.~\eqref{eom_constraint_HS} and \eqref{extra_constraint_HS} the components \eqref{Bianchi_2} of the Bianchi identities without any $z$ index eventually reduce to
\be \label{Bianchi_HS}
    (d-n)\, \mathcal{F}^{(n)}_{u(k)i(s-k)} = \partial \cdot \mathcal{F}_{u(k-1)i(s-k)}^{(n-1)} \, . 
\ee
This shows that, in analogy with the spin-one case, when the equations of motion $\cF_{i(s)} = 0$ and $\cF_{zu(s-k-1)i(k)} = 0$ are satisfied, the remaining components of the equation of motion only give the additional algebraic equations \eqref{extra_constraint_HS} and the following evolution equations for the longitudinal components at Coulombic order, arising from $\cF^{(d)}_{u(s-k)i(k)} = 0$:
\be \label{evolution_HS}
\begin{split}
    &(d+k+s-2) \partial_u \varphi^{(d-1)}_{u(s-k)i(k)} - k \partial_i \varphi^{(d-1)}_{u(s-k+1)i(k-1)} + \frac{k(k-1)}{d+2k-4} \delta_{ii} \partial \cdot \varphi^{(d-1)}_{u(s-k+1)i(k-2)} \\
    &\quad - \frac{(d+k+s-2)(d+k+s-1)}{(d+2k) R^2}\, \partial \cdot \varphi^{(d-1)}_{u(s-k-1)i(k)} + \lap \varphi^{(d-2)}_{u(s-k)i(k)} = 0 \,.
\end{split}
\ee

\paragraph{Source and vev.}

Imposing $k = s$ in eq.~\eqref{eom_HS}, one can observe that asking for a finite leading power in the expansion of $\varphi_{i(s)}$ fixes it to be of $\cO(z^{2-2s})$. Eqs.~\eqref{eom_constraint_HS} then imply $\varphi_{u(s-k)i(k)} = \cO(z^{1-s-k})$ for $0 \leqslant k \leqslant s-2$. This mechanism, eliminating most of the longitudinal components of the source, is the counterpart of that detailed for $s = 2$ in section~\ref{sec:bondi-like_gauge}: the Bondi-like gauge \eqref{Bondi-like} already requires by consistency boundary conditions on the longitudinal components of the fields corresponding to gauge-fixing to zero most of the longitudinal components of the source (cf.~\eqref{huu-huz}). In analogy with the spin-one case, the component of the source sitting in $\varphi_{ui(s-1)}$ at order $z^{2-2s}$ would still be compatible with our on-shell gauge fixing. On the other hand, following \cite{Campoleoni:2017mbt, Campoleoni:2017qot, Campoleoni:2020ejn}, we impose the convenient additional gauge fixing
\be \label{extra-bnd_HS}
\varphi^{(2-2s)}_{ui(s-1)} = 0 \, ,
\ee
which is the counterpart of eq.~\eqref{extra_bnd-cond}. The reachability of this condition for $s = 2$ is proven explicitly by eq.~\eqref{delta_hui}. Eventually, our boundary conditions on the non-vanishing components of the fields thus become
\be
\varphi_{u(s-k)i(k)} = \cO(z^{2-s-k}) \, .
\ee

The analysis of the equations of motion presented below shows that in AdS the independent boundary data in this setup are the free tensor $\varphi^{(2-2s)}_{i(s)}(u,\mathbf{x})$, which corresponds to the partially gauge-fixed source, as well as the free tensor $\varphi^{(d-1)}_{i(s)}(u,\mathbf{x})$ and the aspects $\varphi^{(d-1)}_{u(s-k)i(k)}(u,\mathbf{x})$ with $0 \leqslant k \leqslant s-1$ satisfying the evolution equations \eqref{evolution_HS}. In analogy with the spin-one example, the components appearing at order $z^{d-1}$ encode the AdS/CFT vev, and one can combine them with the source to define a conserved boundary traceless tensor of rank $s$. As for Maxwell fields, in the following, with a slight abuse of terminology, we shall refer to $\varphi^{(2-2s)}_{i(s)}$ as the source and to $\varphi^{(d-1)}_{u(s-k)i(k)}$ as the vev.

\paragraph{Solution space.}

Eq.~\eqref{eom_HS} at $k = s$ has a structure very close to that we already encountered in sections~\ref{sec:solution0} and \ref{sec:solution1}. Indeed, eliminating from it $\varphi^{(n)}_{u(2) i(s-2)}$ and $\varphi^{(n)}_{u i(s-1)}$ using \eqref{eom_constraint_HS}, for $n \neq d-1$ and $n \neq d$ we get
\be \label{recursion_HS}
\begin{split}
    \varphi^{(n)}_{i(s)} \;&=\; \frac{R^2}{(n - d + 1)(n - 2 + 2s)}\Bigg\{\,\big(-2(n-1)+d-2s\big)\,\partial_u\varphi^{(n-1)}_{i(s)} - \lap\varphi^{(n-2)}_{i(s)} \\
    &\quad - \, \frac{s\left(d + 2s - 4\right)}{(n - d )(n - 3 + 2s)} \,\partial_{i}\partial\!\cdot\!\varphi^{(n-2)}_{i(s-1)}
    + \,  \frac{s(s-1)}{(n - d)(n - 3 + 2s)}\,\delta_{ii}\,\partial\!\cdot\!\partial\!\cdot\!\varphi^{(n-2)}_{i(s-2)}\Bigg\} \,, 
\end{split}
\ee
which fixes all $\varphi^{(n)}_{i(s)}$ in terms of previous ones in the $z$-expansion, except for $n = d-1$ and $n = 2 -2s$. Note that the second line automatically implements the traceless projection of $\partial_{i}\partial\!\cdot\!\varphi^{(n-2)}_{i(s-1)}$. 

Before moving to a detailed analysis of this relation, let us notice that, again in analogy with the spin-one case, eqs.~\eqref{eom_constraint_HS} and the $k = s$ instance of eq.~\eqref{eom_HS} evaluated at $n = d-1$, a priori, might constrain the source. On the other hand, we prove in appendix~\ref{sec:source_constraints} that this is not the case: for instance, the expression for the fields $\varphi^{(d-2)}_{u(s-k)i(k)}$ in terms of the source is automatically divergenceless. Moreover, let us notice that the fields $\varphi^{(1-2k)}_{u(s-k)i(k)}$ are forced to vanish by eqs.~\eqref{eom_constraint_HS} evaluated at $n = 1-2k$. 

Instead of working directly with the tensors $\varphi^{(n)}_{i(s)}$, to study eq.~\eqref{recursion_HS} it is more convenient to work with a field $\Phi^{(n)}(u,\mathbf{x};v)$, defined by contracting the indices with an auxiliary vector $v^i$:
\be
    \Phi^{(n)}(u,\mathbf{x};v) := \varphi^{(n)}_{i_1 i_2 \cdots i_s}(u,\mathbf{x}) \, v^{i_1}v^{i_2}\cdots v^{i_s} \, .
\ee
This allows one to rewrite the recursion relation \eqref{recursion_HS}, for $n \neq d-1$, in the following index-free form:
\be \label{HS_recursion}
    \Phi^{(n)} = R^2 \left( f^{(n)} \Phi^{(n-1)} + g^{(n)}\Phi^{(n-2)} \right) +  \frac{ R^2 s \, \delta_{n,d} }{d+2s-2} \left[ v\cdot \partial - \frac{1}{d+2s-4} \abs{v}^2 (\partial\!\cdot\!\partial_v) \right] \Phi^{(d-1)}_{u}\,,  %\textrm{for } n \neq d-1.
\ee
where 
\be
\Phi^{(d-1)}_{u(k)}(u,\mathbf x;v) := \varphi^{(d-1)}_{u\cdots u i_{k+1} \cdots i_s}(u,\mathbf x) \, v^{i_{k+1}} \cdots v^{i_s}\,,
\ee
and where we introduced the operators
\begin{subequations} \label{HS_fg}
\begin{align}
    f^{(n)} & = \,-2\,b_{n}\left(n -\frac{d}{2}+s-1\right) \partial_u \, , \\
    g^{(n)} & = -\,b_{n}\left(
    \,\lap
    + \,b_{n-1}\,(d+2s-4)\,(v\!\cdot\!\partial)\,(\partial\!\cdot\!\partial_v)
    - \,b_{n-1}\,\abs{v}^2 (\partial\!\cdot\!\partial_v)^2 \right) , 
\end{align}
\end{subequations}
with
\be \label{HS_b_n}
    b_n = \frac{1}{(n-2+2s)(n-d+1)} \, .
\ee
Eq.~\eqref{HS_recursion} takes into account that \eqref{eom_constraint_HS} evaluated at $n = d-1$ and $k = s-1$ implies $\partial\cdot \partial_v \Phi^{(d-2)} = 0$, so that one can substitute the ill-defined $g^{(d)}_{ij}$ in \eqref{HS_fg} with $g^{(d)} = - \frac{1}{d+2s-2}\, \lap$. 

The recursion relation \eqref{HS_recursion} implies that the $\Phi^{(n)}$ satisfy an analogue of the mirror relations that we encountered in the cases of lower spin:
\be  \label{mirror_HS}
    \Phi^{\left(\frac{d}{2} - s +k\right)} = R^{2k} \left(\prod_{l=0}^{k-1} g^{\left(\frac{d}{2} - s +k-2l\right)} \right)\Phi^{\left(\frac{d}{2} - s -k\right)} \quad \text{for } 1 \leqslant k \leqslant \frac{d}{2} - 2 + s \,.
\ee
A proof of this relation is given in appendix~\ref{sec:mirror_anyspin} and we stress that it is instrumental to prove that the source is completely free, as discussed in appendix~\ref{sec:source_constraints}. Note that the ordering of the operators $g^{(n)}$ is immaterial, as they mutually commute: $[g^{(n)} , g^{(m)}] = 0$. This property follows from the fact that $g^{(n)}$ is a linear combination of two operators $g^{(n)} = -b_n (\lap + b_{n-1} \hat{g})$, where $\hat{g} = (d+2s-4) (v \cdot \partial)(\partial \cdot \partial_v) - |v|^2 (\partial \cdot \partial_v)^2$ and one can verify that $[\lap, \hat{g}] = 0$. 

For any value of $s$, the mirror relation admits a useful rewriting analogous to that in eq.~\eqref{Mirror_Spin_1_new}. To this end, it is convenient to introduce the traceless projector 
\be \label{traceless_proj}
    P_s = \sum_{n=0}^{\lfloor \frac{s}{2} \rfloor } a_n \abs{v}^{2n} \left(\partial_v \cdot \partial_v \right)^n  \quad \textrm{with} \quad  a_n = -\frac{a_{n-1}}{2n(d+2s-2n-2)} \quad\textrm{and}\quad a_0 = 1 \,,
\ee
which is built in such a way that if
$\Psi := \psi_{i_1 \cdots i_s} v^{i_1}\cdots v^{i_s}$, then $(\partial_v \cdot \partial_v) P_s \Psi = 0$. The mirror relation can then be rewritten as
\be \label{Improved_mirror_HS}
    \Phi^{\left(\frac{d}{2} - s +k\right)} =R^{2k} \left[ \sum_{l=0}^{k}(-2)^l \binom{k}{l}  \frac{\left( \frac{d}{2} + s - k - 1 \right)_{2k-l}}{\left[ \left( \frac{d}{2} + s - k - 1 \right)_{2k} \right]^2} \lap^{k-l} P_s (v \cdot \partial )^l (\partial \cdot \partial_v)^l \right] \Phi^{\left(\frac{d}{2} - s -k\right)}
\ee
for $1 \leqslant k \leqslant \frac{d}{2} - 2 + s$.
This is proven in the paragraph \textit{Identity 4} of appendix~\ref{sec:identities}.

%%%%%%%%%%%%%%%%%%%%%%%%%%%%%%%%%%%%%%%%%%%%%%%%%%%%%%%
\subsection{Flat limit of the AdS solution space}

Following the scalar and spin-one examples, we expand the source and the vev in powers of the AdS radius $R$:
\begin{subequations} \label{R_expansion_HS}
\begin{align} 
    \Phi^{(2-2s)} & = \sum_{k = 0}^{\frac{d}{2} + s-2} \frac{1}{R^{2k}}\, \Phi^{(2-2s,\,k)} + \sum_{k = \frac{d}{2} + s-1}^{\infty}\frac{1}{R^{2k}}\, \Lambda^{(k)} \,, \\
    \Phi^{(d-1)} & = \sum_{k = 0}^{\infty} \frac{1}{R^{2k}}\, \Phi^{(d-1,\,k)} \,.
\end{align}
\end{subequations}
Compared to eqs.~\eqref{R_expansion} and \eqref{R_expansion_maxwell}, we introduced here an infinite expansion for the source, but we assume that the $\Lambda^{(k)}$ are pure-gauge modes. As we shall see, a mere counting of the powers of $R^{-2}$ shows that the $\Lambda^{(k)}$ fields can only enter the vev Chthonians in the limit $R \to \infty$ and, being pure-gauge, their presence does not affect the group-theoretical discussion of section~\ref{sec:group-theory}.
On the other hand, they are necessary by consistency since, differently from the spin-one case, in general the gauge parameter admits an infinite expansion in powers of $R^{-2}$. This is easily seen in the spin-two case since the conditions \eqref{constr-gauge_spin2} defining the residual symmetries of the Bondi-like gauge \eqref{Bondi-like} supplemented by the boundary condition \eqref{extra-bnd_HS} can be diagonalised into
\be
\partial_u^2 v_i = \frac{1}{R^2 d}\, \partial_i \partial \cdot v \,, \qquad
\partial_u^2 f = \frac{1}{R^2 d}\, \lap f \,,
\ee
whose solutions admit an infinite expansion in powers of $R^{-2}$.

The remaining components of the vev, i.e.\ the $\varphi^{(d-1)}_{u(s-k)i(k)}$ with $k < s$, satisfy the evolution equations \eqref{evolution_HS}. As a result, the independent information they contain is encoded in the $u$-independent ``integration constants''. As for spin-one, we assume that the latter be of order $R^0$ and this condition is consistent for $d > 2$ because in this case the vev is gauge-invariant. The remaining components of the vev will therefore be substituted in the following by the explicit solution of the coupled system of equations \eqref{eom_constraint_HS}. For instance, still for $s = 2$, eqs.~\eqref{eom_constraint_HS} become
\be \label{Aspects_spin2}
\begin{split} 
    \mathcal{F}^{(d)}_{uu} & =  d \, \partial_u h^{(d-1)}_{uu} - \frac{d+1}{R^2}\, \partial \cdot h^{(d-1)}_{u}   + \lap h^{(d-2)}_{uu} = 0 \,, \\
    \mathcal{F}^{(d)}_{u i} & = (d + 1) \partial_u h^{(d-1)}_{u i}- \frac{d + 1}{R^2}\, \partial \cdot h^{(d-1)}_{i}- \partial_i h^{(d-1)}_{uu}  +  \lap h^{(d-2)}_{u i}  = 0 \,.
\end{split}
\ee
We can solve these coupled differential equations by
\be \label{solution_phi-u}
\begin{split}
h^{(d-1)}_{uu} &= M(\mathbf x) - \frac{1}{d} \partial_u^{-1} \lap h^{(d-2)}_{uu} + \mathcal O(R^{-2}) \,, \\
h^{(d-1)}_{ui} &= N_i(\mathbf x) + \frac{1}{d+1} \partial_u^{-1} \left( \partial_i M(\mathbf x) - \frac{1}{d} \partial_u^{-1} \partial_i \lap h^{(d-2)}_{uu}  - \lap h^{(d-2)}_{ui} \right) + \mathcal O(R^{-2}) \,,
\end{split}
\ee
where we assume that we have already set to zero the divergent terms in $h^{(d-2)}_{uu}$ and $h^{(d-2)}_{ui}$ and where we defined
\be
M = h^{(d-1)}_{uu} \eval_{u=0} \,,\quad N_i = h^{(d-1)}_{ui} \eval_{u=0} \,.
\ee
We stress that $M$ and $N_i$ do not contain any $R$-dependence, which allows one to smoothly take the $R \to \infty$ limit of the expressions \eqref{solution_phi-u}. As in the rest of the paper, we defined for convenience the ``integration constant'' appearing when solving eq.~\eqref{evolution_HS} on a cut of $\scri^+$ at $u = 0$.

As in sections~\ref{sec:flat-limit_scalar} and \ref{sec:flat-limit_maxwell}, substituting the previous expansions in the general asymptotic solution \eqref{HS_recursion} of the equations of motion gives both divergent and regular contributions in the limit $R \to \infty$,
\be \label{def_notation_HS}
\Phi^{(n)} = \sum_{k} \frac{1}{R^{2k}}\,\Phi^{(n,\,k)} \,,\quad \text{for}\;\; 3-2s\leqslant n\leqslant d-2\;\; \text{or}\;\; n\geqslant d \,,
\ee
while the explicit solution of the evolution equations \eqref{evolution_HS} (cf.~\eqref{solution_phi-u}) implies that the components of the vev only admit regular contributions,
\be
\Phi_{u(l)}^{(d-1)} = \sum_{k=0}^\infty \frac{1}{R^{2k}}\,\Phi_{u(l)}^{(d-1,\,k)} \,.
\ee
Imposing the vanishing of the divergent part of the on-shell field forces the regular components $\Phi^{(n,\,0)}$ to satisfy the Minkowski equations of motion also for arbitrary values of $s$. Indeed, eq.~\eqref{HS_recursion} implies
\be
\begin{split}
\label{recursion_divergences_HS}
    \Phi^{(n,\,k)} &=  f^{(n)} \Phi^{(n-1,\,k+1)} + g^{(n)}\Phi^{(n-2,\,k+1)}\\
    &\quad+  \frac{  s \, \delta_{n,d} }{d+2s-2} \left[ v\cdot \partial - \frac{1}{d+2s-4} \abs{v}^2 (\partial\!\cdot\!\partial_v) \right] \Phi^{(d-1,\,k+1)}_{u(1)} 
\end{split}
\ee
for $n \neq d-1$.
It is therefore enough to impose $\Phi^{(n+1,\,-1)} = 0$ to cancel all divergences in $R^2$. The condition
\be \label{flat_constraints_HS}
\begin{split}
    & \Phi^{(n+1,\,-1)} \propto (2n-d+2s) \partial_u \Phi^{(n,\,0)} - s \, \delta_{n,d-1} \left[ v\cdot \partial - \frac{\abs{v}^2 (\partial\!\cdot\!\partial_v)}{d+2s-4} \right] \Phi^{(d-1,\,0)}_{u} \\
    & \quad + \left[ \lap + \frac{d+2s-4}{(n-2+2s)(n-d+1)} \left( (v \!\cdot\! \partial) (\partial \!\cdot\! \partial_v)- \frac{|v|^2 (\partial \!\cdot\!\partial_v)^2}{d+2s-4} \right)\right] \Phi^{(n-1,\,0)} \overset{!}{=} 0
\end{split}
\ee
then implies that the regular components $\Phi^{(n,\,0)}$ obey the Minkowski equations of motion.

\paragraph{Over-leading and pure-gauge modes.} 
We now start our analysis of the relation between the regular components $\Phi^{(n,\,0)}$ ---~corresponding to the terms in the expansion in powers of $z$ of the on-shell field in Minkowski spacetime~--- and the various contributions in the expansions \eqref{R_expansion_HS} and \eqref{solution_phi-u} of the source and the vev from the terms that appear above radiation, for which $2-2s \leqslant n < \frac{d}{2} - s$. These can be divided in two groups: those with $2-2s \leqslant n \leqslant 1 - s$ and the others. 

We begin with the latter group, corresponding to the Erdmenger--Osborn celestial fields in figure~\ref{nice_graph}, with the goal of showing recursively that assuming suitable fall-offs as $\abs{\mathbf x} \to \infty$ such that (as in sections~\ref{sec:flat-limit_scalar} and \ref{sec:flat-limit_maxwell}) the Laplacian is invertible, the fields $\Phi^{(n,\,0)}$ in this range vanish in the limit $R \to \infty$ (we shall discuss their properties when this hypothesis is relaxed at the end of this section). To this end, it is convenient to consider $s$ divergences 
of the mirror relation \eqref{mirror_HS}:
\be \label{s_div}
    \left(\partial\!\cdot\!\partial_v \right)^s \Phi^{\left(\frac{d}{2} - s +k\right)} = R^{2k}\frac{\left(\frac{d}{2} - k - 1\right)_{2k}}{\left[\left(\frac{d}{2} + s - k - 1\right)_{2k}\right]^2} \lap^k  \left(\partial\!\cdot\!\partial_v \right)^s\Phi^{\left(\frac{d}{2} - s -k\right)} \quad \text{for } 1 \leqslant k \leqslant \frac{d}{2} - 2 + s\,.
\ee
The factor on the right hand side of \eqref{s_div} does not vanish in the range $1 \leqslant k \leqslant \frac{d}{2} - 2 $. As a result, assuming $\lap \Phi = 0 \implies \Phi = 0$ and demanding the regularity of the flat limit, imposes the constraint 
\be
    \left(\partial\!\cdot\!\partial_v \right)^s\Phi^{\left(\frac{d}{2} - s - k ,\, 0\right)} = 0 \quad \text{for } 1 \leqslant k \leqslant \frac{d}{2} - 2 \,.
\ee
We now show that the mirror relation actually allows one to set to zero recursively all divergences, and eventually the fields themselves.
Assuming that the $s-p+1$ divergence of $\Phi^{\left(\frac{d}{2} - s -k ,\, 0\right)}$ vanishes (for $0 \leqslant p \leqslant  s$),  
i.e., assuming
\be \label{assumption_hs}
    \left(\partial\!\cdot\!\partial_v \right)^{s-p+1}\Phi^{\left(\frac{d}{2} - s -k ,\, 0\right)} = 0 \quad \text{for } 1 \leqslant k \leqslant \frac{d}{2} - 2 \, ,
\ee
taking the $s-p$ divergence of the mirror relation we get
\be \label{s-p_div}
    \left(\partial\!\cdot\!\partial_v \right)^{s-p} \Phi^{\left(\frac{d}{2} - s +k\right)} = R^{2k}  \frac{\left(\frac{d}{2} - k + p - 1\right)_{2k}}{\left[\left(\frac{d}{2} + s - k - 1\right)_{2k}\right]^2} \lap^k \left(\partial\!\cdot\!\partial_v \right)^{s-p}\Phi^{\left(\frac{d}{2} - s -k ,\, 0\right)} \, .
\ee
In the range $ 1 \leqslant k \leqslant \frac{d}{2} +p-2$, which always contains $ 1 \leqslant k \leqslant \frac{d}{2} -2$, the assumption \eqref{assumption_hs} is valid and this implies the constraint 
\be
    \left(\partial\!\cdot\!\partial_v \right)^{s-p}\Phi^{\left(\frac{d}{2} - s -k ,\,0\right)} = 0 \quad \text{for } 1 \leqslant k \leqslant \frac{d}{2} - 2 \, .
\ee
Therefore, in this range, we can recursively set all divergences to zero. Actually this relation also holds for $p = s$, thus implying the vanishing of all these fields in the flat limit
\be
    \Phi^{\left(\frac{d}{2} - s -k ,\,0\right)} = 0 \quad \text{for } 1 \leqslant k \leqslant \frac{d}{2} - 2 \,.
\ee

On the other hand, within the range $\frac{d}{2} - 1 \leqslant k \leqslant \frac{d}{2} + s - 2$, the fields $\Phi^{\left(\frac{d}{2} - s -k ,\,0\right)}$ cannot be identically set to zero, and they correspond to the pure-gauge modes corresponding to the residual symmetries of the Bondi-like gauge \eqref{Bondi-like} supplemented by the boundary condition \eqref{extra-bnd_HS}. Demanding finiteness of the mirror relation \eqref{mirror_HS} in the flat limit imposes indeed the following constraint on the regular part
\be \label{mirror_const_HS}
    \left(\prod_{l=0}^{k-1} g^{\left(\frac{d}{2} - s +k-2l\right)} \right)\Phi^{\left(\frac{d}{2} - s -k ,\, 0\right)}  = 0\quad \text{for } \frac{d}{2} - 1 \leqslant k \leqslant \frac{d}{2} + s - 2 \, .
\ee
We can shift the label $n$ and represent it as $\Phi^{(1-2s+m,\, 0)}$ for $1\leqslant m\leqslant s$ and express it purely in terms of the source. For instance,
\be  
    \Phi^{(3-2s,\, 0)} =  f^{(3-2s)}\Phi^{(2-2s,\, 1)} \,.
\ee
In general, $\Phi^{(1-2s+m,\, 0)}$ is given by summing over all diagrams connecting it to the source, which can be schematically expressed as 
\be
\Phi^{(1-2s+m,\, 0)} = \sum_{n=0}^{\lfloor \frac{m-1}{2} \rfloor }  (f)^{m-1-2n} (g)^{n} \, \Phi^{(2-2s ,\, m-1-n)} \,,
\ee
where we refrained from presenting the precise products of $f^{(n)}$ and $g^{(m)}$ operators entering this expression to only stress how many of them appear.
Note that the field associated with the highest power of $R^{-2}$ in the expansion \eqref{R_expansion_HS} of the source which appears in the regular contribution (i.e., in $\Phi^{(1-2s+m ,\, 0)}$) 
is $\Phi^{(2-2s ,\, m-1)}$. We will now demonstrate that these fields possess a gauge symmetry of the form
\be \label{gauge_sym}
    \delta \Phi^{(1-2s+m ,\, 0)} = P_s(v\!\cdot\!\partial)^m \xi_{s-m} \quad \text{for } 1 \leqslant m \leqslant s \, .
\ee
For example, in the spin $s = 2$ case, 
translating this into the familiar index notation gives
\be \label{Spin2_gauge_sym}
    \delta h^{(-2,\, 0)}_{ij} = \partial_{(i} v_{j)} -\frac{1}{d}\, \delta_{ij} \partial \cdot v \,, \qquad \delta h^{(-1 ,\, 0)}_{ij} = - \left( \partial_i \partial_j - \frac{1}{d}\, \delta_{ij} \lap \right) f \,,
\ee 
which represent the vector and scalar gauge symmetries, respectively.  
To show that these components possess a gauge symmetry, we need to show that the $\delta \Phi$ in eq.~\eqref{gauge_sym} solves \eqref{mirror_const_HS}, i.e., we have to show that
\be \label{mirror_const_HS_m}
    \left(\prod_{l=0}^{\frac{d}{2} + s - m - 2} g^{(d - m - 1 - 2l)} \right)\delta\Phi^{(1 - 2s + m ,\, 0)}  = \left(\prod_{l=0}^{\frac{d}{2} + s - m - 2} g^{(d - m - 1 - 2l)} \right)P_s(v\!\cdot\!\partial)^m \xi_{s-m} = 0
\ee
for $1\leqslant m \leqslant s$. This is proven in appendix~\ref{sec:gauge_symmetry}. 

The gauge symmetries we identified then allow us to set to zero the divergences of these fields that cannot be set to zero using \eqref{s_div}. Assuming the invertibility of the Laplacian, one can eventually proceed as in the previous analysis of the Erdmenger--Osborn fields to prove recursively that the entire portion of these fields that is not pure-gauge is set to zero. For instance, for $s=2$,
if we use the vector and scalar gauge-symmetry parameters in \eqref{Spin2_gauge_sym} to set, respectively, $\partial^j h_{ij}^{(-2,\,0)}$ and $\partial^i \partial^j h^{(-1,\,0)}_{ij}$ to zero, the constraint \eqref{mirror_const_HS} implies (using the improved mirror relation \eqref{Improved_mirror_HS} evaluated at $k = \frac{d}{2}$)
\begin{equation}
\label{Improved_mirror_HS_case1}
    \lap^{\frac{d}{2}} h_{ij}^{(-2,\,0)} = 0 \implies h_{ij}^{(-2,\,0)} = 0 \, .
\end{equation}
Computing the divergence of $\eqref{mirror_const_HS}$ evaluated at $k = \frac{d}{2} -1$ gives
\begin{equation}
\frac{1}{(d-1)^2 (d-2)!}\, \lap^{\frac{d}{2}-1} \partial^j%\cdot
h_{ij}^{(-1,\, 0)} = 0 \implies \partial^j%\cdot
h_{ij}^{(-1,\, 0)} = 0 \, ,
\end{equation}
which in turn can be substituted in \eqref{mirror_const_HS} at $k = \frac{d}{2}-1$ to give
\be
\lap^{\frac{d}{2}-1} h_{ij}^{(-1,\, 0)} = 0 \implies h_{ij}^{(-1,\, 0)} = 0 \,,
\ee
which shows these are pure-gauge modes.

\paragraph{Radiation.} 
In the limit $R \to \infty$, one obtains 
\be
    \Phi = \sum_{m = 1}^s z^{1-2s+m } \Phi^{(1-2s+m ,\, 0)} + \sum_{k=0}^{\frac{d}{2} + s -2}z^{\frac{d}{2} - s + k}\Phi^{(\frac{d}{2} -s +k,\, 0)} + \cO(z^{d-1}) \, ,
\ee
where $\Phi^{(\frac{d}{2} -s,\, 0)}$ plays the role of the shear tensor in the flat limit. It is obtained by summing over diagrams, and takes the schematic form
\be \label{radiation_HS}
\begin{split}
    \mathbf{C}(u,\mathbf{x}) :=  \Phi^{(\frac{d}{2} -s,\, 0)} &= \frac{(-1)^{\frac{d}{2} + s - 2}}{(d + 2s - 5)!!}\, \partial_u^{\frac{d}{2} + s - 2} \Phi^{(2-2s,\, \frac{d}{2}+s -2)}\\
    &\quad+  \sum_{n=1}^{\lfloor \frac{1}{2}\left(\frac{d}{2}+s -2\right) \rfloor }  (f)^{\frac{d}{2}+s -2-2n} (g)^{n} \, \Phi^{(2-2s ,\, \frac{d}{2}+s -2-n)}\,.
    \end{split}
\ee
Here we have written the first term with all single-path contribution explicitly, because $\Phi^{(2-2s,\, \frac{d}{2}+s -2)}$ contains a free function of $u$, while the additional contributions, for which we only emphasised the number of $f^{(n)}$ and $g^{(m)}$ operators entering the products, are necessary to guarantee the gauge invariance of $\Phi^{(2-2s,\,0)}$ for $d > 2$. Working under the hypothesis of invertibility of the Laplacian, they depend on pure-gauge modes while, in general, they contain combinations of Erdmenger--Osborn fields. When the dimension is high enough, these products always contain a vanishing coefficient, so that the second line in eq.~\eqref{radiation_HS} only appears in a finite number of spacetime dimensions.  

As for spin-one, we now prove that $\Phi^{(2-2s,\, \frac{d}{2}+s -2)}$ remains completely unconstrained. To this end, one can notice that the regular contribution from the mirror relation \eqref{mirror_HS} gives
\be \label{Source_cht_HS}
    \Phi^{\left(\frac{d}{2} - s +k ,\, 0\right)} = (-1)^{\frac{d}{2}+s-k-2} \frac{\left(\frac{d}{2}+s-k-1\right)_{2k}}{2^k k! (d+2s-5)!!}  \prod_{l=0}^{k-1} g^{\left(\frac{d}{2} - s +k-2l\right)} \partial_u^{\frac{d}{2}+s-k-2}\Phi^{(2-2s,\, \frac{d}{2}+s -2)} + \cdots
\ee
where the dots represent terms coming from lower order terms in the $R^{-2}$ expansion: $\Phi^{(2-2s ,\, \frac{d}{2}+s -2-n)}$ for $1 \leqslant n \leqslant \lfloor \frac{1}{2}\left(\frac{d}{2}+s -2 -k \right) \rfloor$. Solving the equations of motion \eqref{HS_recursion}, summing over all diagrams connected to the source shows that the divergent term $\Phi^{\left(\frac{d}{2} - s +k,\, -1\right)}$ only contains contributions coming from $\Phi^{(2-2s,\, \frac{d}{2}+s -3-n)}$ for $n \geqslant 0$ which do not include the term under scrutiny. The only other place the term $\Phi^{(2-2s ,\, \frac{d}{2}+s -2)}$ appears is at order $\cO(z^d)$:
\begin{align} \label{HS_d}
    \Phi^{(d)} = - R^2 \Bigg\{ &\partial_u \Phi^{(d-1,\,0 )} - \frac{s}{d+2s-2} \left[ v\cdot \partial - \frac{1}{d+2s-4}\, \abs{v}^2 (\partial\!\cdot\!\partial_v) \right] \Phi^{(d-1,\,0)}_{u} \nn \\
    &+ \frac{1}{d+2s-2}\, \lap \Phi^{(d-2,\,0)} \Bigg\} - \partial_u \Phi^{(d-1,\,1 )} \nn \\& + \frac{s}{d+2s-2} \left[ v\cdot \partial - \frac{1}{d+2s-4}\, \abs{v}^2 (\partial\!\cdot\!\partial_v) \right] \Phi^{(d-1,\,1)}_{u} + \cO(R^{-2}) \,.
\end{align}
The diverging contribution determines $\partial_u \Phi^{(d-1 ,\,0 )}$ in terms of the aspect $\Phi^{(d-1,\,0)}_{u}$ and the source (via $\Phi^{(d-2,\,0)}$). As such, as for scalars and Maxwell fields, this relation cannot impose any constraint on $\Phi^{\left(2-2s ,\, \frac{d}{2}+s -2\right)}$. 
  
\paragraph{Source Chthonians.} From \eqref{Source_cht_HS} we notice that in the regular contribution at each order after the radiative one, there is one less retarded time derivative $\partial_u$ acting on $\Phi^{\left(2-2s ,\, \frac{d}{2}+s -2\right)}$ implying that it can be written as
\be\label{vev_Chth_HS_full}
    \Phi^{\left(\frac{d}{2}-s+k,\,0\right)} = \frac{(-1)^k}{2^k k!}\sum_{l=0}^{k}(-2)^l\binom{k}{l}\frac{\left(\frac{d}{2}+s-k-1\right)_{2k-l}}{\left(\frac{d}{2}+s-k-1\right)_{2k}}\,\lap^{k-l} P_s\,(v\cdot\partial)^l(\partial\cdot\partial_v)^l\phi^{(k)} + \cdots
\ee
with 
\be\label{phi_k_HS}
    \phi^{(k)}(u,\mathbf{x}) := \partial_u^{-k}\mathbf{C} + \frac{(-1)^{\frac{d}{2}+s-2}}{(d+2s-5)!!}\sum_{l=0}^{k-1}\frac{u^l}{l!}\,\partial_u^{\frac{d}{2}+s-k-2+l}\Phi^{(2-2s,\,\frac{d}{2}+s-2)}\eval_{u=0} \,,
\ee
and where the dots in \eqref{vev_Chth_HS_full} represent terms coming from lower order terms in the $R^{-2}$ expansion: $\Phi^{\left(2-2s ,\, \frac{d}{2}+s -2-n\right)}$ for $n \geqslant 1 $. By imposing $\partial_u^k \Phi^{(2-2s,\,n)}\eval_{u=0} = 0$ for $k < n$ as for scalars (see also appendix~\ref{app:scalar_chthonians}), we then recover the Minkowski space source Chthonians in the flat limit as
\begin{align}
    &\Phi^{\left(\frac{d}{2}-s+k,\,0\right)}\eval_{u=0}= \frac{(-1)^{\frac{d}{2}+s+k-2}  }{2^k k!(d+2s-5)!!}
    \\ &\times\sum_{l=0}^{k}(-2)^l\binom{k}{l}\frac{\left(\frac{d}{2}+s-k-1\right)_{2k-l}}{\left(\frac{d}{2}+s-k-1\right)_{2k}}\,\lap^{k-l} P_s\,(v\cdot\partial)^l(\partial\cdot\partial_v)^l\; \partial_u^{\frac{d}{2}+s-k-2}\Phi^{(2-2s,\,\frac{d}{2}+s-2)}\eval_{u=0}\,.\nn
\end{align}
The source Chthonians are gauge-invariant and do not depend on the $\Lambda^{(k)}$ as a result of the mirror relation. 
For instance, for $s=2$, the residual gauge transformations $f(u,\mathbf x)$ and $v_i(u,\mathbf x)$ left by the gauge fixing discussed in section~\ref{sec:bondi-like_gauge} induce the following transformations of the coefficients in the expansion \eqref{R_expansion_HS}:
\begin{align}
    \label{induced_gauge_spin2_1}
   \delta h_{ij}^{(-2,\,0)} &= \partial_{(i} v_{j)}^0 - \frac{1}{d}\,\delta_{ij}\partial \cdot v^0 \,, \\
 \label{induced_gauge_spin2_2}
   \delta h_{ij}^{(-2,\,n)} &= \frac{u^{2n-1}}{d^{n-1}(2n-1)! } \left[\partial_i \partial_j -  \frac{\delta_{ij}}{d}\, \lap \right] \lap^{n-1} \left[ f^0 + \frac{u}{2dn}\, \partial \cdot v^0 \right]  \quad \textrm{for } 1 \leqslant n \leqslant \frac{d}{2} \,,
\end{align} 
where $v^0_i = v_i(u = 0, \mathbf{x})$ and $f^0 = f(u = 0, \mathbf{x})$ and we stress that these quantities are precisely of order $R^0$. This implies
\be 
\delta h_{ij}^{\left( \frac{d}{2}-2+k,\,0 \right) }\eval_{u=0} \propto  \partial_u^{\frac{d}{2}-k} \delta h_{ij}^{\left(-2,\,\frac{d}{2}\right)}\eval_{u=0} = 0 \,,
\ee
which shows the spin-$2$ source Chthonians are gauge-invariant.

\paragraph{Celestial conserved currents.} Taking $m$ divergences of the mirror relation \eqref{mirror_HS} gives
\begin{eqnarray}
    &&(\partial \cdot \partial_v)^m\Phi^{\left(\frac{d}{2} - s +k\right)}\\
    &&= R^{2k}\left[\sum_{l=0}^k (-2)^l \binom{k}{l} \frac{\left( \frac{d}{2} + s - m - k - 1 \right)_{2k-l}}{\left[ \left( \frac{d}{2} + s - k - 1 \right)_{2k} \right]^2} \lap^{k-l}P_{s-m} (v \cdot \partial)^l(\partial \cdot \partial_v)^{l+m}  \right] \Phi^{\left(\frac{d}{2} - s -k\right)}\,. \nn
\end{eqnarray}
This relation is proved in appendix~\ref{sec:identities} and it vanishes in the range $\frac{d}{2} + s - m - 1 \leqslant k \leqslant \frac{d}{2} + s - 2$. In the flat limit, this implies the existence of the relations
\be \label{conserved_current_HS}
    (\partial \cdot \partial_v)^m \Phi^{\left(d-m-1,\, 0\right)}  = 0 \quad \textrm{for } 1\leqslant m \leqslant s \,,
\ee
identifying the celestial conserved currents in figure~\ref{nice_graph}. For $s=2$, this result follows from the expressions
\begin{align}
h_{ij}^{(d-3)} & = \frac{R^{d-2}}{(d-1)!}\, \lap^{\frac{d}{2}-3} \Big( \lap^2 h_{ij}^{(-1)} - 2\, \frac{d-2}{d-1}\, \lap \partial_{(i} \partial^k %\cdot
h^{(-1)}_{j)k} \nonumber \\
&\qquad\qquad\qquad\qquad + \frac{d-4}{d-1}\, \partial_i \partial_j \partial^k %\cdot
\partial^l %\cdot
h^{(-1)}_{kl} + \frac{1}{d-1}\, \delta_{ij} \lap \partial^k %\cdot 
\partial^l %\cdot
h^{(-1)}_{kl} \Big) \,, \label{d-3spin2}\\
h_{ij}^{(d-2)} & = \frac{R^d}{d!}\, \lap^{\frac{d}{2}-2} \Big(\lap^2 h^{(-2)}_{ij} - 2\, \lap \partial_{(i} \partial^k %\cdot
h^{(-2)}_{j)k} \nonumber \\
&\qquad\qquad\qquad + \frac{d-2}{d-1}\, \partial_i \partial_j \partial^k %\cdot
\partial^l %\cdot
h^{(-2)}_{kl} + \frac{d}{d-1}\, \delta_{ij} \lap \partial^k %\cdot 
\partial^l %\cdot
h^{(-2)}_{kl} \Big) \,, \label{celestial_stress_tensor}
\end{align}
that one can obtain by evaluating the mirror relation \eqref{Improved_mirror_HS} for $s=2$ and $k = \frac{d}{2}-1$ and $k = \frac{d}{2}$ respectively. These combinations satisfy
\be
\partial^i \partial^j h_{ij}^{(d-3)} = 0 \, , \qquad \partial^j h^{(d-2)}_{ij} = 0 \, . 
\ee
Moreover, the combinations \eqref{d-3spin2}-\eqref{celestial_stress_tensor} are left invariant by transformations of the form \eqref{Spin2_gauge_sym}, as already proven in general in eq.~\eqref{mirror_const_HS_m}. Note that $h^{(d-2)}_{ij}$ corresponds to the linearisation of the Bach tensor (which is the only Weyl and diffeomorphism invariant operator with $d$ derivatives) where we also used the Weyl invariance to gauge-fix the trace of the metric perturbation to zero.

\paragraph{Aspects.}
The fields $\Phi^{(d-1)}_{u(s-k)}$ obey the evolution equations \eqref{evolution_HS}, that we rewrite here in index-free notation:
\begin{align} \label{Aspects_HS}
    \mathcal{F}^{(d)}_{u(s-k)} & = (s+k+d-2)\, \partial_u \Phi^{(d-1)}_{u(s-k)}  - \frac{(d+k+s-2)(d+k+s-1)}{(k+1)(d+2k)R^2}\, \partial\!\cdot\!\partial_v \Phi^{(d-1)}_{u(s-k-1)}  \nn \\
    &\quad - k\left[  v\cdot \partial  - \frac{1}{d+2k-4} \abs{v}^2 \partial\!\cdot\!\partial_v \right]\Phi^{(d-1)}_{u(s-k+1)}   + \lap \Phi^{(d-2)}_{u(s-k)} = 0  \quad \textrm{for }  k < s \, .
\end{align}
This relation is a coupled, first-order $u$-evolution equation for $\Phi^{(d-1)}_{u(s-k)}$, driven by the adjacent components $\Phi^{(d-1)}_{u(s-k-1)}$ and $\Phi^{(d-1)}_{u(s-k+1)}$. As already discussed at the beginning of this section, integrating this with respect to $u$ introduces an arbitrary function on the celestial sphere (at say $u = 0$) for each $k$ in the range $0 \leqslant k < s$. These celestial integration functions constitute the free data in AdS. Crucially, the definition of these free data remains identical whether we are in AdS or taking the flat-space limit $R \to \infty$: the data reside at exactly the same order $(n=d-1)$ in the expansion. The physical distinction lies entirely in how these data evolve. While the AdS equations dictate a mutually coupled evolution, the flat-space limit reduces the dynamics to a one-way hierarchy. Note also that, as we shall see in the next paragraph, each order in the $R^{-2}$-expansion of the fully-transverse parts of the vev $\Phi^{(d-1,\,n)}$ verifies an evolution equation in the limit $R \to \infty$, so that altogether they do not constitute fully independent data as in AdS.

For instance, for the case of spin $s=2$ one gets eqs.~\eqref{Aspects_spin2}.
In AdS, the system is coupled: the $u$-evolution of $h^{(d-1)}_{uu}$ is driven by $h^{(d-1)}_{u i}$, and conversely, the evolution of $h^{(d-1)}_{u i}$ is driven by $h^{(d-1)}_{uu}$ (alongside the source and the purely spatial vev). But in the flat limit $R \to \infty$, the evolution of $h^{(d-1)}_{uu}$ decouples from the higher components and is dictated purely by the source, while the evolution of $h^{(d-1)}_{ui}$ continues to be driven by the now-autonomous $h^{(d-1)}_{uu}$ field.

\paragraph{Vev Chthonians.} 
The structure of terms below order $z^{d-1}$ in the vev follows the same pattern established in the scalar and spin-one analyses. The regular terms $\Phi^{(d-1+n,\,0)}$ with $n > 0$ satisfy the flat-space evolution equations \eqref{flat_constraints_HS}.
For each integer $n \geqslant 0$, this leaves precisely one integration function on the transverse plane, generating a second family of Chthonians, those originating from the vev, which are distinct from the Chthonians arising from the source expansion. Through the evolution equation for $\Phi^{(d-1,\,0)}$, the evolution equations receive recursive input from both the radiation $\mathbf{C}(u,\mathbf{x})$ and the source Chthonians; however, the source cannot contribute directly to the integration constants below the vev. This follows from a simple power-counting argument in $R^2$: source contributions can only enter through divergent terms $\Phi^{(n,\,k<0)}$, which vanish by virtue of the recursion relation \eqref{flat_constraints_HS} once $\cO(R^2)$ of \eqref{HS_d} is set to zero. 

Had the expansion of the source in eq.~\eqref{R_expansion_HS} been carried beyond order $R^{-d-2s+4}$ involving terms that are not pure-gauge, additional source contributions to the vev Chthonians would have appeared. These would, however, be spurious, for the same reason detailed in the scalar case in appendix~\ref{app:scalar_chthonians}: at each order below $z^{d-1}$, a new contribution from the vev expansion also arises, so that the physically independent boundary data in Minkowski spacetime are encoded in a linear combination of source and vev contributions: the two never appearing independently in the solution space. It is therefore natural to eliminate the redundant source contributions by truncating the first sum in the expansion \eqref{R_expansion_HS} as done above, a choice that is furthermore consistent with the group-theoretical analysis of the limit carried out in section~\ref{sec:group-theory}. Concerning the pure-gauge contributions of eq.~\eqref{R_expansion_HS} that are necessary for the consistency of the $R$-expansion of the residual gauge symmetries, these also do not contribute to the vev Chthonians and, therefore, do not alter their gauge-invariant character.
For $s=2$, this follows from the observation that expanding eq.~\eqref{celestial_stress_tensor} in powers of $R^2$ one gets
\be
h^{(d-2,\,k)}_{ij} = \frac{\lap^{\frac{d}{2}-2}}{d!} \Big(\lap^2 \Lambda^{(k)}_{ij} - 2 \lap \partial_{(i} \partial \cdot \Lambda^{(k)}_{j)} + \frac{d-2}{d-1} \partial_i \partial_j \partial \cdot \partial \cdot \Lambda^{(k)} + \frac{d}{d-1} \delta_{ij} \lap \partial \cdot \partial \cdot \Lambda^{(k)} \Big) ,
\ee
for $k \geqslant 1$. As emphasised when we introduced \eqref{celestial_stress_tensor}, this tensor is left invariant by the transformations \eqref{Spin2_gauge_sym} and therefore it trivially vanishes on the pure-gauge contributions encoded in $\Lambda_{ij}^{(k)}$. Since the vev Chthonians (and their evolution equations in the flat limit) only depend on the source through $h_{ij}^{(d-2,\,k)}$ for $k \geqslant 1$ (with the exception of the evolution equation for $h_{ij}^{(d-1)}$, see eq.~\eqref{HS_d}), this proves that the vev Chthonians are independent of the pure-gauge contributions. This statement generalises to higher spins in a similar manner: the vev Chthonians depend on the $\Lambda^{(k)}$ parts of the source only through $\Phi^{(d-2)}$, but the latter involves gauge-invariant generalisations of the Bach tensor which exactly vanish on pure-gauge contributions (see also eq.~\eqref{mirror_const_HS_m} and appendix~\ref{sec:gauge_symmetry}).

To express the contribution to $\Phi^{(d-1+n,\,0)}$ coming from the vev coefficients $\Phi^{(d-1,\,k)}$, one applies the recursion \eqref{recursion_divergences_HS}, summing over diagrams connecting $d-1$ to $d-1+n$. As in the spin-one case, the explicit expressions grow increasingly involved as $n$ increases. Nevertheless, the Chthonians can be identified by noting that a single-path contribution is always present among the diagrams:
\be \label{vev_Chth_HS}
    \Phi^{(d-1+n,\,0)} = \prod_{k=1}^n f^{(d-1+k)} \Phi^{(d-1,\,n)} + \cdots \propto \partial_u^n \Phi^{(d-1,\,n)} + \cdots \,,
\ee
where $(\cdots)$ denotes contributions involving at least one factor of $g^{(n)}$ and strictly fewer retarded-time derivatives.

At each order $n$, the expression for $\Phi^{(d-1+n,\,0)}$ involves the new coefficient $\Phi^{(d-1,\,n)}$, entering with $\partial_u^n$ derivatives, together with the earlier coefficients $\Phi^{(d-1,\,k < n)}$, each appearing with fewer retarded-time derivatives (cf.~\eqref{vev_Chth_HS}). The evolution equation for $\Phi^{(d-1+n,\,0)}$ then constrains the $\partial_u^{n+1}$ derivative of $\Phi^{(d-1,\,n)}$, so that the latter effectively contributes only one new function on the transverse plane. This function can be identified with the Chthonian $\Phi^{(d-1+n,\,0)}\eval_{u=0}$ by imposing $\partial_u^k \Phi^{(d-1,\,n)}(u=0,\mathbf x) = 0$ for all $0 \leqslant k \leqslant n-1$. Following the analysis of appendix~\ref{app:scalar_chthonians} for the scalar case, this condition is consistent: these lower-derivative terms always appear alongside the leading term $\partial_u^n \Phi^{(d-1,\,n)}$, so that relaxing the condition would merely introduce a redundancy in the parametrisation of the solution space.

With this condition in place, one obtains the following compact expression for the flat-space Chthonians below order $z^{d-1}$:
\be \label{eq_vev-Chthonians_realisation_HS}
    \Phi^{(d-1+n,\,0)}\eval_{u=0} =\frac{(-2)^n}{n!} \frac{\left(\frac{d}{2} + s - 1\right)_n}{(d + 2s - 2)_n}\, \partial_u^n\, \Phi^{(d-1,\,n)}(u=0,\mathbf x) \,.
\ee

\paragraph{Erdmenger--Osborn fields.} 
If we relax the assumption that the Laplacian is invertible, it is no longer true that in the limit $R \to \infty$ only the pure-gauge modes survive from orders $z^{2-2s}$ to $z^{1-s}$. In analogy with the case of the vev Chthonians discussed above, a new integration function depending only on the transverse coordinates thus arises for each $2-2s \leqslant n \leqslant \frac{d}{2} - s-1$. On the other hand, the mirror relation \eqref{Improved_mirror_HS} implies that the regular term must satisfy the condition
that guarantees the cancellation of divergences of order $R^{2k}$ in the terms below radiation:
\be \label{eq_erd-osborn_HS}
    \left[ \sum_{l=0}^{k}(-2)^l \binom{k}{l}  \frac{\left( \frac{d}{2} + s - k - 1 \right)_{2k-l}}{\left[ \left( \frac{d}{2} + s - k - 1 \right)_{2k} \right]^2} \lap^{k-l} P_s (v \cdot \partial )^l (\partial \cdot \partial_v)^l \right] \Phi^{\left(\frac{d}{2} - s -k,\,0\right)} = 0\,,
\ee
for $1 \leqslant k \leqslant \frac{d}{2} - 2 + s$.
Moreover, in analogy with the discussion of the vev Chthonians, at each subleading order a new term in the expansion of the source enters with the maximal number of retarded-time derivatives,
\be
    \Phi^{(2-2s+n,\,0)} = \prod_{k=1}^{n} f^{(2-2s+k)} \Phi^{(2-2s,\,n)} + \cdots \propto \partial_u^n \Phi^{(2-2s,\,n)} + \cdots
\ee
By adopting the same strategy as for the Chthonians, i.e.\ imposing $\partial_u^k \Phi^{(2-2s,\,n)}\eval_{u=0} = 0$ for $k < n$, we can conclude that the $z$ expansion in Minkowski spacetime also contains at the orders $z^{n}$ with $2-2s \leqslant n \leqslant \frac{d}{2} - s-1$ the coefficients
\be \label{eq_erd-osborn_realisation_HS}
    \Phi^{(2-2s+n,\,0)}\eval_{u=0} = \frac{(-2)^n}{n!} \frac{\left(\frac{d}{2} + s - n - 1\right)_n}{(d + 2s - n - 3)_n}\, \partial_u^n \Phi^{(2-2s,\,n)}\eval_{u=0} \,,
\ee
satisfying however also eq.~\eqref{eq_erd-osborn_HS}. These are the Erdmenger--Osborn celestial fields in figure~\ref{nice_graph}. Note that the celestial fields appearing from orders $z^{2-2s}$ till $z^{1-s}$ still display the gauge symmetry \eqref{gauge_sym}, but one should now consider all solutions of eq.~\eqref{eq_erd-osborn_HS}, and not only the pure-gauge modes that we discussed above.

%%%%%%%%%%%%%%%%%%%%%%%%%%%%%%%%%%%%%%%%%%%%%%%%%%%%%%%
\section{Group theoretical perspective}\label{sec:group-theory}

In this section, we study the decomposition of the irreducible representations of the AdS isometry group corresponding to the source and the vev in irreducible representations of the Lorentz subgroup. To this end, in section~\ref{sec:group-theory-branching} we provide branching rules for the characters and we show that the decomposition agrees with the structure of the Minkowski solution space in terms of celestial primary fields.
In section~\ref{Carrollianlimitoftheconformalscalar}, this is made more explicit for the vev by an analysis 
singling out descendants of the conformal primary that are themselves celestial primaries and taking a flat limit for spins $s=0,1,2$.

\subsection{Branching rules of the source and the vev}\label{sec:group-theory-branching}

Primary fields of scaling dimension $\dim$ taking values in the totally symmetric and traceless spin-$s$ representation of the Lorentz group $SO(d,1)$, together with all their descendants, span a lowest-weight representation of the $(d+1)$-dimensional conformal group with Lorentzian signature $SO(d+1,2)$. More precisely, they span a (generalised) Verma module of the conformal algebra $\mathfrak{so}(d+1,2)$. The corresponding character of $\mathfrak{so}(d+1,2)$ will be denoted $\mathcal V(\dim,s)$. For the sake of simplicity, we will sometimes be sloppy and identify modules with their characters (explicit computations are done in terms of the latter, in order to derive results valid for the former).

\paragraph{Vev (aka conformal conserved currents).} Unitary irreducible representations of the group $SO(d+1,2)$ are given by lowest-weight representations which satisfy the unitarity bound $\dim \geqslant s+d-1$ for $s>0$. The saturation of this bound corresponds to the case where there is an invariant subspace in the Verma module
(in CFT language, a descendant becomes a primary of zero norm). We can quotient by this submodule, which corresponds to a shortening of the degrees of freedom contained in this irreducible representation. This quotient corresponds to the vev we encountered in section~\ref{sec:HS}. In more explicit terms, if $j_{ a_1 \cdots  a_s}$ is a totally-symmetric traceless tensor of rank $s$ with scaling dimension $\dim_+ = s+d-1$, the submodule in question is described by setting the 
divergence to zero 
\be 
    \partial^{a_1} j_{a_1 \cdots a_s} = 0 \,,
\ee 
where the letter at the beginning of the Latin alphabet will be boundary indices ranging from $0$ to $d$.
This divergence is a primary field of scaling dimension $s+d$ and a totally-symmetric traceless tensor of rank $s-1$.
The $\mathfrak{so}(d+1,2)$ character of the irreducible quotient module describing the spin-$s$ conformal conserved current is the difference of the corresponding characters
\be \label{eq: spin s singleton}
    \mathcal D(s+d-1, s) := \mathcal V(s+d-1,s) - \mathcal V(s+d,s-1) \,.
\ee 
We are interested in the branching rules for the restriction $SO(d+1,2) \downarrow SO(d+1,1)$ of the AdS$_{d+2}$ isometry group to its Lorentz subgroup. In terms of the boundary CFT$_{d+1}$, this means that the components of the primary and descendant fields experience the branching rules for the restriction $SO(d,1)\downarrow SO(d)$ of the boundary Lorentz group to its rotation subgroup. More explicitly, this corresponds to decomposing the indices inside the components $j_{a_1 \cdots a_s}$ into time and space, i.e. $a_n = (u,i_n)$ for all $n$. Remember the convention that letters like $a,b,c,\ldots$ range from $0$ to $d$, while $i,j,k,\ldots$ range from $1$ to $d$. 
This gives $(s+1)$ totally-symmetric traceless $SO(d)$-irreducible tensors of spin $s$ down to $0$. The conservation equation itself is branched to give rise to $s$ conservation equations of rank $s-1$ down to $0$.

In terms of Verma modules, we can use the following lemma (see appendix~\ref{Vermabranch} for a proof)
\be \label{lemma}    
    \mathcal V(\dim,s)  \;\overset{\mathfrak{so}(d+1,2)}{\underset{\mathfrak{so}(d+1,1)}{\downarrow}}\;
    \bigoplus_{n=0}^\infty \bigoplus_{r=0}^s \mathcal V(\dim+n,r) \,.
\ee 
Applied to the character \eqref{eq: spin s singleton} of the vev gives
\be  
\begin{split}
\mathcal D(s+d-1,s) 
    \quad\;\overset{\mathfrak{so}(d+1,2)}{\underset{\mathfrak{so}(d+1,1)}{\downarrow}}\;&\quad\bigoplus_{n=0}^\infty \bigoplus_{r=0}^s  \mathcal V(s+d-1+n,r) - \bigoplus_{n=0}^\infty \bigoplus_{r=0}^{s-1} \mathcal V(s+d+n,r) \\
    &\quad = \bigoplus_{r=0}^s \mathcal V(s+d-1,r) + \bigoplus_{n=1}^\infty \bigoplus_{r=0}^s \mathcal V(s+d-1+n,r)\\
    &\qquad-  \bigoplus_{n=0}^\infty \bigoplus_{r=0}^{s-1} \mathcal V(s+d+n,r) \,.
\end{split}
\ee 
Therefore, the branching rule of the vev reads
\be \label{eq: spectrum Chthonian}
    \mathcal D(s+d-1,s) \;\;\overset{\mathfrak{so}(d+1,2)}{\underset{\mathfrak{so}(d+1,1)}{\downarrow}}\;\; \bigoplus_{r=0}^{s-1} \mathcal V(s+d-1,r)\,\oplus\,  \bigoplus_{n=0}^\infty \mathcal V(s+d-1+n,s) \,.
\ee 
The first sum corresponds to celestial fields of the same scaling dimension $\dim = d-1+s$ for each spin from $0$ to $s-1$, while the second sum corresponds to fields of increasing scaling dimension $\dim \geqslant d-1+s$ and the same spin $s$. Comparing with the spectrum that we obtained in section~\ref{sec:HS}, the first (finite) tower corresponds to fields at Coulombic order, i.e. the generalisation to any spin of the charge, mass and angular momentum aspects, while the second (infinite) tower describes the vev Chthonian degrees of freedom. This reproduces the vev spectrum on the right-hand side of figure~\ref{nice_graph}. Note that the above branching rule is valid for any dimension $d$ (irrespective of its parity) so one expects the same to be true for the flat limit of the solution space. 

\paragraph{Source (aka conformal gauge field).} The shadow fields of conformal conserved currents $j^{a_1\cdots a_s}$  are conformal gauge fields $h_{ a_1\cdots a_s}$ since the pairing $\int j^{ a_1\cdots a_s}h_{ a_1\cdots a_s}$ is invariant (up to boundary terms) under linearised Weyl-like transformations and gauge symmetries, of the form $\delta h_{ a_1\cdots a_s}=\eta_{( a_1 a_2}\alpha_{ a_3\cdots a_s)}+\partial_{( a_1}\varepsilon_{ a_2\cdots a_s)}$. These conformal gauge fields with dual scaling dimension $\Delta_-=d+1-\Delta_+=(d+1)-(s+d-1)=2-s$ span an irreducible module of the conformal algebra $\mathfrak{so}(d+1,2)$ whose character is \cite{Beccaria:2014jxa}
\be
    \mathcal{S}(2-s,s):=\mathcal{V}(2-s,s)-\mathcal{V}(1-s,s-1)+\mathcal{D}(1-s,s-1)\,.
\ee

Anticipating on the following, for the Euclidean CFT$_d$ one can consider traceless and partially conserved currents, $\partial_{i_1}\cdots\partial_{i_t}j^{i_1\cdots i_s}=0$ where $1\leqslant t\leqslant s$. The shadow fields of these partially conserved currents $j^{i_1\cdots i_s}$ are conformal gauge fields $h_{i_1\cdots i_s}$, since the pairing $\int h_{i_1\cdots i_s}j^{i_1\cdots i_s}$ is invariant under linearised Weyl-like transformations and depth-$t$ gauge symmetries $\delta h_{i_1\cdots i_s}=\delta_{(i_1i_2}\alpha_{i_3\cdots i_s)}+\partial_{(i_1}\cdots\partial_{i_t}\varepsilon_{i_{t+1}\cdots i_s)}$. Such a celestial current $j^{i_1\cdots i_s}$ of spin $s$ with partial conservation law of order $t$ in $d$ dimensions defines an irreducible $\mathfrak{so}(d+1,1)$-module with character 
\be\label{charconscurr}
    \mathcal{D}(s-t+d-1,s):=\mathcal{V}(s-t+d-1,s)-\mathcal{V}(s+d-1,s-t)\,.
\ee
A simple scaling argument on the pairing confirms that the celestial gauge field $h_{i_1\cdots i_s}$ of spin $s$ and depth $t$ has scaling dimension $d-(s-t+d-1)=1+t-s$. Together with all its descendants, it spans an $\mathfrak{so}(d+1,1)$-modules whose character will be denoted $\mathcal{S}(1+t-s,s)$.

From the analyses in \cite{Beccaria:2014jxa,Basile:2018eac}, one can extract the following result about the characters of conformal gauge fields of depth $t$:
\be\label{charshadow}
    \mathcal{S}(1+t-s,s)=\mathcal{V}(1+t-s,s)-\mathcal{V}(1-s,s-t)+\mathcal{D}(1-s,s-t) \,,
\ee
where the first term on the right-hand side corresponds to a traceless tensor $h_{i_1\cdots i_s}$ while the second corresponds to the subtraction of the degrees of freedom of the gauge parameters $\varepsilon_{i_1\cdots i_{s-t}}$, and finally the third term corresponds to conformal Killing tensor fields (solutions of $\partial_{(i_1}\cdots\partial_{i_t}\varepsilon_{i_{t+1}\cdots i_s)}=\delta_{(i_1i_2}\tilde\alpha_{i_3\cdots i_s)}$) in order to account for the oversubtraction, since these gauge transformations have no effect on the gauge potential.
Applying the lemma \eqref{lemma} to the formulae \eqref{charshadow}, one finds the branching rule of the source in the crude form:
\be\label{eq: spectrum source reducible}
    \mathcal{S}(2-s,s) \;\;\overset{\mathfrak{so}(d+1,2)}{\underset{\mathfrak{so}(d+1,1)}{\downarrow}}\;\; \bigoplus_{t=1}^{s} \mathcal{S}(1+t-s,s)\,\oplus\,  \bigoplus_{n=0}^\infty \mathcal V(2+n,s) \,.
\ee 
This branching rule is valid in any dimension, including the odd $d$ case for which there is nothing to improve.
However, the modules on the right-hand side are reducible for $d$ even. Equivalently, in more physical terms, all the corresponding fields in \eqref{eq: spectrum source reducible} are off-shell. Therefore, this decomposition should be improved for $d$ even.

On-shell, a conformal gauge field in even dimension $d$ obeys Bach-like equations of order $2(s-t-1)+d$. They are of the form $B_{i_1\cdots i_s}=\lap^{s-t-1+\frac{d}2}h_{i_1\cdots i_s}+\cdots$. The Noether identities on this Bach-like tensor are partial conservation laws, $\partial_{i_1}\cdots\partial_{i_t}B^{i_1\cdots i_s}=0$, and this tensor $B^{i_1\cdots i_s}$ is a descendant field of the primary field $h_{i_1\cdots i_s}$ with the same properties as a conformal partially-conserved current. An on-shell conformal gauge field of spin-$s$ and depth-$t$ on $S^d$ defines an irreducible $\mathfrak{so}(d+1,1)$-module whose character is
\be\label{charonshellFT}
    \mathcal{D}(1+t-s,s):=\mathcal{S}(1+t-s,s)-\mathcal{D}(s-t+d-1,s)\,,
\ee
where the first term corresponds to the gauge field itself, while the second is its gauge-invariant conformal descendant (the Bach tensor) that is set to zero by the equation of motion.
Note that the conformally-invariant differential operators defining the Bach-like tensors also exist for values $t>s$ for which there is no gauge symmetry any more. In fact, a totally-symmetric  primary tensor field in $d$ dimensions of rank $s$ and integer scaling dimension $N$ ranging from $2$ to $\frac{d}2-1$ admits a conformally-invariant differential operator of order $d-2N$; the space
of solutions is an irreducible $\mathfrak{so}(d+1,1)$-module whose character is 
\be\label{charEO}
    \mathcal{D}(N,s):=\mathcal{V}(N,s)-\mathcal{V}(d-N,s)\,.
\ee
For $N=\frac{d}2-1$, they correspond to the fields discussed in \cite{Erdmenger:1997wy}. 

Using the definitions \eqref{charonshellFT} -- \eqref{charEO}, one can rewrite the right-hand side of \eqref{eq: spectrum source reducible} only in terms of irreducible modules and obtain the branching rule of the source
\be\label{eq: spectrum source}
\begin{split}
    \mathcal{S}(2-s,s) \;\;\overset{\mathfrak{so}(d+1,2)}{\underset{\mathfrak{so}(d+1,1)}{\downarrow}}&\quad \bigoplus_{t=1}^{s} \mathcal{D}(1+t-s,s)\,\oplus\,  \bigoplus_{N=2}^{\frac{d}2-1} \mathcal D(N,s)\,\oplus\,\bigoplus_{n=0}^\infty \mathcal{V}(\tfrac{d}2+n,s) \\
    &\quad\oplus\,\bigoplus_{N=2}^{\frac{d}2-1} \mathcal V(d-N,s)\,\oplus\,\bigoplus_{t=1}^{s} \mathcal{D}(s-t+d-1,s)\,.
\end{split}
\ee 
The first sum on the right-hand side of the first line corresponds to the on-shell celestial gauge fields with the various depths, the second sum corresponds to the celestial Erdmenger--Osborn fields, seen as on-shell celestial fields with depth exceeding the spin. The third sum is an infinite tower of celestial fields that can be interpreted as the Taylor series of the radiative modes expanded in powers of the retarded time coordinate at the boundary. In the second line, the first sum corresponds to the finite tower of source Chthonians (the shadows of the Erdmenger--Osborn fields) with scaling dimension from $\frac{d}2+1$ until $d-2$; the second sum corresponds to the finite tower of celestial conserved currents with scaling dimension from $d-1$ until $s+d-2$ (the shadows of the celestial gauge fields). 

Note that the above decomposition does not change when the conformal representative of the transverse space is not a plane, but a sphere. In the latter case, the conformal Laplacian (aka GJMS operator) becomes invertible when acting on smooth fields in our weight range. This means that the fields solving the equations of motion associated with modules represented in the first sum can be expressed as pure-gauge solutions, while the fields solving the equations of motion described in the second sum vanish entirely.\footnote{This is not in contradiction with the above decomposition since (generalised) Verma modules only correspond to local information about primary fields (their Taylor series at a given point) while the above statements are only valid for globally-defined fields on the sphere.} This is in agreement with \cite{Campoleoni:2020ejn}, where the Fronsdal equations in flat spacetime were solved working with a round Bondi metric instead of the flat one.

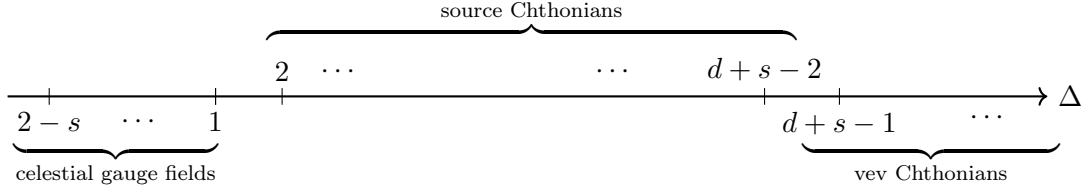
\begin{figure}[h!] 
\centering
\tikzset{
  myabove/.style={above,yshift=2pt},
  mybelow/.style={below,yshift=-2pt}
  }
\begin{tikzpicture}[scale=1.1]
  % Axis
  \draw[->, thick] (-1.5,0) -- (11,0) node[right] {$\Delta$};

  % Major tick positions and labels
  \foreach \x/\label/\pos in {
    -1.0/{$2-s$}/mybelow,
    1.0/{$1$}/mybelow,
    1.8/{$2$}/myabove,
    7.6/{$d+s-2$}/myabove,
    8.5/{$d+s-1$}/mybelow,
  }{
    \draw (\x,0.1) -- (\x,-0.1);
    \node[\pos] at (\x,0) {\label};
  }

  % Ellipses between groups
  \node at (0.1,-0.3) {$\cdots$};
  \node at (2.5,+0.3) {$\cdots$};
  \node at (5.8,+0.3) {$\cdots$};
  \node at (10.3,-0.25) {$\cdots$};

  % Underbraces to indicate grouped ranges
  \node[below=10pt] at (-0.2,-0.1)
    {$\underbrace{\hspace{2.7cm}}_{\text{celestial gauge fields}}$};
  \node[above=12pt] at (4.8,+0.15)
    {$\overbrace{\hspace{7.0cm}}^{\text{source Chthonians}}$};
  \node[below=10pt] at (9.6,-0.1)
    {$\underbrace{\hspace{3.4cm}}_{\text{vev Chthonians}}$};
\end{tikzpicture}
\caption{The spin-$s$ celestial fields of  scaling dimension $\Delta$ expected to appear 
in the asymptotic expansion of the transverse components of a massless field of integer spin $s>0$
in spacetime dimension $d+2$ with $d$ odd. \label{somewhat_boring_graph}}
\end{figure}

As a brief digression on the case of odd $d$, remember from above that the branching rule \eqref{eq: spectrum source} is to be replaced in this case with \eqref{eq: spectrum source reducible}, while the branching rule \eqref{eq: spectrum Chthonian} suggests that the flat limit of the vev remains exactly the same (i.e. aspects plus vev Chthonians) through the dimensions and that figure~\ref{boring_graph} is unchanged. However, the source is very different and there is a substantial reorganisation of the representation, e.g. for odd $d$ the celestial gauge fields and the Erdmenger--Osborn fields would be replaced with celestial primary fields that are not subject to any equation and, accordingly, should not be pure-gauge or vanishing, respectively (even if some global regularity conditions are imposed). Furthermore, the flat limit of an unconstrained source now includes an infinite tower of Chthonians which overlap with the vev Chthonians. To get an unconstrained source one has to include logarithms in the $z$-expansion, and this may explain the doubling of Chthonian degrees of freedom. On the other hand, the flat limit of a source with vanishing Fefferman-Graham obstruction (i.e.\ a source which is on-shell) would only include the finite tower of these source Chthonians, as summarised in figure~\ref{somewhat_boring_graph} that is meant to substitute figure~\ref{nice_graph} when $d$ is odd. Radiation cannot instead be recovered in this naive way since it sits at half-integer order in $z$ while the source sits at integer order.

\subsection{Carrollian limit of the vev}\label{Carrollianlimitoftheconformalscalar}

The effective speed of light\footnote{Note, however, that it is distinct from the speed of light in the bulk, which will be kept fixed while $c = \frac{1}{R}$ will be sent to zero. See e.g. \cite{Ciambelli:2018wre, Campoleoni:2023fug, Alday:2024yyj} for a discussion on this point.} on the boundary will be denoted by $c$. The conformal boundary of AdS$_{d+2}$ is a Lorentzian manifold of dimension $d+1$ endowed with the Minkowski metric $\eta_{ ab} = \text{diag}(-c^2,1,\dots,1)$ with $ a,b = 0, 1, \dots, d$. The Lie brackets of the conformal algebra $\mathfrak{so}(d+1,2)$ spanned by $\{J_{ ab}, P_ a, K_ a, D\}$ read
\be \label{eq: conformal algebra}
\begin{split}
    [J_{ ab}, J_{cd}] &= \eta_{bc} J_{ ad} - \eta_{ ac} J_{bd} - \eta_{bd} J_{ ac} + \eta_{ ad} J_{bc} \,,\\[5pt]
    [J_{ ab}, P_c] &= \eta_{bc} P_a - \eta_{ ac} P_b \,,\quad
    [J_{ ab}, K_c] = \eta_{bc} K_a - \eta_{ ac} K_b \,,\\[5pt]
    [D, P_a] &= + P_a \,,\quad
    [D, K_a] = - K_a \,,\\[5pt]
    [K_a, P_b] &= - 2 J_{ ab} + 2 \eta_{ ab} D \,.
\end{split}
\ee

\paragraph{Carrollian limit of the conformal algebra.} In the following, we will distinguish the Lorentz subalgebra $\mathfrak{so}(d+1,1)\subset\mathfrak{so}(d+1,2)$, spanned by the generators $\{J_{ij}, P_i, K_i, D\}$ (with $i,j = 1, \dots, d$) which can be thought of as the generators of conformal transformations on the celestial plane $\mathbb R^d$. The linear complement $\mathbb{R}^{d+2}$ is spanned by the generators $\{B_i := J_{0i}, H := P_0, K := K_0 \}$. The Lie brackets \eqref{eq: conformal algebra} split accordingly as
\be \label{eq: conformal algebra split}
\begin{split}
    [J_{ij}, J_{kl}] &= \delta_{jk} J_{il} - \delta_{ik} J_{jl} - \delta_{jl} J_{ik} + \delta_{il} J_{jk} \,,\\[5pt]
    [J_{ij}, B_k] &= \delta_{jk} B_i - \delta_{ik} B_j \,,\quad
    [J_{ij}, P_k] = \delta_{jk} P_i - \delta_{ik} P_j \,,\quad
    [J_{ij}, K_k] = \delta_{jk} K_i - \delta_{ik} K_j \,,\\[5pt]
    [B_i, B_j] &= c^2 J_{ij} \,,\quad
    [B_i, P_j] = \delta_{ij} H \,,\quad
    [B_i, K_j] = \delta_{ij} K \,,\\[5pt]
    [B_i, H] &= c^2 P_i \,,\;\quad
    [B_i, K] = c^2 K_i \,,\\[5pt]
    [D, P_i] &= + P_i \,,\quad
    [D, H] = + H \,,\quad
    [D, K_i] = - K_i \,,\quad
    [D, K] = - K \,,\\[5pt]
    [K_i, P_j] &= - 2 J_{ij} + 2 \delta_{ij} D \,,\quad
    [K_i, H] = 2 B_i \,,\\[5pt]
    [K, P_i] &= -2 B_i \,,\quad
    [K,H] = -2\,c^2 D\,.
\end{split}
\ee
In the Carrollian limit $c \to 0$, the relativistic conformal algebra contracts into the Carrollian conformal (aka Poincar\'e) algebra: $\mathfrak{so}(d+1,2)\stackrel{c\to 0}{\rightarrow}\mathfrak{iso}(d+1,1)$, where the Lorentz subalgebra $\mathfrak{so}(d+1,1)$ remains intact while the remaining generators abelianise into the translation subalgebra $\mathbb{R}^{d+2}$. The remaining nontrivial Lie brackets are the following 
\be\label{confCarrollcomm}
\begin{split}
    [J_{ij}, J_{kl}] &= \delta_{jk} J_{il} - \delta_{ik} J_{jl} - \delta_{jl} J_{ik} + \delta_{il} J_{jk} \,,\\[5pt]
    [J_{ij}, B_k] &= \delta_{jk} B_i - \delta_{ik} B_j \,,\quad
    [J_{ij}, P_k] = \delta_{jk} P_i - \delta_{ik} P_j \,,\quad
    [J_{ij}, K_k] = \delta_{jk} K_i - \delta_{ik} K_j \,,\\[5pt]
    [B_i, P_j] &= \delta_{ij} H \,,\quad
    [B_i, K_j] = \delta_{ij} K \,,\\[5pt]
    [D, P_i] &= + P_i \,,\quad
    [D, H] = + H \,,\quad
    [D, K_i] = - K_i \,,\quad
    [D, K] = - K \,,\\[5pt]
    [K_i, P_j] &= - 2 J_{ij} + 2 \delta_{ij} D \,,\quad%\\
    [K_i, H] = 2 B_i \,,\quad
    [K, P_i] = -2 B_i \,.
\end{split}
\ee

\paragraph{Primary and descendants of the conformal algebra.} Generalised Verma modules of $\mathfrak{so}(d+1,2)$ are given by lowest-weight representations with primary vectors denoted $|\Delta, \vec s\,\rangle$ and labelled by the scaling dimension $\Delta \in \mathbb R$ and the ``spin'' $\vec s$ of an irreducible representation $\Sigma_{ ab}$ of $\mathfrak{so}(d,1)$. The latter is a dominant integral $\mathfrak{so}(d,1)$-weight  
$\vec s\equiv(s_1,\ldots,s_\frac{d}2)$ where the entries are non-increasing integers ($s_1 \geqslant \ldots \geqslant s_{\frac{d}2} \geqslant  0$).
The primary vectors $|\Delta, \vec s\,\rangle$ must verify the conditions
\be\label{primaryconds}
    K_a |\Delta, \vec s\,\rangle = 0 \,,\quad D |\Delta, \vec s\,\rangle = \Delta |\Delta, \vec s\,\rangle \,,\quad J_{ab} |\Delta, \vec s\,\rangle = \Sigma_{ab} |\Delta, \vec s\,\rangle \,,
\ee
where $\Sigma_{ ab} |\Delta, \vec s\,\rangle$ stands for the action of $\mathfrak{so}(d,1)$ on the primary $|\Delta, \vec s\,\rangle$ in the representation labelled by $\vec s$. For totally-symmetric representations (i.e. $s_2=\cdots=s_\frac{d}2=0$) of spin $s\geqslant 1$, the unitarity bound reads $\Delta \geqslant d-1+s$, while for scalar fields it is $\Delta \geqslant \frac{d-1}{2}$. The action of the remaining generators $P_a$, gives rise to descendants. These generators can be used to translate the fields from the origin to any point in $\mathbb R^{d,1}$.

\subsubsection{Spin $0$}

Although our main target is the scalar vev, i.e. the boundary data of the bulk scalar with $\Delta=d-1$ analysed in section~\ref{sec:spin0}, it is fairly easy to discuss the case of a generic scaling dimension $\Delta$ (above the unitarity bound), so the present subsection addresses the general case.

\paragraph{Scalar descendants of a scalar primary.}
We start with the simplest case of a primary: a scalar primary of the conformal algebra $\mathfrak{so}(d+1,2)$. This corresponds to taking the trivial representation $\Sigma_{ ab} = 0$ (i.e. $\vec s=\vec 0$) of the Lorentz subalgebra $\mathfrak{so}(d,1)$. This primary will be noted simply by $|\Delta \rangle$, where $\Delta$ is a generic scaling dimension, which we only assume to be above the unitarity bound. Instead of working with characters as in section~\ref{sec:group-theory-branching}, we will branch explicitly this $\mathfrak{so}(d+1,2)$-module into a collection of $\mathfrak{so}(d+1,1)$-modules by separating time from the other directions.
The condition \eqref{primaryconds} reads
\be
K |\Delta\rangle = 0 \,,\quad    K_i |\Delta\rangle = 0 \,,\quad D |\Delta\rangle = \Delta |\Delta\rangle \,,\quad B_i |\Delta\rangle = 0\,,\quad J_{ij} |\Delta\rangle =0 \,.
\ee
There is an infinite tower of scalar descendants obtained by successive action of the generator $H$ on the primary, with the addition of an extra term for later purposes,
\be\label{descendantsDelta}
    |\Delta+1\rangle := \frac{1}{c^2} H |\Delta \rangle \,,\quad |\Delta+2\rangle := \frac{1}{c^2} \big(\,H |\Delta+1 \rangle - \alpha_1 P_i P^i |\Delta \rangle\, \big)\,, \quad \dots
\ee
The motivation for the global factor $1/c^2$ will be explained soon; furthermore the coefficient $\alpha_1$ will  be determined shortly.
Let us consider the first two such scalar descendants at level one and two.

The level-one scalar descendant is an $\mathfrak{so}(d+1,1)$-primary since it verifies the identities
\be
    K_i |\Delta+1 \rangle = 0 \,,\quad J_{ij} |\Delta+1\rangle = 0 \,,\quad D |\Delta+1 \rangle = (\Delta+1) |\Delta+1 \rangle \,.
\ee
However, it is 
not an $\mathfrak{so}(d+1,2)-$primary because it satisfies 
\be\label{nonprimary}
    K |\Delta+1\rangle = - 2 \Delta |\Delta\rangle \,,\quad B_i |\Delta+1\rangle = P_i |\Delta\rangle \,.
\ee
The normalisation \eqref{descendantsDelta} was chosen in order to have no $c^2$ factor on the right-hand sides of \eqref{nonprimary}.\footnote{A distinct Carrollian limit can also be considered where one does not include the factors $1/c^2$ in the scalar descendants. The structure of the resulting $\mathfrak{iso}(d+1,1)$-module is rather different. It is briefly discussed in appendix~\ref{Lemma 1}.}

Similarly, the level-two descendant verifies $J_{ij} |\Delta+2\rangle = 0$, $D |\Delta+2\rangle = (\Delta+2) |\Delta+2\rangle$ as well as
\be
    K_i |\Delta+2\rangle = \frac{2}{c^2} \Big(1 - \alpha_1(2\Delta+2-d)\Big) P_i |\Delta\rangle \,,
\ee
We can see that in order to have an $\mathfrak{so}(d+1,1)$ primary, i.e. $K_i |\Delta+2\rangle = 0$, we must require
\be
    \alpha_1 = \frac{1}{2\Delta+2-d} \,.
\ee
This is fine as long as the scalar field is above the unitarity bound.
Even so, the descendant is -- in general -- not an $\mathfrak{so}(d+1,2)$-primary since it also satisfies
\be
    K |\Delta+2\rangle = - 2(2\Delta+1-\alpha_1 d) |\Delta+1\rangle \,,\quad B_i |\Delta+2\rangle = 2\left(1-\alpha_1\right) P_i |\Delta+1\rangle \,.
\ee
Note that an interesting case is when the scaling dimension $\Delta = \frac{d-1}{2}$ saturates the unitarity bound. For this value of the scaling dimension, if one takes $\alpha_1 = 1$, then the level-two descendant in \eqref{descendantsDelta} is itself a primary reading $|\Delta+2\rangle = (\frac{1}{c^4}H^2-\frac{1}{c^2}P_i P^i) |\Delta\rangle= -\frac{1}{c^2}P_a P^a |\Delta\rangle$. The on-shell conformal scalar (aka singleton) corresponds to quotienting by this null vector, thereby effectively imposing the shortening condition $P_a P^a |\frac{d-1}{2}\rangle = 0$.

The above results for the first two scalar descendants generalise as follows.

\begin{lemma}\label{Lemma_1} For each level $n+1 >0$, there exists a unique scalar $\mathfrak{so}(d+1,2)$-descendant
\be \label{eq: recursion scalar}
    |\Delta + n +1 \rangle = \frac{1}{c^2}\Big( H |\Delta+n\rangle - \alpha_n P_i P^i |\Delta+n-1\rangle \Big)\,,
\ee
that is an $\mathfrak{so}(d+1,1)$-primary, in the sense that
\be\label{so(d+1,1)-primary}
    K_i |\Delta+n\rangle = 0 \,,\quad J_{ij} |\Delta + n \rangle = 0\,, \quad D |\Delta + n\rangle = (\Delta + n) |\Delta+n \rangle\,.
\ee
Moreover, we have
\be\label{so(d+1,2)-notprimary}
    K |\Delta+n\rangle = \kappa_n  |\Delta+n-1\rangle \,,\quad B_i |\Delta+n\rangle = \beta_n P_i |\Delta+n-1\rangle \,.
\ee
\end{lemma}
The expression of the coefficients $\alpha_n$, $\beta_n$ and $\kappa_n$, as well as the proof of this lemma can be found in appendix~\ref{Lemma 1}.

Note that the descendants $|\Delta+n\rangle$ are the analogue of the fields $\varphi^{(n)}$ that would appear in the asymptotic expansion for a free scalar field of mass-squared $m^2 =-\Delta(d-\Delta)/R^2$ on AdS$_{d+2}$ with flat Bondi coordinates. In fact, the definition  \eqref{eq: recursion scalar} is somewhat the analogue of \eqref{scalar_eom}.

\paragraph{Magnetic limit of a scalar primary.} On the one hand, the $c\to 0$ limit of the definition \eqref{eq: recursion scalar} of the descendants, gives the relations
\be \label{eq: Chthonians scalar}
    H|\Delta\rangle = 0 \,,\quad H|\Delta+n\rangle = \alpha_n P_i P^i |\Delta + n-1 \rangle \,,
\ee
for $n \geqslant1$. On the other hand, the remaining conditions \eqref{so(d+1,1)-primary}-\eqref{so(d+1,2)-notprimary} are unaffected. The $\mathfrak{iso}(d+1,1)$-module generated by the vectors $|\Delta+n\rangle$ ($n=0,1,2,\ldots$) obeying the relations \eqref{so(d+1,1)-primary}-\eqref{eq: Chthonians scalar} will be denoted $\mathcal{V}^{\mathfrak{iso}(d+1,1)}(\Delta,0)$ and called the \textit{magnetic limit of the Verma module} $\mathcal{V}^{\mathfrak{so}(d+1,2)}(\Delta,0)$ spanned by the primary $|\Delta\rangle$ and all its descendants. 
One important remark is that the value of the quadratic Casimir operator $C_2:=B_i B^i + H K$ of $\mathfrak{iso}(d+1,1)$ vanishes on $\mathcal{V}^{\mathfrak{iso}(d+1,1)}(\Delta,0)$. Therefore, we are describing a massless representation in the sense that
\be
    (B_i B^i + H K) |\Delta +n\rangle = 0 \,,
\ee
for all $n \geqslant 0$. Note that this massless representation $\mathcal{V}^{\mathfrak{iso}(d+1,1)}(\Delta,0)$ is \textit{not} unitary in general, so it does not correspond to a massless unitary irreducible representation of Wigner (more details can be found, e.g., in \cite{Bekaert:2024tkv}).
Let us analyse the structure of this Poincar\'e module and start with the following result.

\begin{proposition}
The vectors $P_{i_1}\cdots P_{i_k}|\Delta\rangle$ ($k=0,1,2,\ldots$) span an unfaithful $\mathfrak{iso}(d+1,1)$-submodule of $\mathcal{V}^{\mathfrak{iso}(d+1,1)}(\Delta,0)$. Effectively, the latter submodule is merely the Verma module $\mathcal{V}^{\mathfrak{so}(d+1,1)}(\Delta,0)$ because all the generators of the abelian ideal $\mathbb{R}^{d+2}$ act trivially on this submodule.
\end{proposition}
The proof is by direct computation. The first equation in \eqref{eq: Chthonians scalar} implies that the $\mathfrak{so}(d+1,1)$-primary $|\Delta\rangle$ is annihilated by all bulk translation generators spanned by the generators $H$, $B_i$ and $K$. Furthermore, one can show by induction, using the commutation relations \eqref{confCarrollcomm}, that all its descendants, $P_{i_1}\cdots P_{i_k}|\Delta\rangle$, are also annihilated by these generators. The module $\mathcal{V}^{\mathfrak{so}(d+1,1)}(\Delta,0)$ corresponds to a celestial primary field.

On the contrary, each descendant $|\Delta+n\rangle$ for $n \geqslant 1$ alone does not generate an $\mathfrak{iso}(d+1,1)$-submodule.
In fact, one can see from \eqref{so(d+1,1)-primary}-\eqref{eq: Chthonians scalar} that the action of the bulk translation generators $\{K,B_i,H\}$ on the collection $\left\{|\Delta+n\rangle\right\}_{n \geqslant 1}$ is not trivial in general.\footnote{This is true in particular for the case considered in section~\ref{sec:spin0}, i.e. $\Delta=d-1$, for which the coefficients $\alpha_n$, $\beta_n$ and $\kappa_n$ are different and do not vanish for $n>0$.} If one quotients $\mathcal{V}^{\mathfrak{iso}(d+1,1)}(\Delta,0)$ by the submodule $\mathcal{V}^{\mathfrak{so}(d+1,1)}(\Delta,0)$, then one gets an $\mathfrak{iso}(d+1,1)$-module that is isomorphic to the one generated by $|\Delta +1\rangle$ since it is now annihilated by all generators of $\mathbb{R}^{d+2}$:
\be
    \mathcal{V}^{\mathfrak{iso}(d+1,1)}(\Delta,0)\,\big/\,\mathcal{V}^{\mathfrak{so}(d+1,1)}(\Delta,0)\,\simeq\,\mathcal{V}^{\mathfrak{iso}(d+1,1)}(\Delta+1,0)\,.
\ee

\paragraph{Magnetic limit of the singleton.} The coefficients $\beta_n$ may vanish for some specific values of the scaling dimension $\Delta$. In particular, this happens at the unitarity bound $\Delta=\tfrac{d-1}{2}$. 
For the on-shell conformal scalar mentioned above, we can consider the following definition 
\be\label{rescaleddefdescsingl}
    |\tfrac{d+1}{2}\rangle := \frac{1}{c^2} H |\tfrac{d-1}{2} \rangle
\ee 
of the scalar descendant and the shortening condition
\be\label{eomsimpleton}
  H |\tfrac{d+1}{2} \rangle = P_i P^i |\tfrac{d-1}{2} \rangle \,.
\ee
The scalar descendant $|\tfrac{d+1}{2} \rangle$ satisfies the properties
\be
    K_i |\tfrac{d+1}{2} \rangle = 0 \,,\quad J_{ij} |\tfrac{d+1}{2}\rangle = 0 \,,\quad D |\tfrac{d+1}{2} \rangle = \tfrac{d+1}{2} |\tfrac{d+1}{2} \rangle \,,
\ee
and
\be
    K |\tfrac{d+1}{2}\rangle = -(d-1) |\tfrac{d-1}{2} \rangle \,,\quad B_i |\tfrac{d+1}{2}\rangle = P_i |\tfrac{d-1}{2}\rangle \,.
\ee
The $c\to0$ limit of the relation \eqref{rescaleddefdescsingl} gives
\be \label{eq:Chthonians scalar}
    H|\tfrac{d-1}{2}\rangle = 0 \,,
\ee
and $|\tfrac{d+1}{2}\rangle$ becomes independent. However, note that the primary $|\tfrac{d-1}{2}\rangle$ appears on the right-hand-side of the shortening condition \eqref{eomsimpleton}. The system of two equations \eqref{eomsimpleton} and \eqref{eq:Chthonians scalar} mirrors the equations of motion of the magnetic simpleton found in \cite{Bekaert:2024itn}. In its present incarnation, the magnetic simpleton is the ultra-short representation of the Poincar\'e algebra spanned by the vectors $P_{i_1}\cdots P_{i_k}|\tfrac{d\pm1}{2}\rangle$.

We now want  to discuss how to obtain the aspects out of relativistic conformal conserved currents in the Carrollian limit. We will only treat the case of spin $1$ and $2$.

\subsubsection{Spin $1$}

To start with, we consider the suitable matrices $\Sigma_{ ab}$ in order to describe the unitary irreducible representation of $\mathfrak{so}(d+1,2)$ corresponding to a conformal conserved current of spin $s=1$ and scaling dimension $\Delta = d$. We will denote the corresponding primary $|d\rangle_a$. This means that the matrices $\Sigma_{ab}$ must satisfy the commutation relations
\be
    [\Sigma_{ ab}, \Sigma_{cd}] = \eta_{bc} \Sigma_{ ad} - \eta_{ ac} \Sigma_{bd} - \eta_{bd} \Sigma_{ ac} + \eta_{ ad} \Sigma_{bc} \,,
\ee
and act on $|d\rangle_a$ in the following way
\be
    J_{ ab} |d\rangle_c = \Sigma_{ ab} |d\rangle_c = 2 \eta_{c[b} |d\rangle_{ a]} \,.
\ee

For this specific value of the scaling dimension $\Delta = d$ saturating the unitarity bound, one has the shortening condition\footnote{More specifically, like the scalar singleton, the descendant $P_a |d\rangle^a$ is itself a primary with zero norm and one obtains the unitary module by quotienting the  Verma module by the corresponding submodule.}
\be
    P_a |d\rangle^a = 0 \,,
\ee
describing the conservation law associated with a current in the CFT$_{d+1}$. We first begin by considering
\be
    |d\rangle^i \,,\quad |d\rangle := |d\rangle^{}_0 = - c^2 |d\rangle^0 \,.
\ee
The shortening condition reads
\be\label{shorteningspinone}
    \frac{1}{c^2} H |d\rangle = P_i |d\rangle^i \,.
\ee
We now single out the vector descendant obtained via the action of the generator $H$ plus a correction term
\be\label{firstvectordescendant}
    |d+1 \rangle^i := \frac{1}{c^2}\Big( H |d\rangle^i - \alpha^{(1)}_0 P^i |d\rangle\Big) \,,
\ee
where the coefficient $\alpha^{(1)}_0$ will be determined shortly.

\paragraph{Scalar and vector primaries.} The scalar primary $|d\rangle$ verifies
\be\label{primaryscalarspinone}
    K_i |d\rangle = 0 \,,\quad J_{ij} |d\rangle = 0 \,,\quad D |d\rangle = d |d\rangle \,,
\ee
\be
    K |d\rangle = 0 \,,\quad B_i |d\rangle = c^2 |d\rangle_i \,,\quad H |d\rangle = c^2 P_i |d\rangle^i \,,
\ee
while the vector primary $|d\rangle^i$ verifies
\be
    K_i |d\rangle^j = 0 \,,\quad J_{ij} |d\rangle^k = 2 \delta^k_{[j} |d\rangle_{i]} \,,\quad D |d\rangle^i = d |d\rangle^i \,,
\ee
\be
    K |d\rangle^i = 0 \,,\quad B_i |d\rangle^j = \delta_i^j |d\rangle \,.
\ee

\paragraph{First descendant.} The first vector descendant \eqref{firstvectordescendant} is not a primary but verifies 
\be
J_{ij} |d+1\rangle^k = 2 \delta^k_{[j} |d+1\rangle_{i]}\,, \quad D |d+1\rangle^i = (d+1) |d+1\rangle^i\,,    
\ee
as well as
\be\label{firstdesc}
    K_i |d+1\rangle^j = \frac{2}{c^2} \Big(1-\alpha^{(1)}_0 d\Big) \delta_i^j |d\rangle \,,\quad K |d+1\rangle^i = - 2\left(1 - \frac{\alpha^{(1)}_0}{d}\right) D |d\rangle^i \,,
\ee
and
\be
    B_i |d+1\rangle^j = P_i |d\rangle^j + \delta_i^j P_k |d\rangle^k - \alpha^{(1)}_0 \left(\delta_i^j P_k |d\rangle^k + P^i |d\rangle_j\right) .
\ee
One can see that there is a divergence as $c \to 0$ in the first equation of \eqref{firstdesc}, unless
\be
    \alpha^{(1)}_0 = \frac{1}{d} \,.
\ee

\begin{lemma}\label{Lemma 2} For each $n \geqslant 1$, there exists a unique vector descendant of the form
\be \label{eq: recursion spin 1}
    |d+n+1\rangle^i = \frac{1}{c^2} \Big(H |d+n\rangle^i - \alpha^{(1)}_n P^i P_j |d+n-1\rangle^j - \alpha'^{(1)}_n P_j P^j |d+n-1\rangle^i\Big) \,,
\ee
that is an $\mathfrak{so}(d+1,1)$-primary:
\be\label{primaryspinone}
K_i |d+n\rangle^j = 0\,,\quad J_{ij} |d+n\rangle^k = 2 \delta^k_{[j} |d+n\rangle_{i]}\,,\quad D |d+n\rangle^i = (d+n) |d+n\rangle^i\,.
\ee
Moreover, we have
\be \label{eq: conditions recursion spin 1}
\begin{split}
    K |d+n\rangle^i &= \kappa^{(1)}_n  |d+n-1\rangle^i \,,\\
    B_i |d+n\rangle^j &= \beta^{(1)}_n P_i |d+n-1\rangle^j + \beta'^{(1)}_n P^j |d+n-1\rangle_i + \beta''^{(1)}_n \delta_i^j P^{}_k |d+n-1\rangle^k \,.
\end{split}
\ee
\end{lemma}
As before, the proof as well as the expression of the coefficients $\alpha^{(1)}_n$, $\alpha'^{(1)}_n$, $\beta^{(1)}_n$, $\beta'^{(1)}_n$, $\beta''^{(1)}_n$ and $\kappa^{(1)}_n$ can be found in appendix~\ref{Lemma 1}.

\paragraph{Carrollian limit of the conserved current.}
First, we take the limit $c\to 0$ of the shortening condition \eqref{shorteningspinone}, as well as the conditions \eqref{eq: recursion spin 1} on the descendants, and find that
\be
\begin{split}
    H |d\rangle &= 0 \,,\qquad H |d\rangle^i = \frac{1}{d} P^i |d\rangle \,,\\
    H |d+n\rangle^i &= \alpha^{(1)}_n P^i P_j |d+n-1\rangle^j + \alpha'^{(1)}_n P_j P^j |d+n-1\rangle^i \,,
\end{split}
\ee
with $n \geqslant 1$.
The conditions \eqref{primaryscalarspinone} on $|d\rangle$ and the conditions \eqref{primaryspinone} on $|d+n\rangle^i$ for all $n\geqslant 0$ assert that they are \textit{celestial primaries}, i.e. primaries of $\mathfrak{so}(d+1,1)$.
Furthermore,
\begin{subequations}
\begin{align}
    K |d\rangle &= 0 \,,& K |d\rangle^i &= 0 \,,& K |d+n\rangle^i &= \kappa^{(1)}_n  |d+n-1\rangle^i \,,\\
    B_i |d\rangle &= 0 \,,& B_i |d\rangle^j &= \delta_i^j |d\rangle \,, &
\end{align}
\end{subequations}
and
\be
    B_i |d+n\rangle^j = \beta^{(1)}_n P_i |d+n-1\rangle^j + \beta'^{(1)}_n P^j |d+n-1\rangle_i + \beta''^{(1)}_n \delta_i^j P_k |d+n-1\rangle^k \,,
\ee
with $n \geqslant 1$.
Again, one can check that the value of the quadratic Casimir operator corresponds to a massless representation of $\mathfrak{iso}(d+1,1)$, i.e.
\be
    (HK+B_i B^i) |d\rangle = 0 \,,\qquad (HK+B_i B^i) |d+n\rangle^j = 0 \,,
\ee
for all $n \geqslant 0$.

\paragraph{Charge aspect as celestial scalar primary.} A physically important conclusion from the above equations is that the scalar primary $|d\rangle$ together with all its descendants $P_{i_1}\cdots P_{i_k}|d\rangle$ span an unfaithful $\mathfrak{iso}(d+1,1)$-submodule which will be denoted $\mathcal{V}^{\mathfrak{so}(d+1,1)}(d,0)$ because this ultrashort representation is, in practice, the irreducible Verma module of $\mathfrak{so}(d+1,1)$ growing from the scalar primary $|d\rangle$. In the context of the asymptotic analysis of electromagnetism, the latter corresponds to the charge aspect, say at spatial infinity, which is indeed a celestial primary field of scaling dimension $\Delta=d$. 

The vector primary $|d\rangle^i$ and the tower of its vector descendants $|d+n\rangle^i$ for $n\geqslant 1$ are celestial primaries corresponding to the tower of vev Chthonians in section~\ref{sec:spin1}. Together with the scalar primary $|d\rangle$, they generate via the successive action of generators $P_j$ an indecomposable  module of the Poincar\'e algebra that is spanned by $\{P_{i_1}\cdots P_{i_k}|d\rangle,P_{i_1}\cdots P_{i_\ell}|d+n\rangle^j\}$ and will be denoted $\mathcal{D}^{\mathfrak{iso}(d+1,1)}(d,1)$. One can consider the quotient of $\mathcal{D}^{\mathfrak{iso}(d+1,1)}(d,1)$ by its submodule $\mathcal{V}^{\mathfrak{so}(d+1,1)}(d,0)$. This quotient is spanned, in practice, by the representatives $P_{i_1}\cdots P_{i_m}|d+n\rangle^j$ with $m,n\geqslant 0$. This quotient is an indecomposable Poincar\'e module corresponding to the infinite tower of vev Chthonians in the asymptotic expansion of electromagnetism analysed in section~\ref{sec:spin1}. In the quotient, the representative $|d\rangle^i$ is now an $\mathfrak{so}(d+1,1)$-primary annihilated by all bulk translation generators. Similarly to the $\Delta=d-1$ and $s=0$ case in \eqref{Carrollianlimitoftheconformalscalar}, one can show that $P_{i_1}\cdots P_{i_m}|d\rangle^j$ span an unfaithful $\mathfrak{iso}(d+1,1)$-module that effectively is the Verma module $\mathcal{V}^{\mathfrak{so}(d+1,1)}(d,1)$.

\subsubsection{Spin $2$}

The case of the stress-energy tensor can now be addressed.
We start by writing explicitly the action of Lorentz generators on the spin-$2$ primary with scaling dimension $\Delta = d+1$
\be
    J_{a_1b_1} |d+1\rangle^{a_2b_2} = \Sigma_{ a_1b_1}|d+1\rangle^{a_2b_2} = 2 \delta^{a_2}_{[b_1} |d+1\rangle^{}_{a_1]}{}^{b_2} + 2 \delta^{b_2}_{[b_1} |d+1\rangle^{}_{a_1]}{}^{a_2} \,.
\ee
Note that the right-hand side has been made symmetric in ${a_2}$ and $b_2$, and is automatically traceless in ${a_2}$ and $b_2$, as it should be. The shortening condition
\be
    P_a |d+1\rangle^{ ab} = 0 \,,
\ee
translates the fact that the physical interpretation of this primary is as a conserved stress-tensor in a CFT$_{d+1}$. Moreover, we have the traceless property $\eta_{ ab} |d+1\rangle^{ ab} = 0$.

\paragraph{Scalar, vector and tensor primaries.}

We branch $SO(d,1) \downarrow SO(d)$
\be
    |d+1\rangle^{\langle ij \rangle} \,,\quad |d+1\rangle^i := |d+1\rangle^i{}_0 \,,\quad |d+1\rangle := |d+1\rangle_{00} = -c^2 |d+1\rangle^0{}_0 \,,
\ee
where angle-brackets implement a symmetric traceless projection in the spatial indices. Thus, here and in the following, we will make the assumption that whenever we are talking about $|d+1\rangle^{ij}$ or one of its descendants, we now refer to the traceless object. The shortening condition reads
\be
    \frac{1}{c^2} H |d+1\rangle = P_i |d+1\rangle^i \,,\quad \frac{1}{c^2} H |d+1\rangle_i = P_j |d+1\rangle^j{}_i + \frac{1}{c^2} \frac{1}{d} P_i |d+1\rangle \,,
\ee
where in the last equality we implemented the decomposition between the $SO(d)$-traceless part $|d+1\rangle^{ij}$ and the trace $|d+1\rangle$. These verify the conditions
\be\label{Bid+1}
    B_i |d+1\rangle = 2c^2 |d+1\rangle_i \,,\quad B_i |d+1\rangle_j = \frac{d+1}{d} \delta_{ij} |d+1\rangle + c^2 |d+1\rangle_{ij} \,,\quad B_i |d+1\rangle_{jk} = 2 \delta_{i\langle j} |d+1\rangle_{k \rangle} \,,
\ee
and are annihilated by $K_i$ and $K$.

\paragraph{First descendant.}

We now single out the descendant
\be
    |d+2\rangle^{ij} := \frac{1}{c^2} \left(H |d+1\rangle^{ij} - \alpha^{(2)}_0 P^{\langle i} |d+1\rangle^{j\rangle} \right) ,
\ee
with $\alpha^{(2)}_0$ to be determined shortly. Note that, due to the conservation equations, the $H$-descendants of $|d+1\rangle$ and $|d+1\rangle^i$ are not new quantities and can be computed as $P_i$-descendants of $|d+1\rangle^i$, and $|d+1\rangle^{ij}$ and $|d+1\rangle$ respectively. We have
\be
    K_k |d+2\rangle^{ij} = \frac{2}{c^2}\left[2 - \alpha^{(2)}_0 (d+2)\right] \delta^{\langle i}_k |d+1\rangle^{j\rangle} \,,
\ee
so we can see that the condition to be a $\mathfrak{so}(d+1,1)$-primary is that
\be
    \alpha^{(2)}_0 = \frac{2}{d+2} \,.
\ee

\begin{lemma}\label{Lemma 3} For each $n \geqslant 1$, there exists a unique tensor descendant of the form
\be \label{eq: recursion spin 2}
    |d+2+n\rangle^{ij} := \frac{1}{c^2} H |d+1+n\rangle^{ij} - \frac{\alpha_n^{(2)}}{c^2} P^{\langle i} P_k |d+n\rangle^{j\rangle k} - \frac{\alpha'^{(2)}_n}{c^2} P^k P_k |d+n\rangle^{ij}
\ee
that is an $\mathfrak{so}(d+1,1)$-primary:
\be\label{primaryspintwo}
\begin{split}
    J_{ij} |d+2+n\rangle^{kl} &= 2 \delta^k_{[j} |d+2+n\rangle_{i]}{}^l + 2 \delta^l_{[j} |d+2+n\rangle_{i]}{}^k \\
    K_i |d+2+n\rangle^{jk} &= 0\,,\quad D |d+2+n\rangle^{ij} = (d+2+n) |d+2+n\rangle^{ij}\,.
\end{split}
\ee
Moreover, we have
\be \label{eq: conditions recursion spin 2}
\begin{split}
    K |d+1+n\rangle^{ij} &= \kappa^{(2)}_n |d+n\rangle^{ij} \\
    B_k |d+1+n\rangle^{ij} &= \beta^{(2)}_n P_k |d+n\rangle^{ij} + \beta'^{(2)}_n P^{\langle i} |d+n\rangle^{j\rangle}{}_k + \beta''^{(2)}_n \delta^{\langle i}_k P_l |d+n\rangle^{j \rangle l} \,.
\end{split}
\ee
\end{lemma}
As before, the proof as well as the expression of the coefficients $\alpha^{(2)}_n$, $\alpha'^{(2)}_n$, $\beta^{(2)}_n$, $\beta'^{(2)}_n$, $\beta''^{(2)}_n$ and $\kappa^{(2)}_n$ can be found in appendix~\ref{Lemma 1}.

\paragraph{Carrollian limit of the conserved current.}

In the Carroll limit, the conservation equation becomes
\be\label{Hd+1}
    H|d+1\rangle = 0 \,,\quad H|d+1\rangle_i = \frac{1}{d} P_i |d+1\rangle \,,
\ee
while the requirement of finiteness turns the definition of the descendants $|d+1+n\rangle$ for $n \geqslant 1$ into constraint equations
\be
    H|d+1\rangle^{ij} = \alpha^{(2)}_0 P^{\langle i} |d+1\rangle^{j\rangle} \,,\quad H |d+1+n\rangle^{ij} = \alpha^{(2)}_n P^{\langle i} P_k |d+n \rangle^{j\rangle k} + \alpha'^{(2)}_n P^k P_k |d+n\rangle^{ij}\,,
\ee
for all $n \geqslant 1$. These are reminiscent of the evolution equations (in the absence of radiation) for the mass aspect $|d+1\rangle$, angular momentum aspect $|d+1\rangle^i$ and Chthonians $|d+1+n\rangle^{ij}$. Also note that the Carrollian limit of \eqref{Bid+1} gives
\be\label{Bid+1'}
    B_i |d+1\rangle = 0\,,\quad B_i |d+1\rangle_j = \frac{d+1}{d} \delta_{ij} |d+1\rangle \,.
\ee
As before, we can verify our calculations by computing the quadratic Casimir and checking that it vanishes on any $SO(d+1,1)$ primary
\be
    (H K + B_i B^i) |d+1\rangle = 0 \,,\quad (H K + B_i B^i) |d+1\rangle^j = 0 \,,\quad (H K + B_i B^i) |d+1+n\rangle^{jk} = 0 \,,
\ee
for all $n \geqslant 0$.

\paragraph{Mass and angular momentum aspects.}

The mass and angular momentum aspects at spatial infinity are both celestial primary fields, incarnated here by the scalar and vector primaries $|d+1\rangle$ and $|d+1\rangle^i$, respectively. Actually, the scalar primary $|d+1\rangle$ is annihilated by $H$, $K$ and $B_i$, cf. \eqref{Hd+1} and \eqref{Bid+1'}. Therefore, the collection of descendants $P_{i_1}\dots P_{i_k}|d+1\rangle$ spans an unfaithful $\mathfrak{iso}(d+1,1)$-submodule that is, effectively, an irreducible Verma module $\mathcal{V}^{\mathfrak{so}(d+1,1)}(d+1,0)$. Together with the descendants $P_{i_1}\dots P_{i_k}|d+1\rangle^j$ of the vector primary, they form an indecomposable $\mathfrak{iso}(d+1,1)$-submodule decomposing as the direct sum $\mathcal{V}^{\mathfrak{so}(d+1,1)}(d+1,0)\oplus\mathcal{V}^{\mathfrak{so}(d+1,1)}(d+1,1)$ upon restriction to the Lorentz subalgebra. The indecomposability of this $\mathfrak{iso}(d+1,1)$-submodule follows from the triangular structure of the equations in \eqref{Hd+1} and \eqref{Bid+1'}. Accordingly, the mass and angular momentum aspects at spatial infinity form an indecomposable Poincar\'e module which can be called the \textit{aspect multiplet}. This picture for $s=1,2$ is in agreement with analyses (see e.g.~\cite{Donnay:2022aba, Barnich:2022bni, Donnay:2022wvx, Ruzziconi:2024kzo, Nguyen:2025sqk}) of Carrollian conserved currents in the context of flat-space holography. On top of that, the infinite tower of vev Chthonian modes corresponds, in this Carrollian CFT formulation, to the quotient of the space of all descendants of the scalar, vector and tensor primaries by the invariant subspace of descendants from the scalar and vector primaries.

One expects this picture to generalise for all integer spins, as summarised in table~\ref{tablevev}.

\begin{center}
\begin{table}[ht]
    \centering
    \renewcommand*{\arraystretch}{1.7}
    \begin{tabular}{c|c|c}
    Solutions & Poincar\'e & Faithful \\
    \hline\hline
Coulombic order and below & Module & Yes 
\\\hline
Charge aspect & Submodule & No  \\
Aspect multiplet & & Yes 
\\\hline
Chthonian modes & Quotient module & Yes  
\\\hline
\end{tabular}
\caption{Characterisation of relevant (sub)spaces of solutions  starting at Coulombic order, as indecomposable Poincar\'e modules, in the language of Carrollian CFT. The last row refers to the quotient of the first row by a submodule in the row below. \label{tablevev}}
\end{table}
\end{center}

Note that for $d=1$, i.e. for a three-dimensional bulk, massive Poincar\'e/BMS modules were constructed in \cite{Campoleoni:2016vsh} using a different representation of the highest-weight primaries and derived from an ultra-relativistic limit of AdS$_3$ highest-weight modules.

%%%%%%%%%%%%%%%%%%%%%%%%%%%%%%%%%%%%%%%%%%%%%%%%%%%%%%%
\section{Outlook}\label{sec:conclusions}

We conclude this work by presenting a few natural research directions suggested by our results. We begin by discussing how one could go beyond our formal expansions in integer powers of $z$ and what advantages this could bring, especially in understanding the flat limit in odd spacetime dimensions. Our systematic approach to the study of the equations of motion is expected to provide useful insights to analyse these issues, which remain open even for “low-spin” examples, including gravity and its simplest matter couplings.  
We then outline how our findings contribute to ongoing efforts to set the stage for a possible flat-space analogue of higher-spin holography.

\paragraph{Regularity and peeling.} Our study of the free equations of motion was restricted to an asymptotic analysis of \textit{formal} solutions, i.e.\ solutions in the vicinity of infinity, expressed as formal power series in $z$, as in Fefferman--Graham and Anderson's classical results \cite{AST_1985__S131__95_0, Anderson:2004yi, Fefferman:2007rka}. We found two independent branches of solutions: the source and vev in the AdS case, and (roughly) the radiative and Chthonian modes in the flat case. Nevertheless, it is well-known that global regularity conditions relate these two sets of solutions to each other. This feature is particularly transparent for asymptotically AdS spacetimes after Wick rotation, where solving the asymptotic equations of motion can be viewed as a sort of Dirichlet (or Neumann) problem in an asymptotically hyperbolic Einstein space. In this setting, the boundary data are given by the conserved, traceless boundary energy-momentum tensor (or by the conformal class of the boundary metric) and there exists a corresponding unique solution that is regular in the interior and is a small deformation of a reference solution \cite{GrahamLee, Biquard, QingShiWu2012RicciFlowAH}. This implies that the source and vev are related to each other on-shell. Similarly, for asymptotically flat spacetimes, imposing smoothness in the interior ensures the uniqueness of solutions to the characteristic (aka Goursat) problem, where the boundary data are located at null infinity (see, e.g., chapter 18 of \cite{Kroon}). 
This implies that the Chthonian modes are determined in terms of the radiative ones in analogy with the relation between source and vev in the AdS case. In this sense, the Chthonian degrees of freedom are actually frozen in the case of regular solutions, which explains why they are, basically always, implicitly discarded in the mathematical-physics literature. It would be nice to extend our systematic analysis to investigate the effect of such global regularity conditions on the space of solutions for any integer spin. 

\vspace{1mm}

As mentioned in Footnote~\ref{integerpowers}, we also restricted our analysis of the Minkowski solution space to the self-consistent asymptotic expansion in integer powers of the radial coordinate in Bondi gauge. The AdS origin, if any, of the logarithmic terms (that, at least in four dimensions, one has to include in order to describe all relevant Ricci-flat metrics near null infinity) remains elusive. In fact, our analysis focuses on even-dimensional spacetimes, while it is for odd-dimensional asymptotically AdS spacetimes that logarithmic terms must be included in order to avoid the Fefferman-Graham obstruction (a feature related to the existence of conformal anomalies on the even-dimensional boundary). Understanding the AdS origin of logarithmic terms in even-dimensional Minkowski backgrounds thus remains an open question. 

\paragraph{Odd spacetime dimensions.} Another tantalising open issue is the AdS origin of radiative solutions when the spacetime dimension is odd. Radiation always appears at order $z^{\frac{d}2-s}$ in flat spacetime, hence at a half-integer order in odd-dimensional spacetimes. This feature is related to the absence of a smooth conformal compactification for asymptotically flat spacetimes in odd dimensions strictly higher than four \cite{Hollands:2003ie, Hollands:2004ac} and 
it is unclear how to obtain such half-integer powers starting from the standard AdS asymptotic expansion in integer powers of $z$, possibly complemented by logarithmic terms.

Despite these open questions, group theory nevertheless suggests that some features remain the same, irrespectively of the parity of the dimension. In particular, the flat limit of the vev should remain exactly the same as discussed above (i.e.\ aspects plus vev Chthonians) and, accordingly, figure~\ref{boring_graph} would be unchanged. However,  
there is a substantial reorganisation of the representation associated to the source. More precisely, group theory suggests that figure~\ref{nice_graph} should be replaced with figure~\ref{somewhat_boring_graph}. 
Nevertheless, this picture remains to be corroborated by a concrete analysis of the solution space. Note that 
radiation is absent in figure~\ref{somewhat_boring_graph}. A refined method thus seems to be required in order to understand the AdS origin of radiation for odd-dimensional spacetimes. 

The possibility that a different approach may exist for taking the flat limit of the solutions, for which radiation  can be obtained for both parities of the spacetime dimension, is suggested by the following heuristic observation. The closed form solution of the Klein-Gordon equation \eqref{scalareom} for a scalar field on AdS$_{d+2}$ with squared mass $m^2  = -2(d-1)/R^2$ reads, for $d$ even, as 
\be\label{closedform}
\varphi_{\textrm{AdS}}(z,u,\mathbf{x}) = R^{\frac{d+1}{2}} z^{\frac{d+1}{2}} \bigg(I_{\frac{d-3}{2}}\left(R^2z \,\hat{p}\right) C_1(u-R^2z,\mathbf{x}) +  I_{-\frac{d-3}{2}}\left(R^2z\, \hat{p} \right) C_2(u-R^2z,\mathbf{x}) \bigg) ,
\ee
where $\hat{p} := \sqrt{\partial^2_u
- \frac{1}{R^2}
\lap } $ and $I_\nu(y)$ is the modified Bessel function of the first kind. Similarly, the solution for $d$ odd takes the form \eqref{closedform}, except that $I_{-\frac{d-3}{2}}$ is replaced with $K_{\frac{d-3}{2}}$, 
where $K_\nu(y)$ is the modified Bessel function of the second kind. (See, e.g., chapter~10.2 in \cite{AbramowitzStegun1972} on modified Bessel functions.) 

Note that precisely for half-integer values of the parameter, i.e.\ $d$ even, the modified Bessel function admits an explicit expression in terms of elementary functions (cf. \S~8.467 in \cite{GradshteynRyzhik2014}), leading to the form 
\begin{eqnarray}
    \label{closedform2}
\varphi_{\textrm{AdS}}(z,u,\mathbf{x})&=& R^{\frac{d-1}{2}} z^{\frac{d}{2}} \bigg(P_{\frac{d-4}2}(R^{-2}z^{-1} \hat{p}^{-1})\exp(R^2z\, \hat{p})\,C_+(u-R^2z,\mathbf{x}) \\&&\qquad\qquad\qquad+P_{\frac{d-4}2}(-R^{-2}z^{-1} \hat{p}^{-1})\exp(-R^2z \,\hat{p})\, 
C_-(u-R^2z,\mathbf{x}) \bigg) \,,\nonumber
\end{eqnarray}
where $P_n$ is a polynomial of degree $n$:
\be
P_n\!\left(y\right)
= \sum_{k=0}^{n} \frac{(n+k)!}{k!\,(n-k)!}\,\left(-\frac{y}{2}\right)^k\,.
\ee
The two functions $C_\pm$ are the free boundary data and presumably correspond to outgoing/ingoing radiation, since
\be
\varphi_{\textrm{AdS}}(z,u,\mathbf{x})\stackrel{R\to\infty}{\sim}
R^{\frac{d-1}{2}} z^{\frac{d}{2}} \bigg(C_+(u,\mathbf{x}) + C_-(u-2R^2z,\mathbf{x}) \bigg)\,,
\ee
where the second term depends on the advanced time $v:=u-2R^2z$. A natural boundary condition is the absence of ingoing radiation at early time, i.e. $\lim_{v\to-\infty}C_-(v,\mathbf{x})=0$. This condition seems necessary for taking a naive flat limit where one keeps fixed the flat Bondi coordinates (since $v\to-\infty$ when $R\to\infty$ with $u$ and $z$ fixed).

The arbitrary functions $C_1(t,\mathbf{x})$ and $C_2(t,\mathbf{x})$ in \eqref{closedform} are the free data at spatial infinity. They are respectively identified with the source and the vev, since 
\be
I_{\nu}(y) \sim \frac1{2^\nu\Gamma(\nu+1)}\,y^\nu \quad\text{and}\quad
K_{\nu}(y) \sim 2^{\nu-1}\Gamma(\nu)\, y^{-\nu}  \quad\textrm{when} \quad y \to 0^+\,,
\ee
where the behaviour of the modified Bessel function of the second kind $K_{\nu}(y)$ holds for $\nu>0$.
There is actually a subtlety for $d=3$ because in that case $K_0(y) \sim - \ln y$ when $y \to 0^+$. This is related to the fact that the source and vev collide for this value of $d$ and a log branch is needed to avoid putting the source on-shell. As mentioned above, this last feature applies to higher odd-dimensional spacetimes as well. In the closed form expression \eqref{closedform}, it is accounted for by the fact that, for integer values $\nu=n\in\mathbb N$ of the parameter, the modified Bessel function of the second kind behaves more precisely as (cf. \S~8.446 in \cite{GradshteynRyzhik2014})
\be
K_n(y)
=
2^{n-1} (n-1)!\, y^{-n} + \dots + \frac{(-1)^{n+1}}{2^n \,n!}\, y^n \ln y\,, \quad \textrm{when} \quad y \to 0^+ \,,
\ee
where the dots stand for subleading terms in the power expansion that we keep until the order where the logarithmic term becomes relevant.

The asymptotic behaviour of the modified Bessel functions is given by
\be
I_{\nu}(y) \sim \frac{1}{\sqrt{2\pi}}\, y^{-\frac{1}{2}}e^y \quad\text{and}\quad
K_{\nu}(y) \sim \sqrt{\frac{\pi}{2}}\, y^{-\frac{1}{2}}e^{-y}  \quad\textrm{when} \quad y \to \infty \, .
\ee
In the flat limit, we use this asymptotic behaviour to obtain the radiative order in Minkowski spacetime $\mathbb{R}^{d+1,1}$ 
\be
 \varphi_{\textrm{AdS}}(z,u,\mathbf{x})\stackrel{R\to\infty}{\sim} R^{\frac{1-d}2} z^{\frac{d}{2}} \phi(u,\mathbf{x})\,,
\ee
where 
$\phi(u,\mathbf{x}):=\frac{1}{\sqrt{2\pi}}\,  \partial_u^{-1/2} \big(C_1(u,\mathbf{x}) +  \tilde C_2(u,\mathbf{x})\big)$, with $\tilde C_2(u,\mathbf{x})=C_2(u,\mathbf{x})$
when $d$ is even, while for $d$ odd, one has $\tilde C_2(u,\mathbf{x})=0$ if one assumes that $\lim_{u \to -\infty} C_2(u,\mathbf{x})=0$. 
This line of reasoning would deserve to be elaborated further, but it suggests there may be ways to obtain radiation from a flat limit also in odd-dimensional spacetimes.

\paragraph{Higher-spin holography.} A strong
motivation behind our systematic study for any spin is higher-spin holography (see, e.g., \cite{Bekaert:2012ux, Giombi:2016ejx, Sleight:2017krf, Bekaert:2022poo} for some reviews) and, more precisely, the quest for its flat spacetime analogue. In spite of the no-go arguments constraining higher-spin interactions in Minkowski spacetime (see, e.g., \cite{Bekaert:2010hw} for a review), it has been recently pointed out that several building blocks of higher-spin holography do admit a non-trivial flat limit on both the bulk \cite{Campoleoni:2021blr, Boulanger:2023prx} and boundary sides \cite{Bekaert:2022oeh, Bekaert:2024itn} (see also \cite{Ponomarev:2022ryp,Pekar:2023fsu,Serrani:2026azw}). Our description of the flat limit  
of the solutions to the equations of motion might provide a useful guiding principle for further investigating this flat limit at the level of the renormalised on-shell action, along the lines, e.g., of \cite{Hartong:2025jpp,Campoleoni:2025bhn,Ammon:2025avo}. In particular, the known coupling $\int j^{ a_1\cdots a_s}h_{ a_1\cdots a_s}$ between the source $h_{ a_1\cdots a_s}$ and the vev $j^{ a_1\cdots a_s}$ suggests that a similar coupling must be considered between the various celestial fields that appear in the flat limit (cf.~section~\ref{sec:summary}). Investigating these issues might help to obtain a flat analogue of the Gubser-Klebanov-Polyakov-Witten prescription.

One puzzle that arises in the quest for flat-space higher-spin holography is the question \cite{Ponomarev:2022qkx} of what might be the flat analogue of the Flato-Fronsdal theorem \cite{Flato:1978qz, Vasiliev:2004cm}. The latter is considered a cornerstone of higher-spin gravity. This theorem is often stated in a colloquial way as follows: the tensor product of two singletons decomposes into an infinite tower of massless fields in AdS with all integer spins. This theorem is instrumental in the duality since it is usually taken as a prediction for the spectrum of the bulk dual of a free CFT of a large number of conformal scalars living on the boundary of AdS. The flat limit of a bulk singleton was analysed in \cite{Bekaert:2022oeh, Bekaert:2024itn} and dubbed ``simpleton'' (see also \cite{Campoleoni:2021blr, Ponomarev:2021xdq} for earlier discussions). Its physical degrees of freedom sit at null infinity and are encoded in a pair of celestial scalars (obtained from the Carrollian limit of a boundary conformal scalar). Accordingly, the product of two simpletons
cannot account for the degrees of freedom of propagating fields in Minkowski spacetime, a fact which appears to be in tension with the standard reading of the Flato-Fronsdal theorem. However, there is a simple resolution to this puzzle: the decomposition of the product of two singletons actually gives a sum over unitary irreducible representations of the AdS isometry group that, as recalled in the introduction, correspond to the boundary vev of massless fields in AdS.  
As we have seen, the flat limit of the vev does \textit{not} give rise to radiation and, therefore, there is no tension. In fact, our results seem to indicate instead that the flat limit of the Flato-Fronsdal theorem should state that the tensor product of two simpletons decomposes into an infinite tower of aspects and vev Chthonians with all integer spins. We intend to return to this issue in the future.

%%%%%%%%%%%%%%%%%%%%%%%%%%%%%%%%%%%%%%%%%%%%%%%%%%%%%%%
\acknowledgments

We thank A.~Delfante, L.~Donnay and Y.~Herfray for discussions.
The work of A.C.\ was supported by
the Fonds de la Recherche Scientifique – FNRS under the grant T.0047.24. The work of S.P.\ is supported by the \textit{European Research Council (ERC) Project 101076737 -- CeleBH}. Views and opinions expressed are however those of the author only and do not necessarily reflect those of the European Union or the European Research Council. Neither the European Union nor the granting authority can be held responsible for them. S.P. is also partially supported by INFN Iniziativa Specifica ST\&FI.
We thank the Galileo Galilei Institute for Theoretical Physics for the hospitality and the INFN for partial support
during the completion of this work. X.B., S.P.\ and A.R.\ thank the University of Mons and A.C.\ thanks SISSA for hospitality.

\appendix

%%%%%%%%%%%%%%%%%%%%%%%%%%%%%%%%%%%%%%%%%%%%%%%%%%%%%%%
\section{Scalar field}

This appendix contains the proofs of technical results used in section \ref{sec:spin0}.

\subsection{Mirror relation}\label{sec:mirror_scalar}

We wish to prove eq.~\eqref{mirror_scalar}, that is
\be 
\varphi^{\left(\frac{d}{2} + k\right)} = R^{2k} \prod_{l=0}^{k-1} g^{\left(\frac{d}{2}+k-2l\right)} \varphi^{\left(\frac{d}{2} - k\right)} \qquad \textrm{for}\ 1 \leqslant k \leqslant \frac{d}{2}-2 \, .
\ee
We prove it by induction, noticing that it is manifestly true for $k=1$ thanks to the condition $f^{(\frac{d}{2}+1)} = 0$ and that it can be easily checked for $k = 2$.
Assuming it is valid for \( k-1 \) and \( k-2 \), from \eqref{recursion_path} one gets
\be \label{vark_app}
\begin{split}
\varphi^{\left(\frac{d}{2}+k\right)} & = R^2f^{\left(\frac{d}{2}+k\right) }\varphi^{\left(\frac{d}{2}+k-1\right)} + R^2g^{\left(\frac{d}{2}+k\right)}\varphi^{\left(\frac{d}{2}+k-2\right)} \\ 
& = R^{2k} f^{\left(\frac{d}{2}+k\right)} \prod_{l=0}^{k-2} g^{\left(\frac{d}{2}+k-1-2l\right)} \varphi^{\left(\frac{d}{2} - k +1\right)} 
 +  R^{2(k-1)} \prod_{l=0}^{k-2} g^{\left(\frac{d}{2}+k-2l\right)} \varphi^{\left(\frac{d}{2} - k+2\right)} \,.
\end{split}
\ee
Using now the recursion relation \eqref{recursion_path} for $\varphi^{\left(\frac{d}{2}-k+2\right)}$,
\be
\varphi^{\left(\frac{d}{2}-k+2\right)} = R^2f^{\left(\frac{d}{2}-k+2\right)} \varphi^{\left(\frac{d}{2}-k+1\right)} + R^2g^{\left(\frac{d}{2}-k+2\right) } \varphi^{\left(\frac{d}{2}-k\right)}
\ee
eq.~\eqref{vark_app} becomes 
\be
\begin{split}
\varphi^{\left(\frac{d}{2}+k\right)} &= R^{2k} \left[ f^{\left(\frac{d}{2}+k\right)} \prod_{l=0}^{k-2} g^{\left(\frac{d}{2}+k-1-2l\right)} + f^{\left(\frac{d}{2}-k+2\right)} \prod_{l=0}^{k-2} g^{\left(\frac{d}{2}+k-2l\right)} \right] \varphi^{\left(\frac{d}{2} - k +1\right)} \\
& \qquad+ R^{2k}  \prod_{l=0}^{k-1} g^{\left(\frac{d}{2}+k-2l\right)} \varphi^{\left(\frac{d}{2}-k\right)} \,.
\end{split}
\ee
To proceed, we notice that $f^{(n)}$ and $g^{(n)}$ have the same denominator. We therefore introduce
\be
b_{n} = \frac{1}{(n-2)(n-d+1)} \quad \Rightarrow \quad f^{(n)} = - b_{n} (2n-d-2)\, \partial_u \quad \textrm{and} \quad g^{(n)} = - b_{n}\, \lap \, .
\ee
We then have
\begin{align}
&\varphi^{\left(\frac{d}{2}+k\right)} \nn\\ &=(-1)^kR^{2k} (2k-2)  \left[ b_{\left(\frac{d}{2}+k\right)} \prod_{l=0}^{k-2} b_{\left(\frac{d}{2}+k-1-2l\right)} - b_{\left(\frac{d}{2}-k+2\right)} \prod_{l=0}^{k-2} b_{\left(\frac{d}{2}+k-2l\right)} \right] \lap^{k-1} \partial_u\varphi^{\left(\frac{d}{2} - k +1\right)} \nonumber \\
&\qquad + R^{2k} \prod_{l=0}^{k-1} g^{\left(\frac{d}{2}+k-2l\right)} \varphi^{\left(\frac{d}{2}-k\right)} \,.
\end{align}
Using the symmetry $b_{\left(\frac{d}{2}+k-2l-1\right)} = b_{\left(\frac{d}{2}-k+2l+2\right)}$, we relabel the first product and then reverse the index $l \to k-2-l$:
\be
\prod_{l=0}^{k-2} b_{\left(\frac{d}{2}+k-1-2l\right)} = \prod_{l=0}^{k-2} b_{\left(\frac{d}{2}-k+2+2l\right)} = \prod_{l=0}^{k-2} b_{\left(\frac{d}{2}+k-2-2l\right)} \,.
\ee
Multiplying this relabelled product by the $b_{\left(\frac{d}{2}+k\right)}$ prefactor, we can shift the index to absorb it, and pull out the last term instead:
\be
b_{\left(\frac{d}{2}+k\right)} \prod_{l=0}^{k-2} b_{\left(\frac{d}{2}+k-2-2l\right)} = \prod_{l=0}^{k-1} b_{\left(\frac{d}{2}+k-2l\right)} = b_{\left(\frac{d}{2}-k+2\right)} \prod_{l=0}^{k-2} b_{\left(\frac{d}{2}+k-2l\right)} \,.
\ee
The terms inside the bracket then cancel identically, which completes the proof.

\subsection{AdS origin of the Minkowski solution space} \label{app:scalar_chthonians}

Here, we motivate the choice to truncate the expansion of the source in eq.~\eqref{R_expansion} to $\mathcal O(R^{-d+2})$, as well as the choice to select
$\varphi^{(2,n)}$ for $0 \leqslant n \leqslant \frac{d}{2}-3$ and $\varphi^{(d-1,n)}$ for $n \geqslant 0$ such that their $k$-th $u$-derivatives evaluated at $u = 0$ vanish, for $0 \leqslant k < n$. 
In both cases, the argument will be the same: we will first prove that relaxing these conditions would only introduce redundant boundary data that do not affect the space of solutions 
and we then express the solution space in terms of the non-spurious data we introduced. 

Let us begin by relaxing the hypothesis on the bound on the expansion of the source, i.e.\ let us consider $\varphi^{(2,k)} \neq 0$ for $k \geqslant \frac{d}{2}-1$. Using the recursion relation \eqref{recursion_divergences} and the mirror relation \eqref{mirror_scalar}, we can express $\varphi^{(n,0)}$ for $n \geqslant 2$ in terms of the data contained in the expansion of the source and the vev as follows:
\begin{align*}
    \varphi^{(2,0)} &= \varphi^{(2,0)} \quad \text{(definition)} \,,\\
    \varphi^{(3,0)} &= f^{(3)} \varphi^{(2,1)} \,,\\
    \varphi^{(4,0)} &= f^{(4)} f^{(3)} \varphi^{(2,2)} + g^{(4)} \varphi^{(2,1)} \,,\\
    &\dots \\
    \varphi^{(\frac{d}{2},0)} &= f^{(\frac{d}{2})} f^{(\frac{d}{2}-1)} \dots f^{(3)} \varphi^{(2,\frac{d}{2}-2)} + \text{ other terms involving } \varphi^{(2,n<\frac{d}{2}-2)} \,,\\
    &\dots \\
    \varphi^{(d-3,0)} &= g^{(d-3)} g^{(d-5)} \dots g^{(5)} f^{(3)} \varphi^{(2,\frac{d}{2}-2)} \,,\\
    \varphi^{(d-2,0)} &= g^{(d-2)} g^{(d-4)} \dots g^{(4)} \varphi^{(2,\frac{d}{2}-2)} \,,\\
    \varphi^{(d-1,0)} &= \varphi^{(d-1,0)} \quad \text{(definition)} \,,\\
    \varphi^{(d,0)} &= f^{(d)} \varphi^{(d-1,1)} + g^{(d)} \varphi^{(d-2,1)} \,,\\
    &\dots
\end{align*}
where in $\varphi^{(n\geqslant d,0)}$ we can see now the appearance of the coefficients $\varphi^{(d-2,k \geqslant 1)}$. A necessary condition for these to be non-zero is that $\varphi^{(2,k \geqslant \frac{d}{2}-1)}$ are non-zero, for instance,
\be
\varphi^{(d-2,1)} = g^{(d-2)} g^{(d-4)} \dots g^{(4)} \varphi^{(2,\frac{d}{2}-1)} \,.
\ee

We already know that the cancellation of the divergent terms $\varphi^{(n,-1)}$ is enough to ensure that the solution is finite in the limit $R \to \infty$ and reproduces the flat-space equations of motion (see the discussion around eq.~\eqref{flat_constraints}). In order to make sure that the full flat solution space is present, one must verify that at each order in the $z$-expansion, one introduces exactly the correct number of new data (i.e.\ an independent integration ``constant'' only depending on the transverse coordinates, except at the order of radiation where an unconstrained function of retarded time appears). In order to prove that no spurious degrees of freedom are introduced, it is necessary and sufficient to realise the aforementioned solution space in terms of AdS data.

\paragraph{Fields above radiation order (Erdmenger--Osborn fields).}

In $\varphi^{(3,0)}$, the field $\varphi^{(2,1)}$ appears for the first time, with one $u$-derivative. One can therefore express the first flat-space Chthonian entering eq.~\eqref{flat_rec_scalar} as
\be
    \phi^{(3)}(\mathbf x) = \varphi^{(3,0)}(u=0,\mathbf x) = f^{(3)} \varphi^{(2,1)}(u=0,\mathbf x) = -\partial_u \varphi^{(2,1)}(u=0,\mathbf x) \,.
\ee
In $\varphi^{(4,0)}$, the field $\varphi^{(2,2)}$ appears for the first time, with two $u$-derivatives, as well as the field $\varphi^{(2,1)}$, this time with no $u$-derivative. It seems that we have two competing contributions for the definition of the new independent data in Minkowski spacetime (the integration ``constant'') at order $\mathcal O(z^4)$: both $\partial_u^2 \varphi^{(2,2)}(u=0,\mathbf x)$ and $g^{(4)}\varphi^{(2,1)}(u=0,\mathbf x)$ are admissible. We resolve this tension by choosing $\varphi^{(2,1)}(u=0,\mathbf x) = 0$. This is in principle stronger than the condition $g^{(4)}\varphi^{(2,1)}(u=0,\mathbf x) = 0$ that would suffice to solve the issue, but it does not restrict the solution space because $\varphi^{(2,1)}(u=0,\mathbf x)$ will never appear in a position where it will contribute alone to the solution, as it is manifest from table~\ref{table: flat from AdS solution space scalar} below. We thus write
\be
    \phi^{(4)}(\mathbf x) = \varphi^{(4,0)}(u=0,\mathbf x) = f^{(4)}f^{(3)}\varphi^{(2,2)}(u=0,\mathbf x) \,.
\ee
We generalise as follows: for all $0 \leqslant k \leqslant \frac{d}{2}-3$, we choose $\partial_u^n \varphi^{(2,k)}(u=0,\mathbf x) = 0$ with $0 \leqslant n \leqslant k-1$ in order to avoid over-counting. The corresponding Chthonian is therefore given by eq.~\eqref{eq_erd-osborn_realisation}, that is
\be
    \phi^{(2+n)}(\mathbf x) = f^{(2+n)} f^{(1+n)} \cdots f^{(3)} \varphi^{(2,n)}(u=0,\mathbf x) \,.
\ee
Due to eq.~\eqref{eq_erd-osborn_scalar}, these fields satisfy the desired Erdmenger--Osborn field equation, proving that the flat solution space is indeed realised, without redundancy.

\paragraph{Radiation and source Chthonians.}

Contrary to the previous $\varphi^{(2,k)}$ with $1 \leqslant k \leqslant \frac{d}{2}-3$, one should not impose any constraint on $\varphi^{(2,\frac{d}{2}-2)}$. While it is true that its $(\frac{d}{2}-2)$-th $u$-derivative feeds radiation, its lower derivatives (from $\frac{d}{2}-1$ down to $0$) also appear at orders $\mathcal O(z^{\frac{d}{2}+1 \leqslant n \leqslant d-2})$ and are relevant due to the mirror relation. These will feed the so-called source Chthonians displayed in eq.~\eqref{1st-Cht-scalar}
\be
\begin{split}
    \phi^{(\frac{d}{2}+k)}(\mathbf x) &= \varphi^{(\frac{d}{2}+k)}(u=0,\mathbf x) \\
    &= g^{(\frac{d}{2}+k)} g^{(\frac{d}{2}+k-2)} \cdots g^{(\frac{d}{2}-k+2)} \varphi^{(\frac{d}{2}-k,k)}(u=0,\mathbf x) \\
    &= g^{(\frac{d}{2}+k)} g^{(\frac{d}{2}+k-2)} \cdots g^{(\frac{d}{2}-k+2)} f^{(\frac{d}{2}-k)} f^{(\frac{d}{2}-k-1)} \cdots f^{(3)} \varphi^{(2,\frac{d}{2}-2)} (u=0,\mathbf x) \,,
\end{split}
\ee
for $1 \leqslant k \leqslant \frac{d}{2}-2$. It is therefore sufficient to keep $\varphi^{(2,\frac{d}{2}-2)}$ as a whole.

Note that strictly speaking, the parts of $\partial_u^{\frac{d}{2}-2-k} \varphi^{(2,\frac{d}{2}-2)}(u=0,\mathbf x)$ that are in the kernel of $\lap^k$ do not show up in the flat solution for $1 \leqslant k \leqslant \frac{d}{2}-2$, so they should be quotiented in order to avoid double-counting. This is consistent with the fact that we have already identified the Erdmenger--Osborn fields $\phi^{(2+n)}$ for $0 \leqslant n \leqslant \frac{d}{2}-3$ as originating from $\varphi^{(2,n)}$ for $0 \leqslant n \leqslant \frac{d}{2}-3$ as being on-shell for $\lap^{\frac{d}{2}-2-n}$. This is also consistent with the discussion in section~\ref{sec:group-theory-branching} where it can be seen in the split \eqref{charEO}, with the differential operator of order $2k$ for scalars being simply $\lap^k$. 

\paragraph{Fields at the order of the vev and lower (vev Chthonians).}

Next, we turn to the fields at the order of the vev and below. A similar mechanism as for the fields above radiation order is at play: in $\varphi^{(d,0)}$, the field $\varphi^{(d-1,1)}$ appears for the first time, with one $u$-derivative, as well as $g^{(d)}\varphi^{(d-2,1)}$ with no $u$-derivative. Therefore we have again a tension on 
how to define the new Minkowski Chthonian appearing at this order in terms of AdS data. One can resolve it by making the choice $g^{(d)}\varphi^{(d-2,1)}(u=0,\mathbf x) = 0$. However, $\varphi^{(d-2,1)}$ is completely new (it has never seen before in the expansion), so we can actually require the even stronger condition $\varphi^{(2,\frac{d}{2}-1)} = 0$, which implies $\varphi^{(d-2,1)} = 0$. This does not restrict the solution space in any way and the vev Chthonian at this order reads
\be
    \phi^{(d)} = \varphi^{(d,0)}(u=0,\mathbf x) = f^{(d)} \varphi^{(d-1,1)}(u=0,\mathbf x) \,.
\ee
This is generalised to any order as follows: for all $k \geqslant 0$, we choose $\partial_u^n \varphi^{(d-1,k)}(u=0,\mathbf x) = 0$ with $0 \leqslant n \leqslant k-1$, as well as $\varphi^{(2,k \geqslant \frac{d}{2}-1)} = 0$, which implies $\varphi^{(d-2,k \geqslant 1)} = 0$. In that case, the Chthonians read
\be
    \phi^{(d-1+n)}(\mathbf x) = \varphi^{(d-1+n,0)}(u=0,\mathbf x) = f^{(d-1+n)} f^{(d-2+n)} \cdots f^{(d)} \varphi^{(d-1,n)}(u=0,\mathbf x)\,,
\ee
which reproduces eq.~\eqref{eq_vev-Chthonians_realisation}.

\paragraph{Summary.}

The AdS origin of the flat solution space, as well as the constraints on the corresponding fields is summarised in table \ref{table: flat from AdS solution space scalar}.
\begin{center}
\begin{table}[ht!]
\centering
\renewcommand*{\arraystretch}{1.7}
\begin{tabular}{cl|c}
    \multicolumn{2}{c|}{AdS field} & constraint in the flat-space limit (finiteness + requirement) \\
    \hline\hline
    $\varphi^{(2,n)}$ & $n \leqslant \frac{d}{2}-3$ & $\partial_u^{k<n} \varphi^{(2,n)}(u=0,\mathbf x) = 0$, $\lap^{\frac{d}{2}-2-n}\,\partial_u^n \varphi^{(2,n)}(u=0,\mathbf x) = 0$ \\\hline
    $\varphi^{(2,n)}$ & $n = \frac{d}{2}-2$ & unconstrained, only $\lap^{n-k}\,\partial_u^k\varphi^{(2,n)}(u=0,\mathbf x)$ appear ($k < n$) \\\hline
    $\varphi^{(2,n)}$ & $n \geqslant \frac{d}{2}-1$ & $\varphi^{(2,n)} = 0$ \\\hline
    $\varphi^{(d-1,n)}$ & $n \geqslant 0$ & $\partial_u^{k<n} \varphi^{(d-1,n)}(u=0,\mathbf x) = 0$
    \end{tabular}
\caption{AdS origin of the flat solution space. The source $\varphi^{(2)}$ and vev $\varphi^{(d-1)}$ are expanded in powers of the cosmological constant into $\varphi^{(2,n)}$ and $\varphi^{(d-1,n)}$. As a result, the AdS solution space seems to be demultiplied, but the following constraints are imposed in the flat limit so as to reproduce the flat solution space exactly.\label{table: flat from AdS solution space scalar}}
\end{table}
\end{center}
In summary, the independent data contained in the source and vev of AdS are redistributed as follows when taking the flat limit:
\begin{itemize}
    \item $\partial_u^n \varphi^{(2,n)}(u=0,\mathbf x)$ for $0 \leqslant n \leqslant \frac{d}{2}-3$ with $\lap^{\frac{d}{2}-2-n} \partial_u^n \varphi^{(2,n)}(u=0,\mathbf x) = 0$\,;
    \item $\partial_u^{\frac{d}{2}-2} \varphi^{(2,\frac{d}{2}-2)}(u,\mathbf x)$\,;
    \item $\lap^{\frac{d}{2}-2-k} \partial_u^k \varphi^{(2,\frac{d}{2}-2)}(u=0,\mathbf x)$ for $0 \leqslant k \leqslant \frac{d}{2}-3$\,;
    \item $\partial_u^n \varphi^{(d-1,n)}(u=0,\mathbf x)$ for $n \geqslant 0$.
\end{itemize}
As advertised, the extra constraints imposed (i.e., $\partial_u^n \varphi^{(2,k)}(u=0,\mathbf x) = 0$ for $0 \leqslant n < k$ and $\partial_u^n \varphi^{(d-1,k)}(u=0,\mathbf x) = 0$ for $0 \leqslant n < k$) do not restrict the solution space.

%%%%%%%%%%%%%%%%%%%%%%%%%%%%%%%%%%%%%%%%%%%%%%%%%%%%%%%
\section{Arbitrary spin}

\subsection{Equations of motion in flat Bondi coordinates} \label{sec:eom_Bondi}

The non-vanishing Christoffel symbols of the AdS metric in the flat Bondi coordinates \eqref{AdS_Bondi} are
\begin{alignat}{9} 
    \Gamma^z{}_{zz} & = - \frac{2}{z}\,, \qquad &&
    \Gamma^u{}_{ij} = \frac{1}{z}\, \delta_{ij} \,, &&
    \Gamma^i{}_{zj} = - \frac{1}{z}\, \delta^i{}_j \,, && && \\[5pt] 
    \Gamma^u{}_{uu} & = - \frac{1}{R^2 z} \,, \qquad &&
    \Gamma^z{}_{uu} = - \frac{1}{R^4 z} \,, \qquad && 
    \Gamma^z{}_{uz} = \frac{1}{R^2 z} \,, \qquad &&
    \Gamma^z{}_{ij} = \frac{1}{R^2 z}\, \delta_{ij} \,. 
\end{alignat}
Taking this information into account and imposing the gauge-fixing conditions \eqref{Bondi-like_intro}, one can check that all components of the Fronsdal tensor \eqref{Maxwell-like} with at least two $z$ indices vanish identically. The components with a single $z$ index have the same form in both AdS and Minkowski spacetimes and read
\be \label{Fz1}
\begin{split}
\cF_{z u(s-k-1) i(k)} = & - \left[ z^2 \pr_z^2 - (d-2k-2)\, z\pr_z - 2k(d-1) \right] \varphi_{u(s-k)i(k)} \\[5pt]
& - z \left( z\pr_z + 2k \right) \pr\cdot \varphi_{u(s-k-1)i(k)} \, .
\end{split}
\ee
The remaining non-vanishing components are
\be \label{Fz2}
\begin{split}
\cF_{u(s-k) i(k)} = & - z\, \Big[\, (s-k-2)\, z\pr_z - d(s-k-1) - 2k \,\Big] \pr_u \varphi_{u(s-k)i(k)} \\
& + z^2 \Big[\, \Delta \varphi_{u(s-k)i(k)} - k\, \pr_i \pr\cdot \varphi_{u(s-k)i(k-1)} \,\Big] \\
& + \frac{1}{R^2} \Big[\, z^2\pr_z^2 - \left( d-2s+(s-k)(s-k-1) \right) z\pr_z \\
& \phantom{+ \frac{1}{R^2} \Big[} + (d-1)\left( (s-k)(s-k-1) - 2(s-1) \right) \Big] \varphi_{u(s-k)i(k)} \\
& - (s-k)\, z\, \Big[\, z\pr_u + \frac{s-k-1}{R^2} \,\Big] \pr\cdot \varphi_{u(s-k-1)i(k)} \\
& - k\, z\, \Big[ \left( z\pr_z - d + 2 \right) \pr_i \varphi_{u(s-k+1)i(k-1)} - (k-1)\, \delta_{ii} \pr\cdot \varphi_{u(s-k+1)i(k-2)} \,\Big] \\
& + k(k-1) \Big[\, z\pr_z - d + 1 \,\Big]\, \delta_{ii} \varphi_{u(s-k+2)i(k-2)} \, .
\end{split}
\ee 

In the Bondi-like gauge \eqref{Bondi-like_intro}, all components of the Bianchi identities \eqref{Bianchi} with at least two $z$ indices are trivial. The components with a single $z$ index signal instead the following relations between the components of the Fronsdal tensor:
\be \label{Bianchi_1}
\begin{split}
\cB_{zu(s-k-2)i(k)} = - \frac{z}{2}\, \Big\{ & \left( z\partial_z + 2k \right) \cF^{\,\prime}{}_{\!\!\!u(s-k-2)i(k)} - 2\,z\, \partial\cdot \cF_{zu(s-k-2)i(k)} \\
& + 2(d+2k)\, \cF_{zu(s-k-1)i(k)} \Big\} \, ,
\end{split}
\ee
where a prime denotes a contraction with $\delta^{ij}$. 
The remaining components read
\begin{align}
& \cB_{u(s-k-1)i(k)} = z\, \Big\{ \left( z\partial_z - d \right) \cF_{u(s-k)i(k)} + \frac{k(k-1)}{2}\, \delta_{ii} \cF^{\,\prime}{}_{\!\!\!u(s-k)i(k-2)} \nonumber \\
& + z \Big( \partial\cdot \cF_{u(s-k-1)i(k)} - \frac{k}{2}\,\partial_i \cF^{\,\prime}{}_{\!\!\!u(s-k-1)i(k-1)} \Big) - \frac{s-k-1}{2} \Big( z\partial_u + \frac{s-k-2}{R^2}  \Big) \cF^{\,\prime}{}_{\!\!\!u(s-k-2)i(k)} \nonumber \\
& - \Big[  (s-k-2)\,z \partial_u - \frac{1}{R^2}\, \Big( z\partial_z - (s-k-1)(s-k-2) - d \Big) \Big] \cF_{zu(s-k-1)i(k)} \nonumber \\[5pt]
& - z\,k\, \partial_i \cF_{zu(s-k)i(k-1)} + k(k-1)\, \delta_{ii} \cF_{zu(s-k+1)i(k-2)} \Big\} \, . \label{Bianchi_2}
\end{align}

\subsection{Mirror relation}\label{sec:mirror_anyspin}

We wish to prove eq.~\eqref{mirror_HS}, that is
\be
\Phi^{\left(\frac{d}{2} - s + k\right)} = R^{2k} \prod_{l=0}^{k-1} g^{\left(\frac{d}{2} - s + k - 2l\right)} \Phi^{\left(\frac{d}{2} - s - k\right)} \qquad \textrm{for}\ 1 \leqslant k \leqslant \frac{d}{2} - 2 + s \, .
\ee
We prove it by induction, noticing that it is manifestly true for $k=1$ thanks to the condition $f^{(\frac{d}{2}-s+1)} = 0$ and that it can be easily checked for $k = 2$. Assuming it is valid for $k-1$ and $k-2$:
\be
\Phi^{\left(\frac{d}{2}-s+k\right)} = R^2f^{\left(\frac{d}{2}-s+k\right) }\Phi^{\left(\frac{d}{2}-s+k-1\right)} + R^2g^{\left(\frac{d}{2}-s+k\right) }\Phi^{\left(\frac{d}{2}-s+k-2\right)}
\ee
\be \label{appendix_HS_eq1}
= R^{2k} f^{\left(\frac{d}{2}-s+k\right)} \prod_{l=0}^{k-2} g^{\left(\frac{d}{2}-s+k-1-2l\right)} \Phi^{\left(\frac{d}{2} - s - k +1\right)} + R^{2(k-1)} \prod_{l=0}^{k-2} g^{\left(\frac{d}{2}-s+k-2l\right)} \Phi^{\left(\frac{d}{2} - s - k+2\right)}
\ee
Using now the recursion relation \eqref{HS_recursion} for $\Phi^{\left(\frac{d}{2}-s-k+2\right)}$
\be
\Phi^{\left(\frac{d}{2}-s-k+2\right)} = R^2f^{\left(\frac{d}{2}-s-k+2\right)} \Phi^{\left(\frac{d}{2}-s-k+1\right)} + R^2g^{\left(\frac{d}{2}-s-k+2\right) } \Phi^{\left(\frac{d}{2}-s-k\right)} \, ,
\ee
eq. \eqref{appendix_HS_eq1} becomes
\be
\begin{split}
\Phi^{\left(\frac{d}{2}-s+k\right)} &= R^{2k} \left[ f^{\left(\frac{d}{2}-s+k\right)} \prod_{l=0}^{k-2} g^{\left(\frac{d}{2}-s+k-1-2l\right)} + f^{\left(\frac{d}{2}-s-k+2\right)} \prod_{l=0}^{k-2} g^{\left(\frac{d}{2}-s+k-2l\right)} \right] \Phi^{\left(\frac{d}{2} - s - k +1\right)} \\
&\quad + R^{2k} \prod_{l=0}^{k-1} g^{\left(\frac{d}{2}-s+k-2l\right)} \Phi^{\left(\frac{d}{2}-s-k\right)} \, .
\end{split}
\ee
Using \eqref{HS_fg} and \eqref{HS_b_n} to write
\be \label{def_g_appendix}
g^{(n)} = -\,b_{n} \mathbf{B}_{n-1} \quad \text{with}\quad \mathbf{B}_{n} = 
    \,\lap
    + \,b_{n}\,(d+2s-4)\,(v\!\cdot\!\partial)\,(\partial\!\cdot\!\partial_v)
    - \,b_{n}\,\abs{v}^2 (\partial\!\cdot\!\partial_v)^2 \,.
\ee
We then have
\begin{align} \label{prove_to_zero}
&\Phi^{\left(\frac{d}{2}-s+k\right)} = (-1)^k R^{2k} (2k-2) \Bigg[ b_{\left(\frac{d}{2}-s+k\right)} \prod_{l=0}^{k-2} \left(b_{\left(\frac{d}{2}-s+k-1-2l\right)}  \mathbf{B}_{\left(\frac{d}{2}-s+k-2-2l\right)} \right) \nn \\ &- b_{\left(\frac{d}{2}-s-k+2\right)}\! \prod_{l=0}^{k-2} \left(b_{\left(\frac{d}{2}-s+k-2l\right)}  \mathbf{B}_{\left(\frac{d}{2}-s+k-2l-1\right)} \right)\!\! \Bigg] \Phi^{\left(\frac{d}{2} - s - k +1\right)}\! +\! R^{2k} \!\prod_{l=0}^{k-1} g^{\left(\frac{d}{2}-s+k-2l\right)} \Phi^{\left(\frac{d}{2}-s-k\right)} .
\end{align}
We will demonstrate that the term enclosed in the square brackets vanishes identically. Isolating the product on the first line, we reverse the sequence of the product via the index substitution $l \to k-2-l$. Using the reflection symmetries $b_{n} = b_{d+1-2s-n}$ and $\mathbf{B}_{n} = \mathbf{B}_{d+1-2s-n}$, we obtain
\begin{align}
\prod_{l=0}^{k-2}\left( b_{\left(\frac{d}{2}-s+k-1-2l\right)} \mathbf{B}_{\left(\frac{d}{2}-s+k-2-2l\right)} \right) = &\prod_{l=0}^{k-2} \left(b_{\left(\frac{d}{2}-s-k+3+2l\right)} \mathbf{B}_{\left(\frac{d}{2}-s-k+2+2l\right)} \right) \nn \\= &\prod_{l=0}^{k-2} b_{\left(\frac{d}{2}-s+k-2l-2\right)}\prod_{l=0}^{k-2} \mathbf{B}_{\left(\frac{d}{2}-s+k-2l-1\right)} 
\end{align}
Upon multiplying this rearranged product by the prefactor $b_{\left(\frac{d}{2}-s+k\right)}$, we can shift the product index to absorb the prefactor, subsequently factoring out the final term to find
\begin{align}   
&b_{\left(\frac{d}{2}-s+k\right)} \prod_{l=0}^{k-2} b_{\left(\frac{d}{2}-s+k-2l-2\right)}   = \prod_{l=-1}^{k-2} b_{\left(\frac{d}{2}-s+k-2l-2\right)}\nn\\
&\qquad= \prod_{l=0}^{k-1} b_{\left(\frac{d}{2}-s+k-2l\right)} = b_{\left(\frac{d}{2}-s-k+2\right)} \prod_{l=0}^{k-2} b_{\left(\frac{d}{2}-s+k-2l\right)} 
\end{align}
This expression exactly matches the second term in the square brackets \eqref{prove_to_zero}, establishing the cancellation and completing the proof.

\subsection{Useful identities}\label{sec:identities}
Note that the operator $g^{(n)}$ maps traceless states into themselves. Consequently, its action on the space of traceless states $\Psi_s = \psi_{i_1 \cdots i_s} v^{i_1} \cdots v^{i_s}$ can be expressed as
\be
    g^{(n)} \Psi_s = -b_{n} P_s \bigg( \lap + b_{n-1} (d+ 2s - 4) (v\cdot \partial)(\partial \cdot \partial_v) \bigg) \Psi_s \,.
\ee

\paragraph{Identity 1:} 
\be \label{commut_iden_1_appendix} \left[g^{(n)}, P_s(v\cdot \partial)^k(\partial\!\cdot\!\partial_v)^k \right] \Psi_s = 0 \ee
where $\Psi_s$ is traceless, i.e., $(\partial_v \cdot \partial_v) \Psi_s = 0$. To prove this identity we have to compute $\Big[P_s\,,\,P_s(v \cdot \partial)^k(\partial \cdot \partial_v)^k\Big]$ and $\Big[P_s(v \cdot \partial)(\partial \cdot \partial_v)\, , \,P_s(v \cdot \partial)^k(\partial \cdot \partial_v)^k\Big]$. The first commutator
\be
    \Big[P_s\,,\,P_s(v \cdot \partial)^k(\partial \cdot \partial_v)^k\Big] \Psi_s = P_s^2(v \cdot \partial)^k(\partial \cdot \partial_v)^k \Psi_s - P_s(v \cdot \partial)^k(\partial \cdot \partial_v)^k P_s \Psi_s = 0 \,,
\ee
vanishes trivially since $P_s^2 = P_s$ and $P_s \Psi_s = \Psi_s$. Now the second commutator can be expanded as
\begin{align} \label{whole_comm}
    &\Big[P_s(v \cdot \partial)(\partial \cdot \partial_v)\,,\,P_s(v \cdot \partial)^k(\partial \cdot \partial_v)^k\Big] \\
&= P_s \bigg( \Big[P_s \,,\,(v \cdot \partial)^k(\partial \cdot \partial_v)^k\Big](v \cdot \partial)(\partial \cdot \partial_v) - \Big[P_s\,,\,(v \cdot \partial)(\partial \cdot \partial_v)\Big](v \cdot \partial)^k(\partial \cdot \partial_v)^k \bigg) \,. \nn
\end{align} 
We now need to compute 
\begin{align}
 &\Big[P_s \,,\,(v \cdot \partial)^{k+1-n}(\partial \cdot \partial_v)^{k+1-n}\Big](v \cdot \partial)^{n}(\partial \cdot \partial_v)^{n}\nn\\& \quad= \sum_{p=0}^{\lfloor s/2 \rfloor}a_p \Big[\abs{v}^{2p}(\partial_v \cdot \partial_v)^p \,,\, (v \cdot \partial)^{k+1-n}(\partial \cdot \partial_v)^{k+1-n}\Big](v \cdot \partial)^{n}(\partial \cdot \partial_v)^{n} \label{twopartscomm}
\end{align}
for $n = k$ and $n=1$, with $a_p$ given by \eqref{traceless_proj}.
The goal is the following: moving all the $\abs{v}^2$ to the left and $(\partial_v \cdot \partial_v)$ to the right. This is helpful since the entire expression is multiplied from the left by $P_s$ and, $P_s \abs{v}^2 \Psi_{s-2} = 0$ and $(\partial_v \cdot \partial_v)\Psi_s = 0$. The right-hand side of eq.~\eqref{twopartscomm}, under the left action of $P_s$ now becomes
\be \label{n=1andk}
    \sum_{p=1}^{\lfloor s/2 \rfloor}a_p \Big[\abs{v}^{2p} \,,\, (v \cdot \partial)^{k+1-n}(\partial \cdot \partial_v)^{k+1-n}\Big](\partial_v \cdot \partial_v)^p(v \cdot \partial)^{n}(\partial \cdot \partial_v)^{n} \, .
\ee
Let us evaluate the surviving terms for the two cases $n=k$ and $n=1$. For $n=k$, the inner commutator simplifies to $\Big[\abs{v}^{2p}, (v \cdot \partial)(\partial \cdot \partial_v)\Big] = -2p \abs{v}^{2p-2}(v \cdot \partial)^2$. Because $P_s$ annihilates any remaining $\abs{v}^2$ on the left, only the $p=1$ term survives the projection. Our expression becomes, still under the left action of $P_s$
\be \label{term_1}
 P_s \Big[ P_s \,, \, (v \cdot \partial)(\partial \cdot \partial_v) \Big] (v \cdot \partial)^k (\partial \cdot \partial_v)^k = -2 a_1 (v \cdot \partial)^2 (\partial_v \cdot \partial_v) (v \cdot \partial)^k (\partial \cdot \partial_v)^k \,.
\ee
To push $(\partial_v \cdot \partial_v)$ to the right, we use the identity $\Big[\partial_v \cdot \partial_v, (v \cdot \partial)^k\Big] = 2k (v \cdot \partial)^{k-1} (\partial \cdot \partial_v) + k(k-1)(v \cdot \partial)^{k-2} \lap$. Then, the whole expression acting on $\Psi_s$ becomes
\begin{align}
&P_s \Big[ P_s \,, \, (v \cdot \partial)(\partial \cdot \partial_v) \Big] (v \cdot \partial)^k (\partial \cdot \partial_v)^k \Psi_s\nn\\ &\quad= -2 a_1 P_s \Big( 2k (v \cdot \partial)^{k+1} (\partial \cdot \partial_v)^{k+1} + k(k-1) (v \cdot \partial)^k (\partial \cdot \partial_v)^k  \lap \Big) \Psi_s \,.\label{term_1_expanded}
\end{align} 
We now evaluate the commutator \eqref{n=1andk} at $n =1$. Notice that the trace operator acting twice or more on $(v \cdot \partial)(\partial \cdot \partial_v) \Psi_s$ yields $(\partial_v \cdot \partial_v)^2 (v \cdot \partial)(\partial \cdot \partial_v) \Psi_s = 0$, where we have used again $[\partial_v \cdot \partial_v, v \cdot \partial] = 2 \partial \cdot \partial_v$ to push $\partial_v \cdot \partial_v$ to the right. Consequently, in the sum over $p$, only the $p=1$ term gives a non-zero contribution, as before.
We need the commutator $\Big[\abs{v}^2, (v \cdot \partial)^k (\partial \cdot \partial_v)^k \Big]$, which expands via the identity $[\abs{v}^2, (\partial \cdot \partial_v)^k] = -2k (v \cdot \partial)(\partial \cdot \partial_v)^{k-1} - k(k-1)  (\partial \cdot \partial_v)^{k-2}\lap$. Applying this $p=1$ term to $(\partial_v \cdot \partial_v)(v \cdot \partial)(\partial \cdot \partial_v)\Psi_s = 2(\partial \cdot \partial_v)^2 \Psi_s$, we get:
\begin{align}
&P_s \Big[ P_s, (v \cdot \partial)^k (\partial \cdot \partial_v)^k \Big] (v \cdot \partial)(\partial \cdot \partial_v) \Psi_s\nn\\&\quad = -2a_1 P_s \Big( 2k (v \cdot \partial)^{k+1} (\partial \cdot \partial_v)^{k+1} + k(k-1) (v \cdot \partial)^k  (\partial \cdot \partial_v)^k \lap \Big) \Psi_s \,.
\label{term_2}
\end{align}
Comparing \eqref{term_1_expanded} and \eqref{term_2}, we see that both terms evaluate to the exact same expression. Subtracting them in \eqref{whole_comm} yields identically zero
\be
\Big[P_s(v \cdot \partial)(\partial \cdot \partial_v)\,,\,P_s(v \cdot \partial)^k(\partial \cdot \partial_v)^k\Big] \Psi_s = 0 \,.
\ee
This proves the commutator identity \eqref{commut_iden_1_appendix}.

\paragraph{Identity 2:} Any $\Psi_{k-2}$, not necessarily traceless, obeys
\be \label{P_kills_K}
P_k \, \abs{v}^2 \, \Psi_{k-2}= 0
\ee
This can be proven directly by using the definition in eq.~\eqref{traceless_proj},
\be \begin{split}
&P_k \, \abs{v}^2 \, \Psi_{k-2} = \sum_{n=0}^{\lfloor \frac{k}{2} \rfloor } a_n \abs{v}^{2n} \left(\partial_v \cdot \partial_v \right)^n \abs{v}^2 \, \Psi_{k-2}  \\
&=\sum_{n=0}^{\lfloor \frac{k}{2} \rfloor } a_n  \left( \abs{v}^{2(n+1)} \left(\partial_v \cdot \partial_v \right)^{n} + 2n\abs{v}^{2n}\left(d+2n-2 + 2v \cdot \partial_v \right)\left(\partial_v \cdot \partial_v \right)^{n-1} \right)\Psi_{k-2} \,.
\end{split}
\ee
Writing the action of the weight (or Euler) operator $v \cdot \partial_v \left(\left(\partial_v \cdot \partial_v \right)^{n-1}   \Psi_{k-2} \right) =\left( k -2n \right) \left(\partial_v \cdot \partial_v \right)^{n-1} \Psi_{k-2}$ in the sum, we can see that the two terms in the sum cancel each other due to the recursion \eqref{traceless_proj}. There remains only the last term of the first sum, $n = \lfloor \frac{k}{2} \rfloor$, whose action on $\Psi_{k-2}$ also vanishes since there are more $\partial_v$-derivatives than there are $v$ variables. All in all, everything vanishes, proving the identity.

\paragraph{Identity 3:} 
\be \label{Identity_3}
\begin{split}
&\partial \cdot \partial_v \left( P_k(v \cdot \partial)^l(\partial \cdot \partial_v)^l \Psi_k \right) \\
&=\frac{P_{k-1} (v \cdot \partial)^{l-1}}{d + 2k - 4} \Bigg\{ (d + 2k - 4 -2l)   (v \cdot \partial) (\partial \cdot \partial_v) +  l(d + 2k - 3 -l)  \lap   \Bigg\} (\partial \cdot \partial_v)^l\Psi_k  \,.
\end{split}
\ee
where $\Psi_k$ is traceless. To prove this, let $X_k = (v \cdot \partial)^l(\partial \cdot \partial_v)^l \Psi_k$. Since $P_k X_k$ is traceless, its divergence must also be exactly traceless, allowing us to write $(\partial \cdot \partial_v) (P_k X_k) = P_{k-1} (\partial \cdot \partial_v) (P_k X_k)$. Using \eqref{traceless_proj}
\be
(\partial \cdot \partial_v) (P_k X_k) = P_{k-1} \Big[ \partial \cdot \partial_v X_k - \frac{1}{d+2k-4} (v \cdot \partial) (\partial_v \cdot \partial_v) X_k \Big] \,.
\ee
where we have eliminated $\mathcal{O}(\abs{v}^2)$ terms due to the $P_{k-1}$ projection \eqref{P_kills_K}. The divergence yields $\partial \cdot \partial_v X_k = (v \cdot \partial)^l (\partial \cdot \partial_v)^{l+1} \Psi_k + l \lap (v \cdot \partial)^{l-1}(\partial \cdot \partial_v)^l \Psi_k$ and since $\Psi_k$ is traceless, the trace operator yields $(\partial_v \cdot \partial_v) X_k = 2l (v \cdot \partial)^{l-1} (\partial \cdot \partial_v)^{l+1} \Psi_k + l(l-1) \lap (v \cdot \partial)^{l-2} (\partial \cdot \partial_v)^l \Psi_k$. Combining these proves the required identity.
\paragraph{Identity 4:} For $1 \leqslant k \leqslant \frac{d}{2} - 2 + s$ and any traceless $\Psi_s$,
\begin{eqnarray}
 &&\left(\partial \cdot \partial_v\right)^m \left(\prod_{l=0}^{k-1} g^{\left(\frac{d}{2} - s +k-2l\right)} \right)\Psi_s\nonumber\\
 &&\quad=  \sum_{l=0}^{k} (-2)^l \binom{k}{l} \frac{\left( \frac{d}{2} + s - m - k - 1 \right)_{2k-l}}{\left[ \left( \frac{d}{2} + s - k - 1 \right)_{2k} \right]^2} \lap^{k-l}  P_{s-m} (v\cdot \partial)^l(\partial \cdot \partial_v)^{m+l} \Psi_s \,.
\end{eqnarray} 
To prove this, first notice that we can rewrite the multiplication of the operator $g^{(n)}$ as
\be \label{g_mul_anzatz}
    \prod_{i=1}^k g^{(n_i)} = \sum_{l=0}^{k} P_s c_l(n_1,\dots,n_k) (v\cdot \partial)^l(\partial\!\cdot\!\partial_v)^l \,,
\ee
for some coefficients $c_l(n_1,\dots,n_k)$ that may be operator-valued in physical space (i.e. they only depend on $\partial$, not on $v$ or $\partial_v$). The $m$-th divergence $(\partial\!\cdot\!\partial_v)^m$ acting on $g^{(n)}$ is given by
\begin{eqnarray}\label{div_on_g}
&&\!\!\!\!\left(\partial\!\cdot\!\partial_v \right)^m g^{(n)}\\
&&\!\!\!\!= -\,b_{n}\!\left[
    \left(1 - \frac{b_{n-1}}{b_{d-m-1}} \right)\lap
    + b_{n-1}\,(d+2s-4-2m)(v \! \cdot\! \partial)(\partial\!\cdot\!\partial_v)
    - b_{n-1}\abs{v}^2(\partial\!\cdot\!\partial_v)^2
\right] \!(\partial \cdot \partial_v)^m \,. \nn
\end{eqnarray} 
For example, the $s$-th divergence of spin-$s$ fields reads
\be \label{div_on_g_s}
\left(\partial\!\cdot\!\partial_v \right)^s g^{(n)} \Psi_s= -\,b_{n}
    \left(1 - \frac{b_{n-1}}{b_{d-s-1}} 
\right)  \lap(\partial \cdot \partial_v)^s \Psi_s \,.
\ee
Since the divergence operator $\left(\partial \cdot \partial_v \right)$ commutes with the trace $(\partial_v \cdot \partial_v)$, one can write similarly to \eqref{g_mul_anzatz}
\be \label{m-div-g}
\left(\partial \cdot \partial_v \right)^m \prod_{i=1}^k g^{(n_i)} \Psi_s = \sum_{l=0}^{k} P_{s-m} c^m_l(n_1,\dots,n_k) (v\cdot \partial)^l(\partial\!\cdot\!\partial_v)^{m+l} \Psi_s \,.
\ee
Now taking one more divergence in \eqref{m-div-g}
\be
\begin{split}
&\left(\partial \cdot \partial_v \right)^{m+1} \prod_{i=1}^k g^{(n_i)} \Psi_s  = \sum_{l=0}^{k} P_{s-m-1} c^{m+1}_l(n_1,\dots,n_k) (v\cdot \partial)^l(\partial\!\cdot\!\partial_v)^{m+l+1} \Psi_s \\&= \sum_{l=0}^{k}  c^m_l(n_1,\dots,n_k) \frac{P_{s-m-1} (v \cdot \partial)^{l-1}}{d + 2s -2m - 4} \Bigg\{ (d + 2s -2m - 4 -2l)   (v \cdot \partial) (\partial \cdot \partial_v) \\ & \hspace{180pt} + l(d + 2s -2m - 3 -l)  \lap \Bigg\} (\partial \cdot \partial_v)^{l+m} \Psi_{s}  \,.
\end{split}
\ee
Here, the second equality follows from acting with the divergence operator from the left on \eqref{m-div-g} and using \textit{Identity 3} \eqref{Identity_3}. Comparing coefficients we get the recursion 
\be\label{clmrecursion}
\frac{(l+1)(d+2s-2m-4-l)}{d+2s-2m-4}c_{l+1}^m \lap = c^{m+1}_l -  \frac{(d+2s-2m-4-2l)}{d+2s-2m-4}c_l^m \,,
\ee
with the convention $c_{k+1}^m (n_1, \dots, n_k) = 0$. We can apply it to the mirror relation, which is a product of the operator $g^{(n)}$
\be
\left(\partial \cdot \partial_v\right)^m \Phi^{\left(\frac{d}{2} - s +k\right)} = R^{2k} \left(\partial \cdot \partial_v\right)^m \left(\prod_{l=0}^{k-1} g^{\left(\frac{d}{2} - s +k-2l\right)} \right)\Phi^{\left(\frac{d}{2} - s -k\right)} \quad \text{for } 1 \leqslant k \leqslant \frac{d}{2} - 2 + s \,.
\ee
Looking at \eqref{div_on_g} and proceeding by recursion, we can explicitly write what is the term which contains no gradient in \eqref{m-div-g}, i.e. the one for $l=0$ 
\be\label{cm0}
c^m_0  = R^{2k}\prod_{ l=0}^{k-1} \left[ -b_{\frac{d}{2} - s +k-2l}\left( 1 - \frac{b_{\frac{d}{2}-s+k-2l-1} }{b_{d-m-1}} \right) \lap   \right] = R^{2k}\frac{\left( \frac{d}{2} + s - m - k - 1 \right)_{2k}}{\left[ \left( \frac{d}{2} + s - k - 1 \right)_{2k} \right]^2} \lap^k \,.
\ee
Note that this is also the term with the highest power of the Laplacian. Given this initial condition, we can now solve completely the recursion on the $c_l^m$, i.e. from \eqref{clmrecursion} and \eqref{cm0} one gets 
\be \label{m_div_coef}
c_l^m = (-2)^l \binom{k}{l} R^{2k} \frac{\left( \frac{d}{2} + s - m - k - 1 \right)_{2k-l}}{\left[ \left( \frac{d}{2} + s - k - 1 \right)_{2k} \right]^2} \lap^{k-l} \,,
\ee
which readily respects the convention that $c_{l > k}^m = 0$. This vanishes in the range
\be
   c_l^m = 0 \quad \textrm{if} \quad  \frac{d}{2} + s - m - 1 \leqslant k \leqslant \frac{d}{2} + s - 2 \,.
\ee
We can then get a more explicit form of the mirror relation 
\be
    \Phi^{\left(\frac{d}{2} - s +k\right)} = R^{2k} \left[ \sum_{l=0}^{k}(-2)^l \binom{k}{l} \frac{\left( \frac{d}{2} + s - k - 1 \right)_{2k-l}}{\left[ \left( \frac{d}{2} + s - k - 1 \right)_{2k} \right]^2} \lap^{k-l} P_s (v \cdot \partial )^l (\partial \cdot \partial_v)^l \right] \Phi^{\left(\frac{d}{2} - s -k\right)} \,,
\ee
for $1 \leqslant k \leqslant \frac{d}{2} - 2 + s$.

\subsection{Gauge symmetry} \label{sec:gauge_symmetry}
We will prove here that the fields $\Phi^{(1-2s+m ,\, 0)}$ possess a gauge symmetry of the form
\be \label{deltaPhi}
    \delta \Phi^{(1-2s+m ,\, 0)} = P_s(v\!\cdot\!\partial)^m \xi_{s-m} 
    = P_s(v\!\cdot\!\partial)^m (\partial\!\cdot\!\partial_v)^m\lambda_{s}^m \, 
    \quad \text{for } 1 \leqslant m \leqslant s \, .
\ee
where $\lambda_{s}^m$ is an arbitrary spin $s$ tensor, which is not unique, for a given $\xi_{s-m}$. Since $\lambda_s^m$ is traceless, it satisfies the commutator identity
\be \label{commut_iden_1}
    \left[g^{(n)}, P_s(v\cdot \partial)^k(\partial\!\cdot\!\partial_v)^k \right] \lambda_s^m = 0 \,.
\ee
A proof of the vanishing of this commutator is given in appendix~\ref{sec:identities}. Using this identity and using \eqref{m_div_coef} evaluated at $k = \frac{d}{2} +s -m -1$, we see
\begin{align}
    &P_s(v \cdot \partial)^m (\partial \cdot \partial_v)^m\left(\prod_{l=0}^{\frac{d}{2} + s - m - 2} g^{\left(d - m - 1 - 2l\right)} \right)\lambda_{s}^m  =0 
\end{align}
because the operators entering \eqref{deltaPhi} annihilate the terms within parentheses. This proves that eq.~\eqref{mirror_const_HS_m} holds and thus that the fields $\Phi^{(1-2s+m,\,0)}$ possess a gauge symmetry.

\subsection{Absence of source constraints}\label{sec:source_constraints}
We now show that the source $\Phi^{(2-2s)}$ is entirely unconstrained by the equations of motion \eqref{eom_constraint_HS} and \eqref{eom_HS}. As before, we handle the transverse indices by contracting them with auxiliary variables $v^{i_1} \cdots v^{i_k}$. Following this contraction, the equations of motion reduce to
\be \label{eom_1_HSappendix}
\mathcal{F}^{(n)}_{z u(s-k-1)} = - (n+2k)(n-d+1) \Phi^{(n)}_{u(s-k)} - \frac{n+2k-1}{k+1}\, \partial\!\cdot\!\partial_v \Phi^{(n-1)}_{u(s-k-1)} \,,
\ee
and
\begin{align} \label{eom_2_HSappendix}
&\quad\mathcal{F}^{(n)}_{u(s-k)} = \frac{1}{R^2}\Big[\big(n-d+1\big)\big( n-(s-k)(s-k-1)+2(s-1) \big) \Big] \Phi^{(n)}_{u(s-k)} \nn \\
&\quad+ k(k-1)(n-d+1) \abs{v}^2 \Phi^{(n)}_{u(s-k+2)}  - \Big[ (s-k-2)(n-d+1)-(2s+d-4) \Big] \partial_u \Phi^{(n-1)}_{u(s-k)} \nn\\
&\quad - \frac{(s-k)(s-k-1)}{R^2 (k+1)} \partial\!\cdot\!\partial_v \Phi^{(n-1)}_{u(s-k-1)}  - k(n-d+1) v\cdot \partial \Phi^{(n-1)}_{u(s-k+1)}  + k \abs{v}^2 \partial\!\cdot\!\partial_v \Phi^{(n-1)}_{u(s-k+1)} \nn\\
& \quad+ \lap \Phi^{(n-2)}_{u(s-k)} - (v\cdot \partial) (\partial\!\cdot\!\partial_v) \Phi^{(n-2)}_{u(s-k)}  - \frac{s-k}{k+1} \partial_u (\partial\!\cdot\!\partial_v) \Phi^{(n-2)}_{u(s-k-1)} \,.
\end{align}
Among these, the only independent equations are \eqref{eom_1_HSappendix} for all $k$ and \eqref{eom_2_HSappendix} at $k = s$. Observe that for $k = s$, equation \eqref{eom_2_HSappendix} uniquely determines $\Phi^{(n)}$, except at the critical orders $n = 2-2s$ and $n = d-1$. Because all modes below the order of the source ($n < 2-2s$) are set to zero, $\mathcal{F}^{(2-2s)}$ vanishes identically. It just remains to be verified that the order $n = d-1$ imposes no further constraints on the source.

Analogously, equation \eqref{eom_1_HSappendix} leaves the components $\Phi^{(n)}_{u(s-k)}$ undetermined only at the specific orders $n = -2k$ and $n = d-1$. Evaluating the relation at $n = -2k+1$ yields
\be \label{not_const}
    \Phi^{(-2k+1)}_{u(s-k)} = 0 \,.
\ee
Subsequently, evaluating the equation at $n = -2k$ leads to the condition
\be
    \frac{1}{k+1} \partial \cdot \partial_v \Phi^{(-2k-1)}_{u(s-k-1)} = 0 \,,
\ee
which is trivially satisfied due to \eqref{not_const}. Evaluating \eqref{eom_1_HSappendix} at order $n = d-1$ yields
\begin{align}
\mathcal{F}^{(d-1)}_{z u(s-k-1)} &= - \frac{d+2k-2}{k+1} \left(\partial\!\cdot\!\partial_v\right) \Phi^{(d-2)}_{u(s-k-1)} = 0 \,.
\end{align}
Naively, this equation seems to constrain the source. If we recursively use \eqref{eom_1_HSappendix} to solve for this in terms of the purely transverse field components, we find that for $0 \leqslant k \leqslant s-1$
\be \label{HS_appendix_const1}
\mathcal{F}^{(d-1)}_{z u(s-k-1)} = - \frac{k! \, (d+2k-2)(d+2k-1)}{s! \, (s-k-1)! \, (d+k+s-2)} \left(\partial\!\cdot\!\partial_v\right)^{s-k} \Phi^{(d+k-s-1)}= 0 \,.
\ee
Eq.~\eqref{conserved_current_HS} shows that this is identically zero and thus imposes no constraints on the source. Since in \eqref{eom_2_HSappendix} the only non-redundant equation is at $k = s$, evaluating it at order $n = d-1$ yields
\be
\mathcal{F}^{(d-1)} =  (d +2s-4) \partial_u \Phi^{(d-2)}  + \lap \Phi^{(d-3)} - (v\cdot \partial) (\partial\!\cdot\!\partial_v)  \Phi^{(d-3)} = 0 \, .
\ee
Adding \eqref{HS_appendix_const1} at $k = s-2$ which reads $(\partial\!\cdot\!\partial_v)^2  \Phi^{(d-3)} = 0$, so as to force the appearance of a projector, we see that the previous equation is equivalent to
\be
(d +2s-4) \partial_u \Phi^{(d-2)}  + \lap \Phi^{(d-3)} - (v\cdot \partial) (\partial\!\cdot\!\partial_v)  \Phi^{(d-3)} + \frac{\abs{v}^2}{d+2s-4}  (\partial\!\cdot\!\partial_v)^2 \Phi^{(d-3)} = 0 \,.
\ee
Then using the mirror relation twice
\begin{eqnarray}
\Phi^{(d-2)} &=& R^{d+2s-4}\left( \prod_{k=1}^{\frac{d}{2}+s-2} g^{(d-2k)} \right) \Phi^{(2-2s)} \,,\\
\Phi^{(d-3)} &=& R^{d+2s-4} \left( \prod_{k=1}^{\frac{d}{2}+s-3} g^{(d-1-2k)} \right) f^{(3-2s)} \Phi^{(2-2s)} \,,
\end{eqnarray}
the equation reduces to
\be \label{provetozero}
\left[  \prod_{k=1}^{\frac{d}{2}+s-2} g^{(d-2k)} - \frac{1}{d + 2s - 4}\mathbf{B}_{3-2s}  \prod_{k=1}^{\frac{d}{2}+s-3} g^{(d-1-2k)}  \right] \partial_u \Phi^{(2-2s)} = 0 \,,
\ee
where we have used the notation of \eqref{def_g_appendix}, which we recall here
\be \label{def_g_appendix_2}
g^{(n)} = -\,b_{n} \mathbf{B}_{n-1} \,,\quad \text{with} \quad
\mathbf{B}_{n} = 
    \lap + \,b_{n}\,(d+2s-4)\,(v\!\cdot\!\partial)\,(\partial\!\cdot\!\partial_v)
    - \,b_{n}\,\abs{v}^2 (\partial\!\cdot\!\partial_v)^2 \,.
\ee
Again, naively, this equation seems to constrain the source. However, we can rewrite the second term in the bracket of \eqref{provetozero} by relabelling the product index $k \to \frac{d}{2} +s -2 -k$:
\be
    \frac{1}{d + 2s - 4}\mathbf{B}_{3-2s}  \prod_{k=1}^{\frac{d}{2}+s-3} g^{(d-1-2k)} = -b_{d-2}\mathbf{B}_{3-2s}  \prod_{k=1}^{\frac{d}{2}+s-3} \left(-b_{(3 - 2s + 2k)} \mathbf{B}_{(2-2s+2k)} \right) . 
\ee
Using the reflection symmetries $\mathbf{B}_{n} = \mathbf{B}_{d+1-2s-n}$ and $b_{n} = b_{d+1-2s-n}$, we obtain that the right-hand side of the previous equation is
\begin{align}
&-b_{d-2}\mathbf{B}_{3-2s}  \prod_{k=1}^{\frac{d}{2}+s-3} \left(-b_{(d-2-2k)} \right)  \prod_{k=1}^{\frac{d}{2}+s-3}  \mathbf{B}_{(d-1-2k)} \nonumber \\
&\quad =\prod_{k=0}^{\frac{d}{2}+s-3} \left(-b_{(d-2-2k)} \right)  \prod_{k=1}^{\frac{d}{2}+s-2}  \mathbf{B}_{(d-1-2k)} =  \prod_{k=1}^{\frac{d}{2}+s-2} g^{(d-2k)} \,.
\end{align}
In the second equality, we have absorbed the $-b_{d-2}$ and $\mathbf{B}_{3-2s}$ factors into the $k=0$ and $k=\frac{d}{2}+s-2$ boundaries of their respective products. In the final equality, shifting the index of the first product by $k \to k-1$ perfectly aligns the indices to reconstruct the definition of $g^{(d-2k)}$. This confirms that the bracket in \eqref{provetozero} vanishes identically, leaving the source unconstrained.

%%%%%%%%%%%%%%%%%%%%%%%%%%%%%%%%%%%%%%%%%%%%%%%%%%%%%%%
\section{Branching rules of generalised Verma modules}\label{Vermabranch}

The branching rules apply for any complexification, so they can be easily adapted to various signatures. For convenience, they will be expressed in the conformal signature. 

We will work in full generality so one needs to introduce some further notation. In particular, we will have to be more careful in this appendix and distinguish explicitly the modules and their characters;
The finite-dimensional irreducible $\mathfrak{so}(D)$-module labelled
by the dominant integral $\mathfrak{so}(D)$-weight  
$\vec s\equiv(s_1,\ldots,
s_r)$ will be denoted by $\cD_{\mathfrak{so}(D)}(\vec s)$.  Here $r$ denotes the rank of $\mathfrak{so}(D)$ 
(i.e. the integer part of $D/2$). The ``spin'' labels of the weight $r$-vector $\vec s$ are either all integers or all half-integers, and they satisfy
\begin{eqnarray}
&&s_1 \geqslant \ldots \geqslant s_r \geqslant  0\quad \mbox{for}\quad D = 2r +1\;,
\nonumber \\
&&s_1 \geqslant \ldots \geqslant s_{r-1} \geqslant \vert s_r \vert \quad \mbox{for}\quad D = 2r\;.
\end{eqnarray}
When $D=2r$, the last label $s_{r}$ can be positive or negative.

\begin{proposition}\label{branchingVermao(d,2)}
The branching rule of generalised Verma $\mathfrak{so}(D,2)$-modules for their restriction to $\mathfrak{so}(D-1,2)$ reads as
\be
{\cal V}_{\mathfrak{so}(D,2)}(\Delta,\vec s)\,\,\downarrow\,\,\bigoplus\limits_{\vec t
}\bigoplus\limits_{n=0}^\infty{\cal V}_{\mathfrak{so}(D-1,2)}(\Delta+n,\vec t)\,,
\ee
where the direct sum over the $\mathfrak{so}(D-1)$-weights $\vec t$ is identical to one in the classical branching rule
for the restriction $\mathfrak{so}(D)\downarrow \mathfrak{so}(D-1)$.
\end{proposition}
The formula \eqref{lemma} is a particular case of the above proposition (adapted to the Lorentzian signature) where one applies the branching rule of spin-$s$ irreducible representations of the (pseudo-)orthogonal algebras: 
\be
    \cD_{\mathfrak{so}(D)}(s)\,\downarrow\,\bigoplus\limits_{t=0}^s\cD_{\mathfrak{so}(D-1)}(t)\,.
\ee

The general proposition \ref{branchingVermao(d,2)} may be well-known to experts in representation theory but we derive it here by a direct computation of characters. 
The character of the generalised Verma $\mathfrak{so}(D,2)$-module ${\cal V}_{\mathfrak{so}(D,2)}(\Delta,\vec s)$
is equal to (see e.g. \cite{Dolan:2005wy,Basile:2014wua} for details)
\be
\mathcal{A}^{(D,2)}_{[\Delta,\vec{s}\,]} (q,x_1,\ldots,x_r) \,=\, q^{\Delta}\, 
\chi^{(D)}_{\vec{s}\,} (x_1,\ldots,x_r)\,{\cal P}^{(D)}(q,x_1,\ldots,x_r)\;,
\label{Ad,2}
\ee
where $\chi^{(D)}_{\vec{s}\,} (x_1,\ldots,x_r)$ denotes the character of the finite-dimensional irreducible $\mathfrak{so}(D)$-module $V_{\mathfrak{so}(D)}(\vec s)$
and the character
\be 
{\cal P}^{(D)}(q,x_1,\ldots,x_r)= \mathcal{A}^{(D,2)}_{[0,\vec{0}\,]} (q,x_1,\ldots,x_r)=\sum\limits_{m,n=0}^\infty
q^{2m+n}\chi^{(D)}_{(n,0,\ldots,0)\,} (x_1,\ldots,x_r)
\label{AFD}
\ee 
encodes the contribution from all the raising generators.
The classical branching rule  $\mathfrak{so}(D)\downarrow \mathfrak{so}(D-1)$ for finite-dimensional irreducible representations translates in terms of characters
\begin{eqnarray}
\chi^{(D)}_{\vec{s}\,} (x_1,\ldots,x_r)=\sum\limits_{\vec t}\chi^{(D-1)}_{\vec{t}\,} (x_1,\ldots,x_r)
\label{brchi}
\end{eqnarray}
where $x_r$ must be set to $1$ when $D=2r$ and the sum over $\vec t$ is such that 
\begin{alignat}{5}
&s_1 \geqslant t_1 \geqslant\ldots\geqslant s_{r-1} \geqslant t_{r-1} \geqslant s_r \geqslant |t_r| & \quad & \mbox{for}\quad D = 2r +1\;,
\label{br1} \\
&s_1 \geqslant t_1 \geqslant\ldots\geqslant s_{r-1} \geqslant t_{r-1} \geqslant |s_r| & \quad & \mbox{for}\quad D = 2r\;,\label{br2}
\end{alignat}
with entries in $\vec s$ and $\vec t$ which are simultaneously all integers or all half-integers.
In the particular case of a Young diagram made of a single row, one finds
\be 
\chi^{(D)}_{(s,0,\ldots,0)\,} (x_1,\ldots,x_r)=\sum\limits_{t=0}^s\chi^{(D-1)}_{(t,0,\ldots,0)\,} (x_1,\ldots,x_r)
\label{chisym}
\ee 
Combining everything, one eventually is able to prove Proposition \ref{branchingVermao(d,2)}
\begin{eqnarray}
&&\mathcal{A}^{(D,2)}_{[\Delta,\vec{s}\,]} (q,x_1,\ldots,x_r)\,=\, \nonumber\\
&&= q^{\Delta}\,\chi^{(D)}_{\vec{s}\,} (x_1,\ldots,x_r)\,{\cal P}^{(D)}(q,x_1,\ldots,x_r)
\label{char1}\\
&&=\sum\limits_{m,n=0}^\infty q^{\Delta+2m+n}\,\chi^{(D)}_{\vec{s}\,}(x_1,\ldots,x_r)\,\chi^{(D)}_{(n,0,\ldots,0)\,} (x_1,\ldots,x_r)
\label{char2}\\
&&=\sum\limits_{m,n=0}^\infty q^{\Delta+2m+n}\sum\limits_{\vec t}\,\chi^{(D-1)}_{\vec{t}\,} (x_1,\ldots,x_r)\,\sum\limits_{p=0}^n\chi^{(D-1)}_{(p,0,\ldots,0)\,} (x_1,\ldots,x_r)
\label{char3}\\
&&=\sum\limits_{\vec t}\sum\limits_{m,\overline n,p=0}^\infty q^{\Delta+2m+\overline n+p}\,\chi^{(D-1)}_{\vec{t}\,} (x_1,\ldots,x_r)\,\chi^{(D-1)}_{(p,0,\ldots,0)\,} (x_1,\ldots,x_r)
\label{char4}\\
&&=\bigoplus\limits_{\vec t
}\bigoplus\limits_{\overline n=0}^\infty\mathcal{A}^{(D-1,2)}_{[\Delta+\overline n,\vec{t}\,]}(q,x_1,\ldots,x_r)
\label{char5}
\end{eqnarray}
where we respectively used \eqref{Ad,2} in \eqref{char1}, \eqref{AFD} in \eqref{char2}, \eqref{brchi}-\eqref{chisym} in \eqref{char3},
$\overline n:=n-p\geqslant 0$ in \eqref{char4}, and \eqref{Ad,2} in \eqref{char5}.

\section{Carrollian limit of conformal modules}
\label{Lemma 1}

The proof of the first technical lemma in section~\ref{Carrollianlimitoftheconformalscalar} goes as follows.

\paragraph{Proof of Lemma \ref{Lemma_1}:} A quick examination shows that for any $|\Delta+n+1\rangle$ as in~\eqref{eq: recursion scalar}, the properties $J_{ij} |\Delta + n +1 \rangle = 0$ and $D |\Delta + n + 1\rangle = (\Delta+n+1) |\Delta + n + 1\rangle$ are satisfied by immediate recursion. In order to verify the other properties, we assume that they are satisfied for $|\Delta+n\rangle$ and $|\Delta+n-1\rangle$. Then, the conditions $K_i |\Delta + n+1 \rangle = 0$, $B_i |\Delta + n + 1\rangle = \beta_{n+1} P_i |\Delta+n\rangle$ and $K|\Delta+n+1\rangle = \kappa_{n+1} |\Delta +n\rangle$ lead to recursion relations for $\alpha_n$, $\beta_n$ and $\kappa_n$ as well as consistency relations. These are all solved for
\be
    \alpha_n = \frac{\beta_n}{2\Delta+2n-d} \,,\quad\beta_n = n \frac{2\Delta+n-d-1}{2\Delta+2n-d-2} \,,\quad  \kappa_n = -2 (\Delta+n-1)\beta_n \,.
\ee
This concludes the proof. $\hfill \blacksquare$

One might have taken a different Carrollian limit of the conformal module in section~\ref{Carrollianlimitoftheconformalscalar}, without rescaling of the descendants. Let us quickly look at this alternative choice to briefly discuss its pro and contra.
\paragraph{Electric limit of a scalar primary.} A distinct Carrollian limit can also be considered where one does not include the factors $1/c^2$ in the definitions \eqref{eq: recursion scalar} of the scalar descendants. In this case, one would have had the recursive definition
\be
    |\Delta + n +1 \rangle :=  H |\Delta+n\rangle - \alpha_n P_i P^i |\Delta+n-1\rangle\,,
\ee
while the conditions \eqref{so(d+1,1)-primary}-\eqref{so(d+1,2)-notprimary} would imply, in the corresponding Carrollian limit, that each scalar descendant is actually a Carrollian primary, i.e. it is a celestial primary for $\mathfrak{so}(d+1,1)$ which is also annihilated by the special conformal generator $K$. Furthermore, it is also annihilated by the Carrollian boost generators $B_i$. More explicitly, one has
\be
K |\Delta+n\,\rangle = 0 \,,\;    K_i |\Delta+n\,\rangle = 0 \,,\; D |\Delta+n\rangle = (\Delta+n) |\Delta+n\rangle \,,\; B_i |\Delta+n\,\rangle = 0\,,\; J_{ij} |\Delta+n\rangle =0 \,.
\ee
These conditions state that all vectors $|\Delta+n\,\rangle$ are \textit{Carrollian primaries}. So the electric Carrollian limit is well suited for Carrollian CFT (see, e.g., \cite{Bagchi:2019xfx}). 
Note that these conditions imply that the corresponding representation of $\mathfrak{iso}(d+1,1)$ spanned by all possible descendants $P_{i_1}\cdots P_{i_k}|\Delta+n\,\rangle$ is indecomposable, hence it cannot be unitary.

\paragraph{Proof of Lemma \ref{Lemma 2}:} The definition of $|d+1\rangle^i$ is given above. A quick examination shows that for any $|d+n+1\rangle^i$ as in \eqref{eq: recursion spin 1}, the conditions $J_{ij} |d+n+1\rangle^k = 2 \delta^k_{[j} |d+n+1\rangle_{i]}$ and $D |d+n+1\rangle^i = (d+n+1) |d+n+1\rangle^i$ are satisfied by immediate recursion. In order to verify the other properties, we assume that they hold for all $|d+k\rangle^i$, $k \leqslant n$. Then the conditions $K_i |d+n+1\rangle^i = 0$ and \eqref{eq: conditions recursion spin 1} lead to recursion relations for $\alpha^{(1)}_n$, $\alpha'^{(1)}_n$, $\beta^{(1)}_n$, $\beta'^{(1)}_n$, $\beta''^{(1)}_n$ and $\kappa^{(1)}_n$ as well as consistency relations. These are all solved for
\begin{subequations}
\begin{align}
    \alpha^{(1)}_n &= \frac{d-2}{(d+2n-2)(d+2n)} \,,& \alpha'^{(1)}_n &= n\frac{d+n-1}{(d+2n-2)(d+2n)} \,,\\
    \beta^{(1)}_n &= n \frac{d+n-1}{d+2n-2} \,,& \beta'^{(1)}_n &= - n \frac{1}{d+2n-2} \,,\\
    \beta''^{(1)}_n &= \frac{d+n-2}{d+2n-2} \,,& \kappa^{(1)}_n &= - 2n \frac{(d+n)(d+n-2)}{d+2n-2} \,,
\end{align}
\end{subequations}
for $n \geqslant 1$. This concludes the proof. $\hfill \blacksquare$

\paragraph{Proof of Lemma \ref{Lemma 3}:} The definition of $|d+2\rangle^{ij}$ is given above. A quick examination shows that for any $|d+2+n\rangle^{ij}$ as in \eqref{eq: recursion spin 2}, the conditions $J_{ij} |d+2+n\rangle^{kl} = 2 \delta^k_{[j} |d+2+n\rangle_{i]}{}^l + 2 \delta^l_{[j} |d+2+n\rangle_{i]}{}^k$ and $D |d+2+n\rangle^{ij} = (d+2+n) |d+2+n\rangle^{ij}$ are satisfied by immediate recursion. In order to verify the other properties, we assume that they hold for all $|d+1+k\rangle^{ij}$, $k \leqslant n$. Then the conditions $K_i |d+2+n\rangle^{ij} = 0$ and \eqref{eq: conditions recursion spin 2} lead to recursion relations for $\alpha^{(2)}_n$, $\alpha'^{(2)}_n$, $\beta^{(2)}_n$, $\beta'^{(2)}_n$, $\beta''^{(2)}_n$ and $\kappa^{(2)}_n$ as well as consistency relations. These are all solved for
\begin{subequations}
\begin{align}
    \alpha^{(2)}_n &= \frac{2 d}{(d+2 n) (d+2 n+2)} \,,& \alpha'^{(2)}_n &= n\frac{d+n+1}{(d+2 n) (d+2 n+2)} \,,\\
    \beta^{(2)}_n &= n \frac{d+n+1}{d+2 n} \,,& \beta'^{(2)}_n &= -n\frac{2}{d+2 n} \,,\\
    \beta''^{(2)}_n &= 2\frac{d+n}{d+2n} \,,& \kappa^{(2)}_n &= -2n\frac{(d+n-1) (d+n+2)}{d+2 n} \,,
\end{align}
\end{subequations}
for $n \geqslant 1$. This concludes the proof. $\hfill \blacksquare$

%%%%%%%%%%%%%%%%%%%%%%%%%%%%%%%%%%%%%%%%%%%%%%%%%%%%%%%

\bibliographystyle{JHEP}
\bibliography{bibliography}

\end{document}